\documentclass{aa}
\usepackage{natbib}
\usepackage{color}
\usepackage{ragged2e}
\usepackage[draft]{hyperref}
\usepackage[varg]{txfonts}
\usepackage{graphicx,rotating}
\usepackage[normalem]{ulem}
\definecolor{darkblue}{rgb}{0.0,0.0,0.8}
\definecolor{darkred}{rgb}{0.8,0.0,0.0}
\definecolor{darkorange}{rgb}{1,0.4,0}
\definecolor{darkgreen}{rgb}{0.0,0.5,0.0}
\definecolor{brown}{rgb}{0.65,.16,0.16}
\definecolor{grey}{rgb}{0.4,0.5,0.6}
\definecolor{trolleygrey}{rgb}{0.5, 0.5, 0.5}

\newcommand{\darkg}[1]{\textcolor{trolleygrey}{#1}}

\newcommand{\darkbl}[1]{\textcolor{darkblue}{#1}}
\newcommand{\darkr}[1]{\textcolor{darkred}{#1}}

\newcommand{\darkor}[1]{\textcolor{darkorange}{#1}}
\newcommand*\diff{\mathop{}\!\mathrm{d}}
\renewcommand{\d}{{\rm d}}
\newcommand{\mpo}{{\sc MAMPOSSt-PM}}
\newcommand{\ngc}{NGC~6397}
\newcommand{\msun}{\rm M_\odot}
\newcommand{\kms}{\rm km\,s^{-1}}
\newcommand{\IMBHisoSingle}{1}
\newcommand{\IMBHfreebetarbetaSingle}{2}
\newcommand{\IMBHfreebetaSingle}{3}
\newcommand{\IMBHfreebetainSingle}{4}
\newcommand{\IMBHfreebetaoutSingle}{5}
\newcommand{\IMBHisoDouble}{6}
\newcommand{\NoneisoSingle}{7}
\newcommand{\NoneisoDouble}{8}
\newcommand{\IMBHCUOisoSingle}{9}
\newcommand{\CUOisoSingle}{10}
\newcommand{\CUOHernquistisoSingle}{11}
\newcommand{\CUOSersicisoSingle}{12}
\newcommand{\CUOfreebetaSingle}{13}
\newcommand{\CUOisoDouble}{14}
\newcommand{\CUOfreebetaDouble}{15}
\bibpunct{(}{)}{;}{a}{}{,} 

\begin{document}

\title{Does NGC~6397 contain an intermediate-mass black hole or a more diffuse inner subcluster?}
\titlerunning{Does NGC~6397 contain an IMBH or a more diffuse inner subcluster?}
\author{Eduardo Vitral\inst{1}
\and Gary A. Mamon\inst{1}
}
\offprints{Eduardo Vitral, \email{vitral@iap.fr}}
\institute{Institut d’Astrophysique de Paris (UMR 7095: CNRS \& Sorbonne Université), 98 bis Bd Arago, F-75014 Paris, France}

\authorrunning{Vitral \& Mamon}
\date{Accepted}

\abstract{ 
We analyze proper motions from the Hubble Space Telescope (HST) and the second {\sc Gaia} data release along with line-of-sight velocities from the MUSE spectrograph to detect imprints of an intermediate-mass black hole (IMBH) in the center of the nearby, core-collapsed, globular cluster NGC~6397.
For this, we use the new \mpo\ Bayesian mass-modeling code, along with updated estimates of the surface density profile of \ngc. We consider different priors on velocity anisotropy and on the size of the central mass, and we also separate the stars into components of different mean mass to allow for mass segregation. 
The velocity ellipsoid is very isotropic throughout the cluster, as expected in post-core collapsed clusters subject to as strong a Galactic tidal field as \ngc.
There is strong evidence for 
a central dark component of 0.8 to 2\% of the total mass of the cluster.
However, we find robust evidence disfavoring a central IMBH in NGC~6397, preferring instead a diffuse dark inner subcluster of unresolved objects 
with a total mass of 1000 to $2000\,\msun$, half of which is  concentrated within 6 arcsec (2\% of the stellar effective radius).
These results require the combination of HST and {\sc Gaia} data: HST for the inner diagnostics and {\sc Gaia} for the outer surface density and velocity anisotropy profiles.
The small effective radius of the diffuse dark component suggests that it is composed of compact stars (white dwarfs and neutron stars) and stellar-mass black holes, whose inner locations are caused by dynamical friction given their high progenitor masses.
We show that stellar-mass black holes should dominate the mass of this diffuse dark component, unless more than 25 per cent escape from the cluster.
Their mergers in the cores of core-collapsed globular clusters could be an important source of 
the gravitational wave events detected by LIGO.
}

\maketitle

\section{Introduction}
\label{sec: intro}

When the \textsc{Laser Interferometer Gravitational-Wave Observatory} (LIGO) first detected gravitational waves coming from a stellar-mass black hole merger (\citealt{Abbott+16}) and then the \textsc{Event Horizon Telescope} (EHT) released the first image of the supermassive black hole (SMBH) in M87 (\citealt{EventHorizonTelescopeCollaboration+19}), astronomers obtained the most compelling evidence about the existence of those intriguing and particular objects. 
Yet, black holes (BHs) have been treated as more than a theoretical object for a considerable amount of time, starting in \citeyear{Oppenheimer&Snyder39}, when \citeauthor{Oppenheimer&Snyder39} proposed them to be the final step of the life of massive stars ($\gtrsim 10$ M$_\odot$) after their final gravitational collapse into stellar-mass black holes, with masses in between $\sim 3$ M$_\odot$ \citep{Thompson+20} and $\approx 52\,\msun$ \citep{Woosley17}, but also when \cite{Hoyle&Fowler63} identified the then recently discovered quasars \citep{Schmidt63} as SMBHs. This latter class of BHs, which reside in the centers of massive galaxies, are responsible for extremely luminous sources in the Universe, such as quasars and active galactic nuclei (AGN), sometimes unleashing powerful jets of relativistic matter, along with outflows that have a profound impact on star formation and galaxy evolution (e.g., \citealt{Croton+06,Hopkins+06}). In addition, 
some black holes may have formed during the early moments of the Universe (e.g. \citealt{Zeldovich&Novikov66,Hawking71}),  and these primordial BHs may constitute an import mass fraction of black holes below the SMBH mass.

Since there is no theoretical constraint to the mass of a black hole, it would be reasonable to believe that intermediate-mass black holes (IMBHs) could exist, filling the considerable gap between stellar-mass black holes and SMBHs (i.e., with masses between 100 and $10^5\,\msun$).
Furthermore, 
SMBHs are understood to 
grow by mergers, where the first seeds are stellar-mass BHs \citep{Madau&Rees01} or metal-free primordial gas clouds \citep{Loeb&Rasio94}, so IMBHs may be a transitory stage in the growth of BHs. 
However, there  is currently little evidence for IMBHs  (see reviews by \citealt{Volonteri10} and \citealt*{Greene+20}), with some important candidates highlighted (e.g., \citealt{Kaaret+01,Chilingarian+18} in dwarf galaxies, and recently \citealt{Lin+20} in a globular cluster) and one gravitational wave confirmation (\citealt{TheLIGOScientificCollaboration+20}). Furthermore, IMBHs could help to explain many enigmas in astrophysics, such as filling up part of the dark matter mass budget (e.g., \citealt{Haehnelt&Rees93} and \citealt{Loeb&Rasio94}) or providing massive seeds for high redshift quasars, whose high masses at such early times represent a challenge to current theories \citep{Haiman13}. Therefore, great efforts have been undertaken to detect IMBHs to better understand their origin and evolution.

{Globular clusters} (GCs) appear to be a unique laboratory to test the existence of IMBHs.
These quasi-spherical star clusters are known to have old stellar populations, indicating that they formed at early epochs. 
Their high stellar number densities provide an excellent environment to increase stellar interactions that could give birth to compact objects. More precisely, contrary to galaxies, the rates of stellar encounters in the inner parts of GCs containing half their stellar mass are sufficiently high to statistically affect the orbits of their stars by two-body relaxation \citep{Chandrasekhar42}.
Moreover, after several relaxation times, the interplay between the negative and positive heat capacities of the inner core and outer envelope causes transfers of energy. This in turn leads to the gravothermal catastrophe, where the core collapses resulting in a steep inner density profile, while the envelope expands. Roughly one-fifth of GCs are believed to have suffered such core-collapse \citep{Djorgovski&King86}.


Several scenarios have been proposed for the existence of IMBHs in GCs.
One is the direct collapse of population~III stars \citep{Madau&Rees01}, but the link of population~III stars with GCs is not clear. 
Another is the accretion of residual gas on stellar-mass BHs formed in the first generation of stars \citep{Leigh+13}, but the availability of the gas is unclear as the first massive stars will blow it out of the GC by supernova explosions.
Stellar mergers are a popular mechanism for IMBH formation.
\citet{PortegiesZwart&McMillan02} proposed a runaway path to IMBH formation in GCs, where an initially massive star suffers multiple  physical collisions with other stars during the first few Myr of the GC, before they have time to explode as supernovae or simply lose mass.
During these collisions, the most massive star will
lose linear momentum, ending up at the bottom of the gravitational potential well. At the same time, its mass will grow during the successive stellar mergers, to the point that it will end up as a BH, possibly reaching 0.1 per cent of the GC stellar mass.
%
\citet{Miller&Hamilton02} proposed a slower process for IMBH formation, where dynamical friction \citep{Chandrasekhar43} causes the most massive stellar remnant BHs to sink to the center of the gravitational well over Gyr. Thus  a  $\gtrsim 50 $ M$_\odot$ stellar remnant BH, sufficiently massive to avoid being ejected from the GC by dynamical interactions, would grow in mass through mergers with these other massive BHs as well as other typically massive stars, reaching a mass of 1000\,M$_\odot$ over the Hubble time, which they argued 
generates IMBHs in some ten per cent of GCs.
Finally, \cite{Giersz+15} proposed that hard binaries containing stellar-mass BHs merge with other stars and binaries, which can be a fast or slow process.

These models present, however, drawbacks: The short relaxation time needed in the \citeauthor{PortegiesZwart&McMillan02} scenario usually requires primordial mass segregation in order not to eliminate too many GCs candidates, while the assumption by \cite{Miller&Hamilton02} of BH seeds above $\approx 50\,\msun$ is not expected as the massive progenitors are fully exploded in pair-instability supernovae (e.g., \citealt{Woosley17}).

Unfortunately, attempts to detect IMBHs have been somewhat inconclusive: Dynamical modeling is still dependent on the assumptions concerning the confusion between the IMBH and a central subcluster of stellar remnants (e.g., \citealt{denBrok+14}, \citealt{Mann+19} and \citealt{Zocchi+19}).
Furthermore, these analyses usually rely on too few stars inside the sphere of influence of the IMBH, which can lead to false detections \citep{Aros+20}. Besides, searches for signs of accretion indicate no strong evidence for $> 1000$ M$_\odot$ black holes in galactic GCs (e.g., \citealt{Tremou+18}).

In this paper, we analyze the core-collapsed Milky Way globular cluster NGC~6397, and search for kinematic imprints of a central IMBH or {subcluster of unresolved objects} (hereafter, CUO). 
We use very precise and deep \textsc{Hubble Space Telescope} (HST) {proper motions} (PMs) 
from \cite{Bellini+14} and {line-of-sight} (LOS) velocities from the \textsc{Multi Unit Spectroscopic Explorer} (MUSE) instrument on the \textsc{Very Large Telescope}, to which we added 
 PMs from the
 {\sc Gaia} Data Release 2 (hereafter, {\sc Gaia} or {\sc Gaia DR2}).
 

\ngc\ has been broadly investigated in previous studies.
The PMs lead to {plane-of-sky} (POS) velocities with equal radial and tangential dispersions, leading to the appearance of isotropic orbits
 (\citealt{Heyl+12} and \citealt{Watkins+15a} with HST for the inner orbits, and \citealt{Jindal+19} with {\sc Gaia} for the outer orbits). But this projected velocity isotropy can hide three-dimensional (3D) variations of the anisotropy of the 3D velocity ellipsoid.  
\cite{Kamann+16} used the JAM mass-orbit modeling code \citep{Cappellari08} to fit the LOS velocities obtained with the MUSE spectrograph, and found that a $600\pm200$\,M$_\odot$ black hole best fits their data, agreeing with the radio emission upper limit of $610$\,M$_\odot$ for the same cluster provided in \cite{Tremou+18}.  On the other hand, numerical $N-$body modeling by \cite{Baumgardt17} strongly excluded the possibility of an IMBH in NGC~6397, not to say that much of the non-luminous mass measured in this cluster could actually be in the form of unresolved white dwarfs and low-mass stars \citep{Heggie&Hut96}. 

Many of the mass modeling studies performed with this cluster assumed or constrained its surface density parameters to values estimated long ago (e.g., \citealt*{Trager+95}), that may suffer from problems such as radial incompleteness, which is better addressed by missions such as HST and {\sc Gaia}. This lack of accuracy on the surface density can strongly impact the dynamical analysis of NGC~6397.

In the present study, we performed state-of-the-art mass-orbit modeling to analyze the presence of an IMBH or a CUO in the center of \ngc\ and better measure the radial variation of its velocity anisotropy.
Among the different popular mass/orbit modeling (e.g., \citealt*{Mamon+13}, \citealt{Watkins+13}, \citealt{Read&Steger17}, \citealt{Vasiliev19a}; see chap.~5 of \citealt{Courteau+14} for a review), we used an extension of the \textsc{MAMPOSSt}: Modeling Anisotropy and Mass Profiles of Observed Spherical Systems (\citealt{Mamon+13}) Bayesian code, which now takes into account both LOS velocities and PMs (\textsc{MAMPOSSt-PM}, Mamon \& Vitral in prep.).
\textsc{MAMPOSSt-PM} was found to perform extremely well in a code challenge using mock data for dwarf spheroidal galaxies, with many similarities to GCs \citep{Read+20}.

We profit from the versatility of \textsc{MAMPOSSt-PM} to fit not only cases with a single population of stars, but also with two mass populations, thus allowing for mass segregation. We also provide new fits to the surface density parameters of this cluster by jointly modeling HST and {\sc Gaia} data.

\section{Method}
\label{sec: Method}


\subsection{MAMPOSSt-PM}
\label{ssec: MPO-PM}

We summarize here the main aspects of \mpo\ (details in Mamon \& Vitral, in prep.).
\mpo\ is a Bayesian code that fits parametrized forms of the radial profiles of mass and velocity anisotropy to the distribution of observed tracers (here, GC stars) in two-, three- or four-dimensional {projected phase space} (PPS): Projected distances to the GC center (hereafter {projected radii}), combined with LOS velocities and/or PMs.
The mass profile is the sum of observed tracers, as well as a possible central BH, and additional dark components.
By working on discrete stars, \mpo\ can probe the inner regions better than methods where velocity or PM data are binned in radial intervals (e.g., \citealt{vanderMarel&Anderson&Anderson10}), and an analysis of dark matter on mock dwarf spheroidals has shown that the conclusions can depend on how the data are binned \citep{Richardson&Fairbairn14}. 


The probability density that a tracer at projected radius $R$ has a velocity vector $\bf v$ is
\begin{equation}
    p({\bf v}|R) 
    = {g(R,{\bf v}) \over \int g(R,{\bf v}) \,\d {\bf v}}
    = {g(R,{\bf v})\over \Sigma(R)} 
 \ ,
\label{pvofR1}
\end{equation}
where $\Sigma(R)$ is the surface density (SD), while
$g(R,{\bf v})$ is the density in 
PPS, such that
\begin{eqnarray}
\int\!\!\int 2 \pi R\,g(R,{\bf v})\,\d R\, \d{\bf v} &\!\!\!\!=\!\!\!\!&   \Delta N_{\rm
    p} \nonumber \\
&\!\!\!\!=\!\!\!\!&   N_{\rm p}(R_{\rm max})-N_{\rm p}(R_{\rm min})\,, 
\end{eqnarray} 
with $N_{\rm p}(R)=\int_0^R 2\pi\,R'\,\Sigma(R')\,{\rm d}R'$ being the projected number of points (e.g., stars in a star cluster) in a cylinder of radius $R$. The likelihood is then calculated as

\begin{equation}
    {\cal L}  = \prod_i p({\bf v}_i|R_i)\ ,
\end{equation}
where the $i$ indices represent individual points (stars).




The observed stars  
tend to be located in
in the GC. But, inevitably, some observed stars  will be {interlopers} (here {field stars}, FS) mostly foreground in our case, but possibly some background too. {\sc MAMPOSSt-PM} splits the PPS distribution into separate GC and FS components. Eq.~(\ref{pvofR1}) becomes
\begin{eqnarray}
     p({\bf v}|R) &\!\!\!\!=\!\!\!\!& {g_{\rm GC}(R,{\bf v}) + g_{\rm FS}(R,{\bf v})
    \over \Sigma_{\rm GC}(R) + \Sigma_{\rm FS}(R)}  \nonumber \\
     &\!\!\!\!=\!\!\!\!&
     {g_{\rm GC}(R,{\bf v}) + \Sigma_{\rm FS} \,f_{\rm LOS}(v_{\rm LOS})\,f_{\rm
    PM}({\bf PM}) / \eta^2
    \over \Sigma_{\rm GC}(R) + \Sigma_{\rm FS}}
  \ ,
  \label{eq: pilop}
\end{eqnarray}
%
where ${\bf PM}$ is the 
vector of PMs ($\mu_{\alpha,*}$, $\mu_{\delta}$),
$f_{\rm LOS}(v_{\rm LOS})$ and  $f_{\rm PM}({\bf PM})$ are the respective distribution functions of field star LOS velocities and PM vectors,\footnote{The functional form of the field star PM distribution function, $f_{\rm PM}({\bf PM})$, is provided in section~\ref{sssec: pm-filter}.} while
$\eta = 4.7405\,D$ is the conversion of PM in mas\,yr$^{-1}$ to POS velocity in $\rm km \, s^{-1}$ given the distance $D$ to the system, in kpc. The second equality of Eq.~(\ref{eq: pilop}) assumes that the FS surface density is independent of position.





The GC contribution to the PPS density is the mean local
GC {velocity distribution function} $h$, averaged along the LOS
\citep{Mamon+13,Read+20}:
\begin{equation}
  g_{\rm GC}(R,{\bf v}) = 2\,\int_R^\infty h({\bf v}|R,r)\, \nu(r)\, {r\over
    \sqrt{r^2-R^2}}\,\d r \ ,
  \label{eq: gsystem}
\end{equation}
where $r$ is the 3D distance to the system center,
while $\nu(r)$ is the 3D tracer density (here stellar number density) profile.

{\sc MAMPOSSt-PM}
assumes that the local velocity ellipsoid, in other words the local 3D {velocity distribution function} (VDF), is separable along the three spherical coordinates ($r$, $\theta$, $\phi$):
\begin{equation}
  h({\bf v}|R,r) = h(v_{r}|R,r) \, h(v_{\theta}|R,r) \, h(v_{\phi}|R,r) \ .
 \label{eq: hvsep}
\end{equation}
{\sc MAMPOSSt-PM}
 further assumes that the three one-dimensional local VDFs are Gaussian:
\begin{eqnarray} 
  h(v_i|R,r) &\!\!\!\!=\!\!\!\!& {1\over \sqrt{2 \pi
      \sigma_i^2}}\,\exp\left(-{v_i^2\over 2\,\sigma_i^2}\right) \ .
  \label{eq: hvi}
\end{eqnarray} 
This Gaussian assumption for the local VDFs is much better than the very popular Gaussian assumption for the LOS-integrated VDF, $p({\bf  v}|R)$
because velocity anisotropy affects the shape of $p({\bf  v}|R)$, in particular that of $p(v_{\rm LOS}|R)$ (\citealt{Merritt87}). By measuring the shape of the PPS, hence of $p({\bf  v}|R)$, \textsc{MAMPOSSt-PM} gets a good handle on velocity anisotropy (see below), and therefore on the mass profile because of the mass-anisotropy degeneracy \citep{Binney&Mamon82}. \mpo\ considers the measurement errors by adding them in quadrature to the velocity dispersion for the GC component.  

%
%

The velocity anisotropy profile $\beta$ at radius $r$ is defined as
\begin{equation}    \label{eq: anisotropy}
    \beta(r) = 1 - \displaystyle{\frac{\sigma_{\theta}^{2}(r) + \sigma_{\phi}^{2}(r)}{2 \,\sigma_{r}^{2}(r)}} \ ,
\end{equation}
where $\theta$ and $\phi$ are the tangential components of the coordinate system and $r$ is the radial component,
while $\sigma_{i}^2$ stands for the velocity dispersion of the component $i$ of the coordinate system.

When considering spherical symmetry in velocity space, the variances of the composite Gaussian VDF $h({\bf v}|(R,r)$ can be written as
\begin{subequations}
  \begin{align} 
  \sigma_{\rm LOS}^2 (R,r) &= \left[1-\beta(r)\left({R\over
      r}\right)^2\right]\,\sigma_r^2(r) \ , 
\label{eq: siglos}
\\
    \sigma_{\mathrm{POS}_R}^2(R,r) &= \left[1-\beta(r)+\beta(r)\left({R\over
      r}\right)^2\right]\,\sigma_r^2(r) \ ,
    \label{eq: sigPOSr}
\\
    \sigma_{\rm POSt}^2(R,r) &= \left[1-\beta(r)\right]\,\sigma_r^2(r) \ ,
    \label{eq: sigPOSt}
\end{align}
\label{eq: sigproj}
\end{subequations} 
\noindent $\!$where Eq.~(\ref{eq: siglos}) is from \cite{Binney&Mamon82}, while Eqs.~(\ref{eq: sigPOSr}) and (\ref{eq: sigPOSt}) are from \citet{Strigari+07}.
In Eqs.~(\ref{eq: sigproj}),
the radial velocity variance $\sigma_r^2(r)$ is obtained by solving the spherical stationary Jeans equation with no streaming motions (in particular, rotation):
\begin{equation} 
  {{\rm d}\left (\nu\sigma_r^2\right) \over {\rm d}r} + 2\,{\beta(r)\over
    r}\,\nu(r)\sigma_r^2(r) = -\nu(r) {G\,M(r)\over r^2} \ ,
\label{eq: jeans}
\end{equation}
where $M(r)$ is the total mass profile.
The distribution of interlopers in PPS is straightforward:
The spatial density is assumed to be uniform (hence $\Sigma_{\rm FS}$ does not depend on $R$ in eq.~[\ref{eq: pilop}]).
The distribution of LOS velocities, $f_{\rm LOS}(v_{\rm LOS})$ in Eq.~(\ref{eq: pilop}), is assumed Gaussian.
The two-dimensional distribution of PMs, $f_{\rm PM}({\bf PM})$ in Eq.~(\ref{eq: pilop}), is found to have wider wings than a Gaussian (see Sect.~\ref{sssec: pm-filter} and Appendix~\ref{app: PMpdf}).
The POS velocity errors are not added in quadrature to the velocity dispersions because the PM distribution of interlopers has wider tails (Appendix~\ref{app: conv-MPOPM}). Instead, \mpo\ uses an approximation (Mamon \& Vitral, in prep.) for the convolution of PM errors with the PM dispersions.

\subsection{Dark component}
\label{sssec: dark-matter}

{\sc MAMPOSSt-PM} allows for a diffuse dark component with a large choice of analytical density profiles. This dark component could represent {dark matter} or alternatively unseen stars.

Dark matter (DM) may dominate the outskirts of GCs, and perhaps also deeper inside.
Contrary to distant GCs \citep{Ibata+13}, the presence of substantial DM in the outskirts of \ngc\  is unlikely because this GC appears to have an orbit around the Milky Way that takes it from an apocenter of roughly 6.6 kpc to a pericenter of 2.9 kpc in 88 Myr \citep{GaiaHelmi+18}, so it should have suffered from dozens of previous episodes of tidal stripping, leaving little DM \citep{Mashchenko&Sills&Sills05}, if there was any in the first place. In particular, the orbital parameters of \cite{GaiaHelmi+18}, $Z=-0.48\pm0.01\,\rm kpc$ and $W = -127.9\pm3\,\rm km\,s^{-1}$, indicate that \ngc\ just passed through the Milky Way disk only $\approx 4$ Myr ago.\footnote{\ngc\ has just passed through the disk near its apocenter, and probably previously passed through the disk at pericenter, since most  GCs highlighted in Fig.~D.2 of \cite{GaiaHelmi+18} have inclined orbits relative to the Galactic disk, passing through the disk near apocenter as well as near pericenter. Also, the important contributions of the thick disk, bulge/spheroid and dark matter halo to the gravitational potential imply that the tidal effect of the Milky Way is strongest at pericenter.}

One may ask whether there is substantial
DM in the inner regions of GCs like \ngc.
 Using $N$-body simulations of isolated
{ultra-compact dwarf} galaxies, which appear to be larger analogs of GCs, but with  relaxation times longer than the age of the Universe, 
\cite{Baumgardt&Mieske08} found that the inner DM component is heated by dynamical friction from the stars, causing a shallower inner DM distribution than that of the stars.  
\cite{Shin+13} traced the dynamical evolution of NGC~6397 with Fokker-Planck methods, discarding those particles that reach the GC tidal radius. They found that the GC could have contained up to one-quarter of its initial mass in the form of DM and match the present-day surface density and LOS velocity dispersion profiles.

On the other hand, there may be unseen matter in the core of the GC that is not dark matter per se, but composed of unseen stars, for example an inner nuclear cluster of faint stars, possibly white dwarfs, neutron stars, or stellar-mass black holes
\citep{Zocchi+19,Mann+19} and binary stars (\citeauthor{Mann+19}).
%
%
We thus performed  \mpo\ runs including 
such a CUO instead of (or in addition to) an IMBH.


\subsection{Velocity anisotropy profile}
\label{sssec: anis}

With LOS data, {\sc MAMPOSSt} is able  (at least partially) to lift the degeneracy first pointed out by \cite{Binney&Mamon82} between the radial profiles of mass and velocity anisotropy \citep{Mamon+13}.
With the three components of the velocity vector, {\sc MAMPOSSt-PM} is even much more able to disentangle mass and anisotropy profiles (\citealt{Read+20}; Mamon \& Vitral in prep.).  With the data used in this paper, we have thus a unique chance of constraining the mass profile of our sources, which is essential to restrain the estimates on the IMBH.

While GCs are often modeled with isotropic velocities, we performed many runs of \mpo\ with freedom in the anisotropy profile.
Indeed, a central IMBH may modify the orbital shapes.
Furthermore, the outer orbits of isolated GCs are often thought to be quasi radial \citep{Takahashi95}. 
Moreover, the frequent passages of \ngc\ across the Galactic disk (Sect.~\ref{sec: intro}) could alter the orbital shapes of GC stars.

The anisotropic runs of \mpo\ used
the generalization (hereafter gOM) of the Osipkov-Merritt model \citep{Osipkov79, Merritt85} for the velocity anisotropy profile:
\begin{equation}    \label{eq: gOM}
    \beta_{\mathrm{gOM}}(r) = \beta_{0} + (\beta_{\infty} - \beta_{0}) \ \displaystyle{\frac{r^2}{r^2 + r_{\beta}^2}} \ ,
\end{equation}
where $r_{\beta}$ is the anisotropy radius, which can be fixed as the scale radius of the luminous tracer by \textsc{MAMPOSSt-PM}.\footnote{\cite{Mamon+19} found no significant change in models of galaxy clusters when using this model for $\beta(r)$ compared to one with a softer transition: $\beta(r) = \beta_{0} + (\beta_{\infty} - \beta_{0})\,r/(r+r_{\beta})$, first used by \cite{Tiret+07}.} 
%


\section{Global structure and distance of NGC 6397}
\label{sec: structure}

Our mass-orbit modeling of NGC~6397 (Sect.~\ref{sec: Method} below) assumes a distance,  the knowledge of the GC center, spherical symmetry, and the absence of rotation. We investigate these assumptions for NGC~6397.


\subsection{Spherical symmetry}
NGC~6397 appears to be close to spherical symmetry. Its elongation on the sky is 0.07 \citep{Harris10}.
    
\subsection{Center}
\label{ssec: center}
    
\begin{figure}
\centering
\includegraphics[width=0.95\hsize]{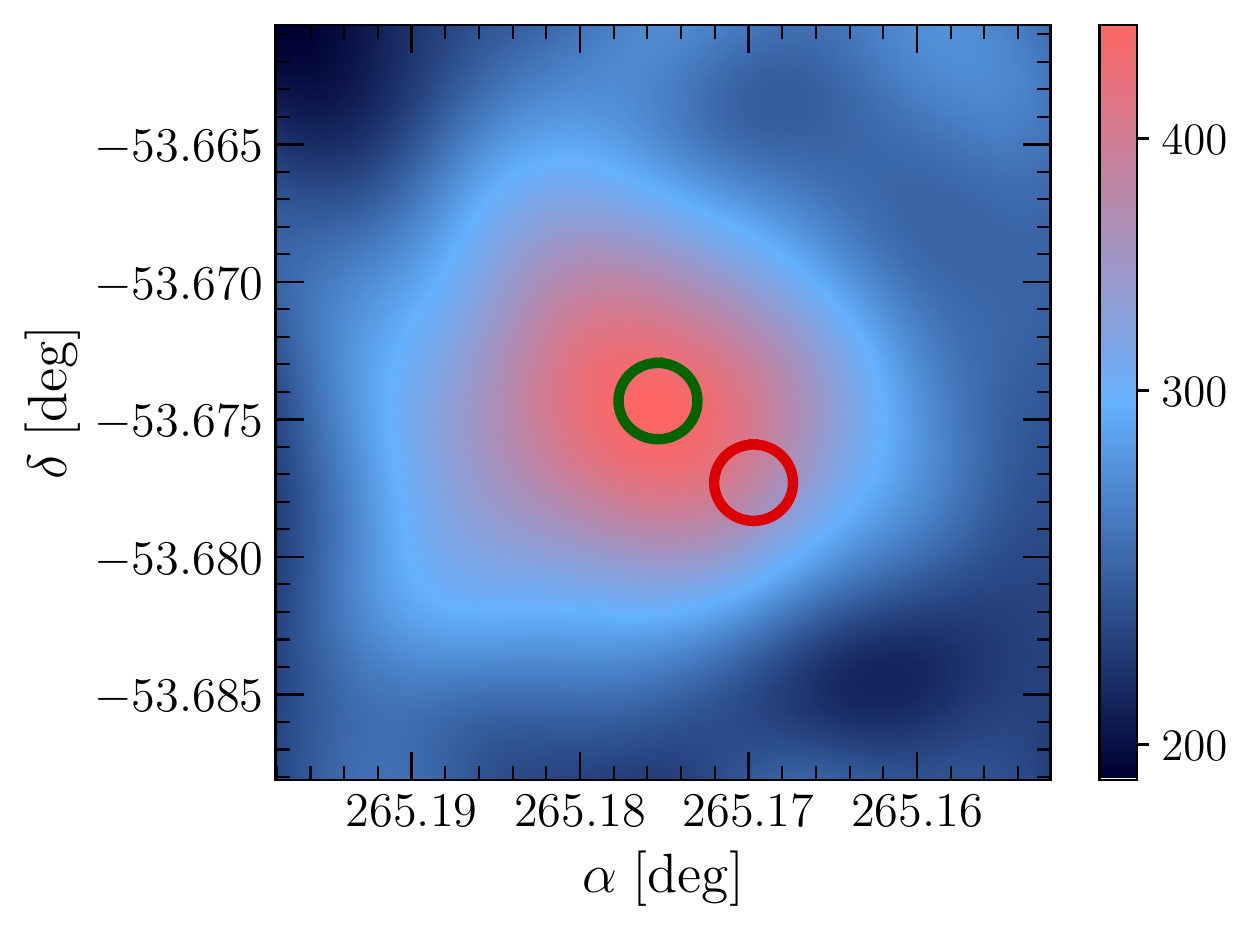}
\raggedright
\caption{
Smoothed star counts map of NGC~6397 from {\sc Gaia}. The image size is $100''$ on the side (half the HST field of view). The star counts are computed in cells of $0\farcs33$ and then smoothed by a Gaussian with $\sigma=0\farcs10$ (using \textsc{scipy.stats.gaussian\_kde} in Python, with a bandwidth of 0.2). 
The color bar provides the star counts per square degree.
The green and red circles  represent the GC center calculated by \protect\cite{Goldsbury+10} and \protect\cite{GaiaHelmi+18}, respectively. 
\label{fig: dist-center}
}
\end{figure}

Our analysis assumes that the IMBH is located at the GC center, so our choice of center is critical.
We have considered two centers: (RA,Dec) =  (265\fdg175375, --53\fdg674333) (\citealt{Goldsbury+10}, using HST data) and
(RA,Dec) =  (265\fdg1697, --53\fdg6773) (\citealt{GaiaHelmi+18}, using {\sc Gaia} data), both in epoch J2000.
We selected the center with the highest local stellar counts.
It is dangerous to use our HST star counts for this because the positions are relative to the center, hence need to be rescaled assuming a given center (see Sect.~\ref{ssec: hst}). We used instead the {\sc Gaia} star counts, which benefit from absolute positional calibration.
Figure~\ref{fig: dist-center} shows a Gaussian-smoothed map of {\sc Gaia} star counts in the inner $100''\times100''$ (half the HST field of view) of NGC~6397, where we overplotted the centers obtained by \cite{Goldsbury+10} and by \cite{GaiaHelmi+18}. 
    %
    %
Clearly, the center of \cite{GaiaHelmi+18} is less well aligned with the density map than the center of \citeauthor{Goldsbury+10}.
We therefore select the \citeauthor{Goldsbury+10} center at
(265\fdg1754, --53\fdg6743)
in J2000 equatorial coordinates.




    

\subsection{Distance}
\label{ssec: distance}
    
\begin{table}
\caption{Distance estimates for NGC~6397}
\begin{center}
\tabcolsep=3.5pt
\begin{tabular}{llll}
\hline
Authors & Method & Observations & Distance \\
& & & (kpc) \\
\hline
Harris 10 & CMD & various & 2.3  \\
Reid \& Gizis 98 & CMD & HST & 2.67$\pm$0.25 \\
Gratton+03 & CMD & VLT & 2.53$\pm$0.05 \\
Hansen+07 & CMD & HST & 2.55$\pm$0.11 \\
Dotter+10 & CMD & HST& 3.0  \\
Heyl+12 & kinematics & HST & 2.0$\pm$0.2 \\
Watkins+15b & kinematics & HST & 2.39$^{+0.13}_{-0.11}$ \\
Brown+18 & parallax & HST & 2.39$\pm$0.07 \\
{\sc Gaia} (Helmi)+18 & parallax & {\sc Gaia} & 2.64$\pm$0.005 \\
Baumgardt+19 & kinematics & N-body & 2.44$\pm$0.04 \\
Shao \& Li 19 & parallax & {\sc Gaia} & 2.62$\pm$0.02 \\
Valcin+20 & CMD & HST & 2.67$^{0.05}_{-0.04}$ \\
\bf this work & \bf kinematics & \bf HST \& MUSE & \bf 2.35$\pm$0.10\\ 
\hline
\end{tabular}
\end{center}
\parbox{\hsize}{{\it Notes}: The columns are: 1) authors; 2) method; 3) observations; 4) distance in kpc. Distances based on kinematics seek dynamical models that match the observed LOS and PM dispersion profiles, where
\citeauthor{Watkins+15b} used Jeans modeling, while \citeauthor{Baumgardt+19} used N-body simulations. The distance of \citeauthor{Baumgardt+19} is a weighted mean with the value given by \cite{Harris96}, while that of \citeauthor{Valcin+20} used the value of \citeauthor{Dotter+10} as a prior. 
}

\label{tab:dist}
\end{table}
\nocite{Harris96}
\nocite{Heyl+12}
\nocite{Reid&Gizis98}
\nocite{Gratton+03}
\nocite{Hansen+07}
\nocite{Dotter+10}
\nocite{Watkins+15b}
\nocite{GaiaHelmi+18}
\nocite{Brown+18}
\nocite{Baumgardt+19}
\nocite{Shao&Li19}
\nocite{Valcin+20}

The adopted distance is important because the mass profile at a given angular radius (e.g., $r$ in arcmin) deduced from 
the Jeans equation of local dynamical equilibrium (Eq.~[\ref{eq: jeans}]) varies roughly as $M \propto r \sigma_v^2$, thus as distance $D$ for given LOS velocities and as $D^3$ for given PMs.
%
Table~\ref{tab:dist} shows a list of distance estimates for NGC~6397.
Recent estimates are bimodal around 2.39 kpc and 2.64 kpc. The CMD-based distance estimates favor the larger distance, the kinematics favor the smaller distance, while the parallax distances point to small (HST) or large ({\sc Gaia}) distances.  

We estimated a kinematical distance by equating the LOS velocity dispersion of stars measured with VLT/MUSE (see Sect.~\ref{ssec: muse}, below) with the HST PM dispersions of the same stars (see Sect.~\ref{ssec: hst}). To avoid Milky Way field stars, we used the cleaned MUSE and HST samples, for which we had 692 matches, among which 445 with separations smaller than $0\farcs1$.\footnote{The distribution of log separations is strongly bimodal with peaks at $0\farcs06$ and $1''$.}
For these 445 stars, we measured a LOS velocity dispersion of $4.91\pm0.16\,\rm\,km\,s^{-1}$, and an HST PM dispersion of 
$0.439\pm0.015\,\rm mas\,yr^{-1}$ in the RA direction and
$0.442\pm0.015\,\rm mas\,yr^{-1}$ along the Dec direction (where the uncertainties are taken as the values divided by $\sqrt{2(n-1)}$).
This yields a kinematic distance of $4.91/[c\,(0.439+0.442)/2] = 2.35\pm0.10\,\rm kpc$, where 
$c  = 4.7405$ is the POS velocity of a star of $\rm PM = 1\,mas\,yr^{-1}$ located at $D=1\,\rm kpc$.

We adopted the lower distance of 2.39 kpc (i.e., a distance modulus of 11.89), for three reasons:
1) We trust more the HST parallax than the {\sc Gaia}-DR2 parallax, given that the former is based on a much longer baseline;
2) Our study of NGC~6397 is based on kinematics, and thus the larger distance would lead to abnormally high POS velocity dispersions compared to the LOS velocity dispersions.
3) It is consistent with our kinematic estimate of the  distance.
We do not adopt our estimated kinematic distance of $2.35\pm0.10\,\rm kpc$ because it is based on the perfect  equality of LOS and POS velocity dispersions, which supposes velocity isotropy, which in turn is not certain.
For our adopted distance of 2.39 kpc, 1 arcmin subtends 0.7 pc.

\section{Data}
\label{sec: data}


\subsection{HST data}
\label{ssec: hst}

The HST data were kindly provided by A. Bellini, who measured PMs for over 1.3 million stars in 22 GCs, including NGC~6397 \citep{Bellini+14}.
The data for NGC~6397 has a 202 arcsec square field of view (Wide Field Camera) and reached down to less than 1 arcsec from the \cite{Goldsbury+10} center (see the right panel of Figure~\ref{fig: hist-rproj}). This minimum projected radius to the center is smaller than the BH radius of influence 
$r_{\rm BH} \sim G\,M_{\rm BH} / \sigma_0^2 \simeq 0.11\,(M_{\rm BH}/600\,\rm M_\odot)\,\rm pc$ (using the LOS velocity dispersion of 
Sect.~\ref{ssec: distance}), which corresponds to  $9\,(M_{\rm BH}/600\,\rm M_\odot)$ arcsec, given our adopted  distance of 2.39 kpc.

This dataset was provided in a particular master frame shape (for details, see Table~29 from \citealt{Bellini+14} as well as \citealt{Anderson+08}), with both GC center and PM mean shifted to zero. 
The first step in our analysis was to convert the positions and PMs to the absolute frame. 

\subsubsection{HST absolute positions}
We applied the \cite{Rodrigues1840} rotation formula  to shift the relative positions back to their original center
to translate the GC stars to their true positions on the sky.
We used the center of \cite{Goldsbury+10}, which was the one considered by \cite{Bellini+14}.
We then rotated the subset with respect to its true center, so that the stars originally parallel to the dataset's increasing $x$ axis remained parallel to the right ascension increasing direction. We then verified our method by matching the stars in sky position with {\sc Gaia} (see Sect.~\ref{ssec: consistency} below).

\subsubsection{HST absolute proper motions}

The HST PMs were measured relative to the bulk PM of the GC. 
We corrected the relative PMs of \cite{Bellini+14} with their provided PM corrections. We just added columns 4 and 5 to columns 31 and 32 from Table~29 of \cite{Bellini+14}, respectively.
We then converted the relative PMs to absolute PMs by computing the bulk PM of NGC~6397 using the stellar PMs provided by {\sc Gaia DR2} as explained in Sect.~\ref{ssec: GaiaDR2} below. 


The small field of view of HST and the few pointed observations do not allow the observation of sufficiently numerous background quasars to obtain an absolute calibration of HST PMs.
On the other hand, the {\sc Gaia} reference frame obtained  with  more  than  half  a  million  quasars provides a median positional uncertainty of 0.12 mas for $G < 18$  stars \citep{GaiaMignard+18} and therefore allows us, by combining its accuracy with HST's precision, to know NGC~6397 PMs with unprecedented accuracy.
We will compare the PMs of stars measured both by HST and {\sc Gaia} in Sect.~\ref{ssec: consistency}.

\begin{figure}
    \centering
    \includegraphics[width=\hsize]{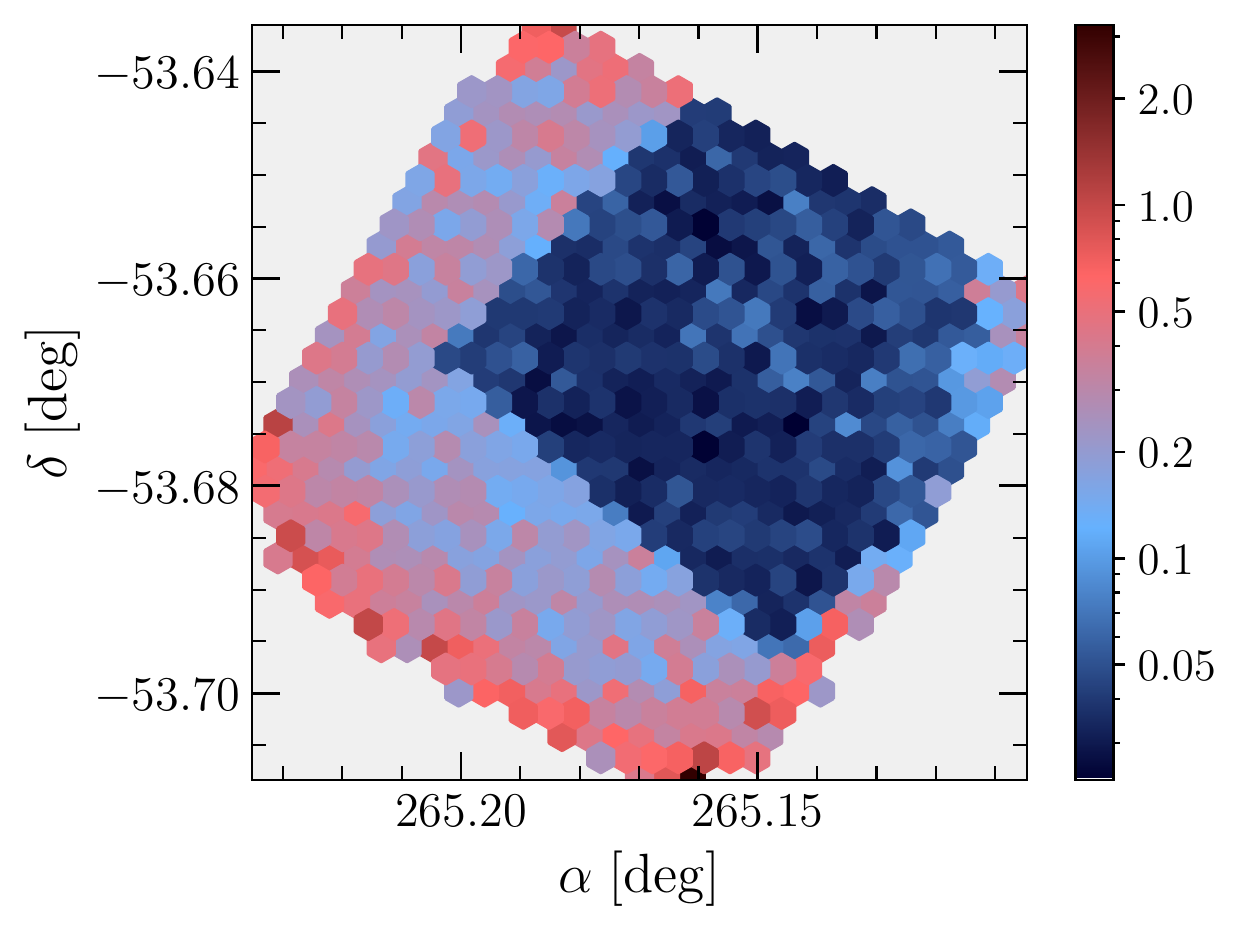}
    \caption{Map of HST PM errors (semi-major axis of error ellipse, in mas\,yr$^{-1}$), color-coded according to the median of each hexagonal cell, using the entire set of HST data (i.e., before any cuts). The total HST field of view is $202''$ on the side.
    }
    \label{fig:pmerrormap}
\end{figure}

Our HST data had 13\,593 stars. The PM precision varies across the field of view because of the different time baselines (principally 1.9 and 5.6 years) in part due to the smaller size of the older HST cameras.
Figure~\ref{fig:pmerrormap} shows that 
the PMs are much more accurate in a large square portion of the field of view, including the western part and extending to the central region, where the baseline is 5.6 years.
At magnitude F606W = 17, the one-dimensional PM precision is 0.02 mas yr$^{-1}$ in the more precise portion of field of view and 0.08 mas yr$^{-1}$ outside.

\subsection{{\sc Gaia} data}
\label{ssec: GaiaDR2}

{\sc Gaia DR2} presented an overall astrometric coverage of stellar velocities, positions and magnitudes of more than $10^9$ stars in the Milky Way and beyond. 
 {\sc Gaia} measured PMs in over 40\,000 stars of magnitude $G < 17$ within a $1^\circ$ cone around \ngc.
These PMs have a typical precision of 0.17 mas/yr at $G = 17$. 
Since the {\sc Gaia} $G$ and HST F606W magnitudes differ by less than 0.1, one sees that {\sc Gaia} PMs are less precise than those of HST by factors of two to eight depending on the region of the HST field of view. 
The {\sc Gaia} data was also used to infer the number density profile out to large distances from the GC center.  
%


\subsection{VLT/MUSE data}
\label{ssec: muse}

We complemented the PMs using
LOS velocities that \cite{Husser+16} acquired with
the MUSE spectrograph on the VLT.
A mosaic of 5$\times$5 MUSE pointings led to an effective square field of view of 5 arcmin on the side. The bulk LOS velocity is $\langle v_{\rm LOS} \rangle = 17.84 \pm 0.07$ km s$^{-1}$ \citep{Husser+16}.
In a companion article, \cite{Kamann+16} assigned membership probabilities to the stars according to their positions in the space of LOS velocity and metallicity, compared to the predictions for the field stars from the Besan\c{c}on model of the Milky Way \citep{Robin+03}. 
The data, kindly provided by S. Kamann, contained 7130 LOS velocities, as well as the  membership probabilities.

\subsection{Consistency between datasets}
\label{ssec: consistency}
\subsubsection{Positional accuracy}
\label{sssec: posacc}
We used \textsc{Tool for OPerations on Catalogues And Tables} (\textsc{TOPCAT}, \citealt{Taylor05}) to check the match positions between catalogs, which we did in our own Python routines that filtered the original datasets (Sect.~\ref{sec: clean} below).
Performing symmetric matches among each pair of datasets, with a maximum allowed separation of 1 arcsec, we obtained median separations  of 0\farcs38 for the 4455 stars in common between {\sc Gaia} and HST, as well as 0\farcs01 for the 4440 stars in common between MUSE and HST.

\subsubsection{Proper motions}

Many of the 4455 stars in common between HST and {\sc Gaia} have very high {\sc Gaia} PM uncertainties (we saw in Sect.~\ref{ssec: GaiaDR2} that HST PM uncertainties were much smaller, but a few have large values).
Limiting to 1179 stars in common with both {\sc Gaia} and HST PM uncertainties below 0.4 mas/yr, we find that the HST PMs in the (RA,Dec)  frame are 
(0.006$\pm$0.812,0.029$\pm$1.239) mas/yr
above those from {\sc Gaia}, where the ``errors'' represent the standard deviation. The uncertainties on the means are $\sqrt{1179} = 34$ times lower: (0.024,0.036) mas/yr. The mean shifts in PMs are (0.3,0.8) times the uncertainties on the mean, and thus not significant.
This possible offset in PMs is of no concern as long as we only use HST data, since we analyze them relative to the bulk PM of the GC. Even  in {\sc MAMPOSSt-PM} runs with the combined HST+{\sc Gaia} dataset, this  shift in PMs is not statistically significant and appears too small to affect the results.



\section{Data cleaning}
\label{sec: clean}

We now describe how we selected stars from each dataset for the mass modeling runs with \mpo. We also used more liberal criteria to select stars in our estimates of the surface density profile (see Sect.~\ref{ssec: surf-dens}).

For each dataset, we discarded all stars with PM errors above half of $0.394\,\rm mas\,yr^{-1}$, corresponding (for our adopted distance) to the one-dimensional PM dispersion of NGC~6397 measured by \cite{Baumgardt+19} for the innermost 2000 {\sc Gaia} stars, whose mean projected radius was 99\farcs25. 
The PM error is computed as the semi-major axis of the error ellipse (eq. B.2 of \citealt{Lindegren+18}):
\begin{eqnarray} \label{eq: err_lind18}
    \epsilon_\mu &=& \sqrt{\frac{1}{2}(C_{33}+C_{44}) + \frac{1}{2}\sqrt{(C_{44}-C_{33})^2 + 4 C_{34}^2}}
    \label{errLindegren} \\
C_{33} &=& \epsilon_{\mu_{\alpha*}}^2 \ ,\\
C_{34} &=&  \epsilon_{\mu_{\alpha*}}\, \epsilon_{\mu_\delta}\,\rho \ ,\\
C_{44} &=& \epsilon_{\mu_\delta}^2 \ ,
\end{eqnarray}
where $\epsilon$ denotes the error or uncertainty and where $\rho$ is the correlation coefficient between $\mu_{\alpha*}$ and $\mu_\delta$.\footnote{We use the standard notation $\mu_{\alpha*} = \cos \delta\,{\rm d}\alpha/{\rm d}t$,
$\mu_\delta={\rm d}\delta/{\rm d}t$.}
Our HST data does not provide $\rho$.
We thus selected stars with 
$\epsilon_\mu < 0.197,\rm mas\,yr^{-1}$, or equivalently for which the LOS velocity error satisfies
$\epsilon_{v_{\rm LOS}} < 0.197\,c\,D = 2.23\,\rm km\,s^{-1}$, for $D = 2.39\,\rm kpc$ and where $c = 4.7405$ 
(see Sect.~\ref{ssec: distance}).

\begin{figure}
    \centering
    \includegraphics[width=\hsize]{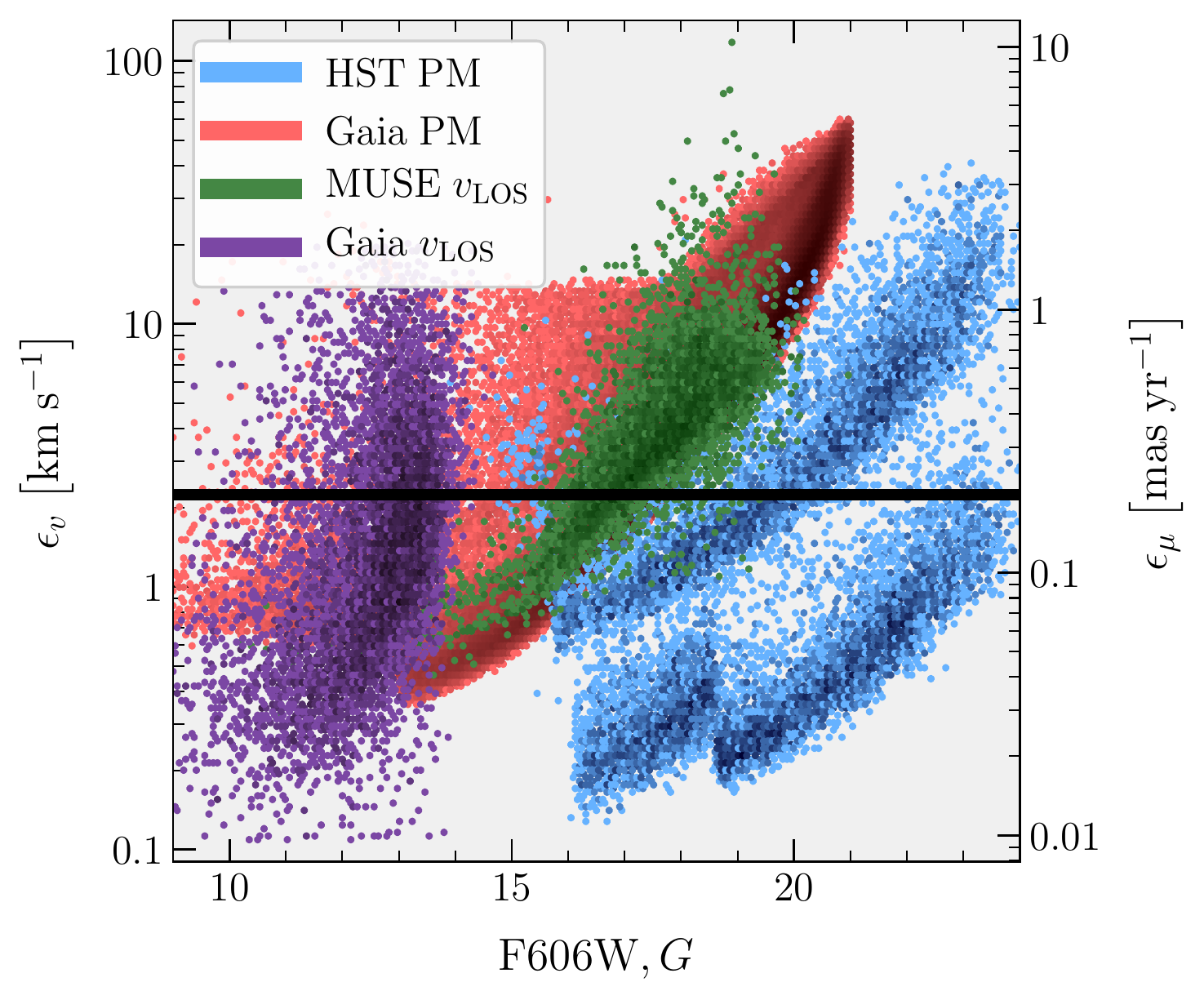}
    \caption{Proper motion errors and LOS velocity errors (converted to PM errors for distance of 2.39 kpc) for the different datasets. The horizontal line displays $\epsilon_\mu = 0.197\,\rm mas\,yr^{-1}$. 
    The three different blue zones correspond to different baselines in the HST PM data.
    We note that magnitudes $G$ and F606W are close to equivalent (they match to $\pm$0.1).
    }
    \label{fig: PMerrs}
\end{figure}

Figure~\ref{fig: PMerrs} shows the PM errors versus magnitude for HST (blue) and {\sc Gaia} (red), as well as the equivalent PM error for the LOS velocity errors of {\sc Gaia} (purple) and MUSE (green).
The figure indicates that our maximum allowed PM errors are reached at magnitudes $m_{\rm F606W} = 19.7$ for the worst HST data (those with the shortest baseline, see Fig.~\ref{fig:pmerrormap}), and for virtually all magnitudes for the other HST stars.
{\sc Gaia} stars reach the maximum allowed PM error at typically $G=17.5$. 
The equivalent LOS velocity error limit is reached at magnitude    $m_{\rm F606W} = 16.8$ for MUSE, but only at $G = 13.4$ for {\sc Gaia}. The {\sc Gaia} radial velocities are thus of little use for our modeling, as too few of them have sufficiently precise values.

\subsection{HST data cleaning}
\label{ssec: hst-clean}
After selecting the stars with low PM errors,
we cleaned our HST data in three ways:
We discarded stars with 
1) PMs far from the bulk PM of the GC;
2) lying off the color-magnitude diagram;
3) associated with X-ray binaries.



\subsubsection{HST proper motion filtering}
 \label{sssec: pm-hst-filter}

%

\begin{figure}
\centering
\includegraphics[width=\hsize]{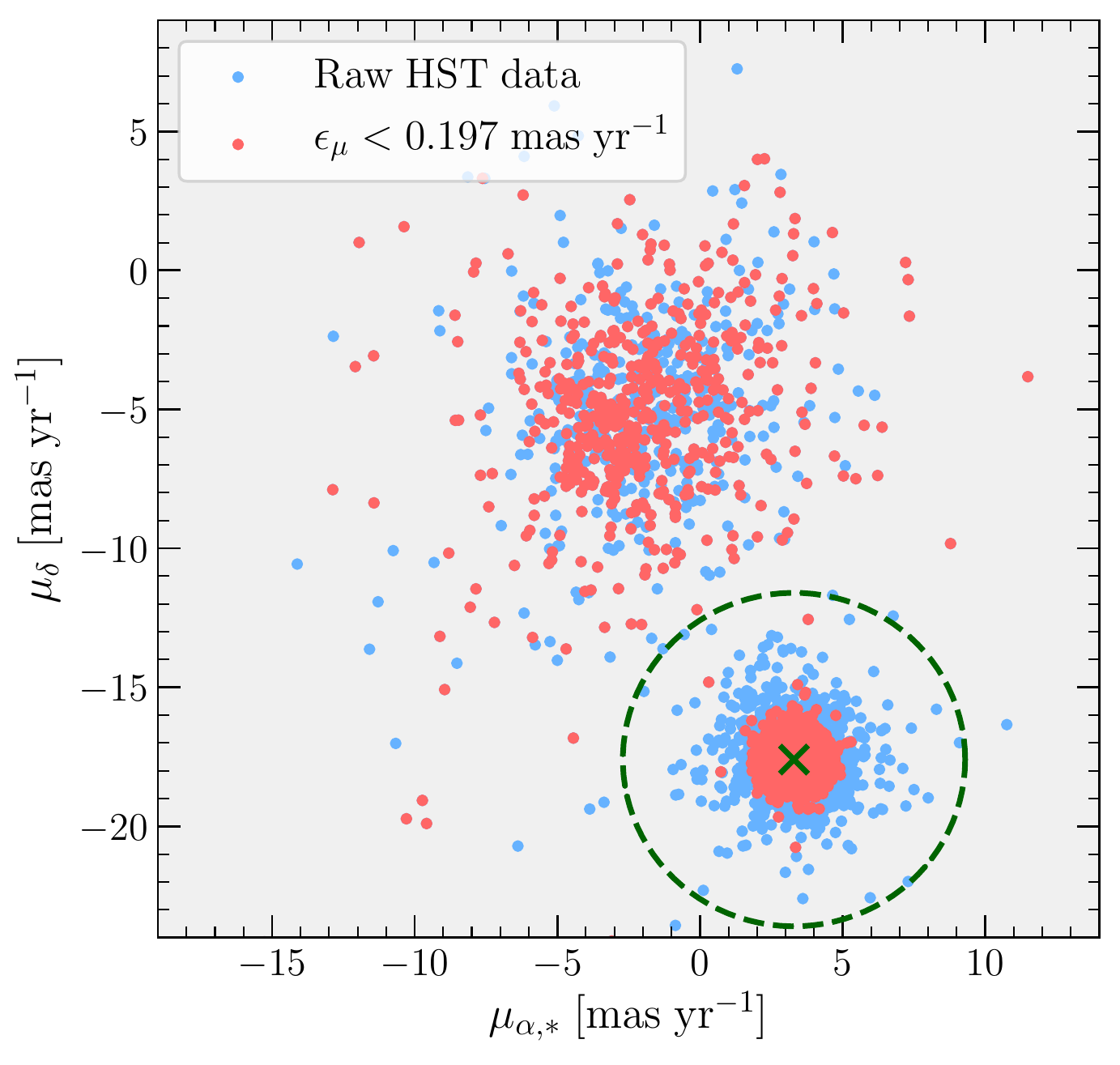}
\caption{
HST proper motions, 
corrected to the mean GC PM shown in Table~\ref{tab: PM 2d}.
The full set of HST stars is shown in \emph{blue}, while the subset obtained after applying the low PM error cut (Sect.~\ref{sssec: pm-hst-filter}) are shown in \emph{red}.
The \emph{dashed green  circle} represents the very liberal hard cut on PMs to remove Milky Way field stars and the \emph{green cross} highlights the mean PM presented in Table~\ref{tab: PM 2d}.
\label{fig: HST-pm-filter}}
\end{figure}



The higher surface number density of stars in the inner regions of NGC~6397 allows us to distinguish GC stars with field stars in PM space, as shown in
Figure~\ref{fig: HST-pm-filter}. 
Moreover, some high velocity stars could in principle be caused by very tight (separations smaller than $\sim 0.1\,\rm AU$) GC binaries.

We made a very liberal cut to select GC stars in PM space, using a circle of radius 6 mas yr$^{-1}$ (green circle in Fig.~\ref{fig: HST-pm-filter}), corresponding to 15 times the PM dispersion measured by \cite{Baumgardt+19}. The PM center of the GC was set at the mean of the values in Table~\ref{tab: PM 2d}. Our cut in PM space also corresponds to a velocity dispersion of $75\,\rm km\,s^{-1}$, which is over 13 times the highest LOS velocity dispersion measured by \cite{Kamann+16}.
We chose such a liberal cut to ensure that we would not miss any high velocity GC members because otherwise our modeling would underestimate the mass.
This left us with 9149 stars among the 9624 with low PM errors.

The filtering of field stars in PM space is not fully reliable because the cloud of field stars in PM space (upper part of Figure~\ref{fig: HST-pm-filter}) may extend into the (smaller) GC cloud (and past it). We thus proceed to another filtering in the color-magnitude diagram (CMD).



\subsubsection{HST color-magnitude filtering}
\label{sssec: cmd-hst-filter}

\begin{figure*}
\centering
\includegraphics[width=0.95\hsize]{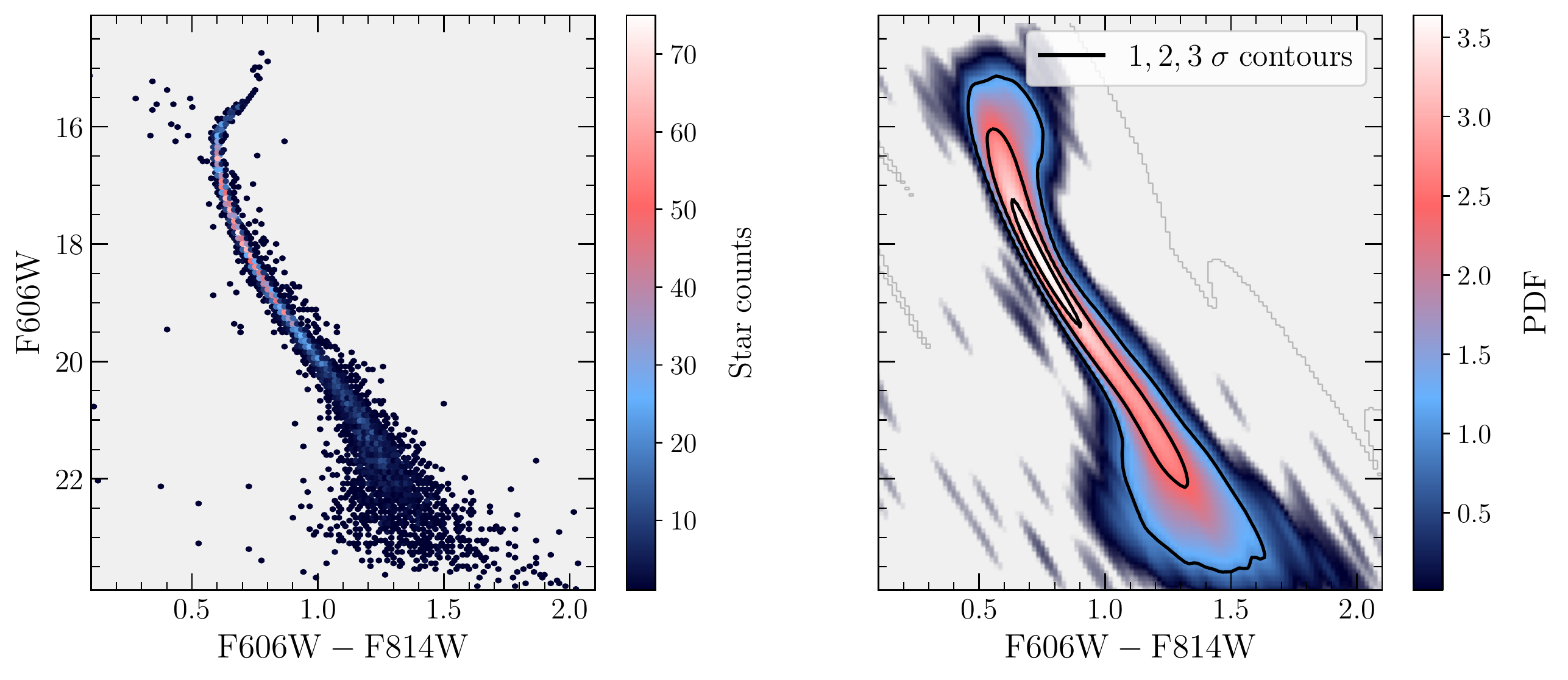}
\caption{HST color-magnitude diagrams.
\textit{Left}: 
CMD
after the PM filtering process explained in section~\ref{sssec: pm-hst-filter}. \textit{Right}: Kernel density estimation of the HST isochrone displayed in the left panel. 
The 1, 2 and 3$\,\sigma$ contours are displayed (from in to out).
\label{fig: cmd-filter}}
\end{figure*}

The left panel of Figure~\ref{fig: cmd-filter} shows the locations in the CMD of the stars that survived the PM error cut and the filtering in PM space. While most stars follow a tight relation, a non-negligible fraction are outliers.
We took a conservative cut of the stars on the CMD using kernel density estimation as displayed in the right panel of Figure~\ref{fig: cmd-filter}.\footnote{We used \textsc{scipy.stats.gaussian\_kde}, setting the bandwidth method to the Siverman's rule (\citealt{Silverman86}).}
This graph displays the 1, 2 and 3$\,\sigma$ contours 
of the kernel density estimation, drawn as black lines. We selected stars inside the 2$\,\sigma$ region because the 3$\,\sigma$ region appears too wide to rule out binaries, while the 1$\,\sigma$ region was too conservative, extracting too few stars for our analysis. 

The CMD filtering not only removes field stars whose PMs coincide by chance with those of GC stars, but also removes GC members that are unresolved binary stars and lie in the edges of the main-sequence, as well as particular Blue Stragglers in the $15 < \rm F606W < 16.5$ range, which are believed to be the result of past 
mergers of GC members (\citealt{Leonard89}).
Removing binaries 
and stars who have gone through mergers
is important because their kinematics are dominated by two-body interactions, while our modeling assumes that stellar motions are dominated by the global gravitational potential of the GC.
As discussed by \cite{Bianchini+16}, three types of binaries need to be considered.
\begin{enumerate}
    \item Resolved (i.e., wide) binaries will produce their own PMs that can be confused with the parallax. But, following \citeauthor{Bianchini+16}, the PMs of such resolved binaries, of order of $a/T$ where $a$ is the semi-major axis of the binary and $T$ is the time baseline, are negligible in comparison to the GC velocity dispersion. Indeed, to be resolved at the distance of NGC~6397, the binary star should be separated by at least $0\farcs1$, which corresponds to 24 AU at the distance of NGC~6397;  the most massive binaries, with mass below $2\,\rm M_\odot$, will have a period of 90 years, thus $a/T=1.26\,\rm km\,s^{-1}$, while less massive binaries will have longer periods, hence lower $a/T$.
    \item Nearly resolved binaries will not be deblended, and their distorted image (in comparison to the PSF) will lead to less precise astrometry, which will be flagged. 
    \item The orbits of unresolved binaries, considered as single entities, will be those of test particles in the gravitational potential. But their higher mass should lead them to have lower velocity dispersions than ordinary stars, in particular in the dense inner regions of GCs where the two-body relaxation time is sufficiently short for mass segregation.  
\end{enumerate}

Our CMD filtering left us with 7259 stars among the 9149 surviving the previous filters.

\subsubsection{Removal of X-ray binaries}
\label{sssec: x-ray}

\cite{Bahramian+20} detected 194 X-ray sources within 200 arcsec from the center of NGC~6397, some of which could potentially be background AGN. The remaining sources are thought to be X-ray binaries, and as such will have motions perturbed by their invisible compact companion.
We therefore deleted the 50 HST stars whose positions coincided within 1 arcsec with the ``centroid'' position of an X-ray source.

\cite{Bianchini+16} found that the effect of unresolved binaries on GC dispersion profiles is  only important  for high initial binary fractions (e.g., $\sim 50\%$), which can induct a difference of $0.1-0.3$ km s$^{-1}$ in the velocity dispersion profile, mainly in the GC inner regions, where the binary fraction should be highest. For lower initial binary fractions, \citeauthor{Bianchini+16} found that unresolved binaries do not significantly affect the PM dispersion profile, but only the kinematics error budget. Therefore, the binary fractions calculated by
\cite{Davis+08} and \cite{Milone+12a} for NGC~6397 are clearly insufficient (i.e., $\lesssim 5\%$ and $\lesssim 7\%$, respectively) to require a special treatment. 

\subsubsection{HST final numbers}
After these cuts, we are left with 7209 stars from the HST observations (among the original 13\,593). 
The combination of  the CMD and PM filtering along with the removal of X-ray binaries  effectively cleans the data of most of the binaries that would affect our modeling, only leaving  binaries of different luminosities that are at intermediate separations ($\approx 0.15$ to 5 AU, causing peculiar motions between 1 and 15 times the GC velocity dispersion). We checked that using a less liberal cut of $7.5\,\sigma$ instead of $15\,\sigma$ affects very little  our results.

\subsection{{\sc Gaia} data cleaning}
\label{sssec: pre-filter}

We followed similar steps in cleaning the {\sc Gaia} data as we did for the HST data.

\subsubsection{Quality flags}
\label{sssec: Gaia-quality}

We filtered the {\sc Gaia} stars with $\epsilon_\mu<0.197\,\rm mas\,yr^{-1}$ using two data quality flags proposed by \cite{Lindegren+18}.
First, we only kept stars whose astrometric solution presented a sufficiently low goodness of fit:
\begin{equation} \label{eq: Gaia_flag1}
    \sqrt{\chi^2\over N-5} < 1.2 \, \mathrm{Max}\left\{1,\exp\left[-0.2 \, (G-19.5)\right]\right\} \  ,
\end{equation}
where $N$ is the number of points (epochs) in the astrometric fit of a given star and 5 is the number of free parameters of the astrometric fit (2 for the position, 1 for the parallax and two for the PM).
Eq.~(\ref{eq: Gaia_flag1}) gives a sharper HR diagram, removing artifacts such as double stars, calibration problems, and astrometric effects from binaries. It is more optimized than the (\url{astrometric_excess_noise}$<1$) criterion, used in \cite{Baumgardt+19} and \cite{Vasiliev19b}, especially for brighter stars ($G\lesssim15$), according to \cite{Lindegren+18}. 

Second, we only kept stars with good photometry.
\begin{equation} \label{eq: Gaia_flag2}
    1.0 + 0.015 \, (G_{\mathrm{BP}} - G_{\mathrm{RP}})^{2} < E < 1.3 + 0.06 \, (G_{\mathrm{BP}} - G_{\mathrm{RP}})^{2}  
     \ ,
\end{equation}
where $E = (I_{\mathrm{BP}}+I_{\mathrm{RP}})/I_{\mathrm{G}}$ is the flux excess factor. 
Eq.~(\ref{eq: Gaia_flag2}) performs an additional filter in the HR diagram, removing stars with considerable photometric errors in the BP and RP photometry, affecting mainly faint sources in crowded areas. 
This  poor photometry broadens the CMD, leading to more confusion with field stars.

Those variables correspond to the following quantities in the {\sc Gaia DR2} archive:
\begin{itemize}
    \item $\chi^{2}$: \url{astrometric_chi2_al} 
    \item $\nu'$: \url{astrometric_n_good_obs_al} 
    \item $E$: \url{phot_bp_rp_excess_factor} 
    \item $G_{\mathrm{BP}} - G_{\mathrm{RP}}$: \url{bp_rp}
\end{itemize}

\subsubsection{Maximum projected radius}
\label{sssec: Gaia-projD}


\begin{figure}
\centering
\includegraphics[width=\hsize]{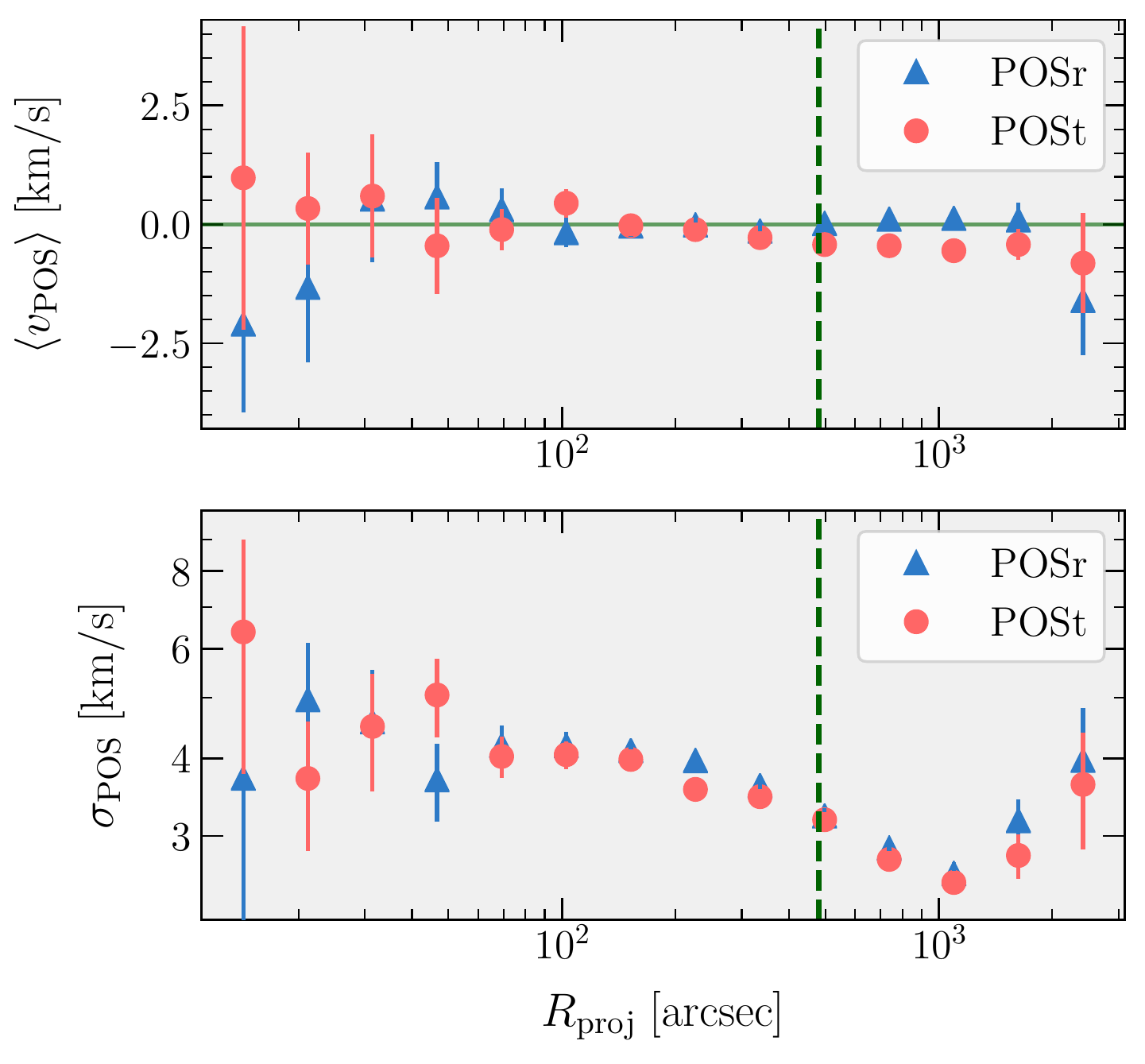}
\caption{
Radial profiles of mean plane of sky velocity (\emph{top}) and velocity dispersion (\emph{bottom}) of \ngc\ from cleaned {\sc Gaia DR2}. The POS motions are split between radial (POSr, \emph{blue triangles}) and tangential (POSt, \emph{red circles}) components. The \emph{dashed green vertical line} displays the $8'$ limit for use in \mpo.
\label{fig: pseudo-anis}}
\end{figure}

\ngc\ likely suffers from tidal heating every time it passes through the Milky Way's disk, in particular during its last passage less than 4 Myr ago (Sect.~\ref{sssec: dark-matter}).
The tidal forces felt by GC stars during passages through the disk will produce velocity impulses that are effective in perturbing the least bound orbits, which typically are those of the stars in the outer envelope of the GC. The {\sc Gaia} data can trace the effects of such tidal disturbances on the GC kinematics.

We searched for anomalous kinematics in \ngc\ using {\sc Gaia DR2} out to  a maximum projected radius of 1 degree with respect to NGC~6397's center.
The top panel of Figure~\ref{fig: pseudo-anis} indicates that the mean POS velocities in the radial and tangential directions differ beyond $8'$.\footnote{\cite{Drukier+98} had noticed a similar effect in the M15 GC.}
We also checked the concordance of the radial and tangential components of the radial profiles of POS velocity dispersion (or equivalently PM dispersion).
The bottom panel of Figure~\ref{fig: pseudo-anis} shows excellent agreement between the two components of the velocity dispersion from $2'$ to $20'$, suggesting that the velocity ellipsoid is nearly isotropic in this range of projected radii. 
We conservatively adopted a maximum projected radius of $8'$ as set by the divergence of the mean velocity profiles beyond that radius.



\subsubsection{{\sc Gaia} proper motion filtering and the bulk proper motion of NGC~6397}
\label{sssec: pm-filter}

As for HST, we filtered {\sc Gaia} stars in PM space to later filter them in CMD space.
Since {\sc Gaia} data extends to much greater projected radii from the GC center, thus to lower GC surface densities, the GC stands out less prominently from the field stars (FS) in PM space.
We therefore first estimated the bulk PM of the GC and we assigned a  first-order probability of membership using a  GC+FS mixture model. 

We were tempted to assign two-dimensional (2D) Gaussian distributions for both GC stars and interlopers.
However, the FS PM distribution has wider tails than a Gaussian. This means that stars on the other side of the GC, relative to the center of the field star component (i.e., its bulk motion) in PM space are more likely to be field stars than assumed by the Gaussian model.
We found that the PM-modulus ``surface density'' profile (the velocity analog of the surface density profile) is well fit by  
a Pearson type VII distribution (\citealt{Pearson16}), as explained in detail in Appendix~\ref{app: PMpdf}. This  distribution relies on two free parameters, a scale radius $a$ and an outer slope $\gamma$, and can be written as:

\begin{equation} \label{eq: PMpdf}
    f_{\mu}(\boldsymbol{\mu}) = - \frac{\gamma + 2}{2 \, \pi \, a^{2}} \, \left[ 1 + \left(\frac{\mu}{ a}\right)^{2} \right]^{\gamma/2} \ ,
\end{equation}
where $\boldsymbol{\mu} = (\mu_{\alpha, *},\mu_{\delta})$ and

\begin{equation} \label{eq: PMdef}
    \mu_{i} = \sqrt{(\mu_{\alpha, *i} - \overline{\mu_{\alpha, *i}})^{2} + (\mu_{\delta,i} - \overline{\mu_{\delta,i}})^{2}} \ ,
\end{equation}
where the suffix $i$ stands for the component analyzed, which in the case of Eq.~(\ref{eq: PMpdf}) is the interlopers (i.e., field stars, hereafter FS).
The reader can verify that, indeed, $\int f_{\mu}(\boldsymbol{\mu}) \, \diff \boldsymbol{\mu} = 1$, with $\diff \boldsymbol{\mu} = 2 \, \pi \, \rm \mu \, \diff \mu$. 

With the respective Gaussian and Pearson~VII distributions of PMs of GC stars and FS, 
we performed a joint fit to the two-dimensional distribution of PMs, which provided us with a precise bulk PM of the GC.
For this,
we considered the convolved expressions of the PM distributions of both GC and interlopers with the errors provided by the {\sc Gaia} archive ($\epsilon_{\mu_{\alpha, *}}$, $\epsilon_{\mu_{\delta}}$ and $\rho_{\mu_{\alpha, *} \mu_{\delta}}$). When passing onto polar coordinates, the uncertainty propagation of Eq.~(\ref{eq: PMdef}) produces
%
\begin{eqnarray} \label{eq: PMerr}
    \epsilon_{\mu,i}^{2} &\!\!\!\!=\!\!\!\!& 
    \left({\mu_{\alpha, *i} - \overline{\mu_{\alpha, *i}} \over \mu_{i}}\right)^{2} \epsilon_{\mu_{\alpha, *i}}^{2} + \left({\mu_{\delta,i} - \overline{\mu_{\delta,i}} \over \mu_{i}}\right)^{2} \epsilon_{\mu_{\delta,i}}^{2} \nonumber \\ &\!\!\!\!\mbox{}\!\!\!\!&
    + \ 2 \, {\left(\mu_{\alpha, *i} - \overline{\mu_{\alpha, *i}}\right) \, \left(\mu_{\delta,i} - \overline{\mu_{\delta,i}}\right) \over \mu_{i}^{2}} \, \epsilon_{\mu_{\alpha, *} \mu_{\delta,i}} \ ,
\end{eqnarray}
%
where $\epsilon_{\mu_{\alpha, *} \mu_{\delta}} = \epsilon_{\mu_{\alpha, *}} \, \epsilon_{\mu_{\delta}} \, \rho_{\mu_{\alpha, *} \mu_{\delta}}$. The convolution with Gaussian errors was straightforward in the case of GC stars since their PM distribution was also modeled as a Gaussian, and thus we just added the errors to the dispersions in quadrature:
\begin{equation}
    \sigma_{\rm GC, new}^{2} = \sigma_{\rm GC}^{2} + \epsilon_{\mu,\rm GC}^{2} \ .
\end{equation}

However, the convolution of the field star distribution  with Gaussian errors cannot be reduced to an analytic function; numerical evaluation of the convolution integrals for each star would dramatically increase the calculation time. 
We therefore used the analytical approximation for the ratio of convolved to raw probability distribution functions of PM moduli,
(which is also incorporated in \mpo),
as briefly described in Sect.~\ref{app: conv-MPOPM} (details are given in Mamon \& Vitral in prep.). 
This allowed us to perform our mixture model fit to the PM data, using  Markov Chain Monte Carlo (MCMC)\footnote{For all MCMC analyses except the one in \mpo, we used the Python package \textsc{emcee}, \citep{ForemanMackey+13}.}
to estimate bulk motions of both the GC and the field stars and assign probabilities of GC membership for each star. 

Table~\ref{tab: PM 2d} displays our estimates of $\mu_{\alpha,*}$ and $\mu_{\delta}$, which show a good agreement with the literature values for the GCs mean PMs. In the same table, we display the overall average of the literature plus our estimates, with its respective uncertainty $\epsilon$, calculated as $\epsilon^2 = \left< \epsilon_{i} \right>^2 + \sigma^2$, where $\epsilon_{i}$ stands for the uncertainties on the estimated values, and  $\sigma$ stands for their standard deviation. This overall average and uncertainties were also later used as priors for the \mpo\  mass modeling analysis.

\begin{table}

\centering
\tabcolsep=2.4pt
\caption{Comparison of estimates on $\mu_{\alpha, *}$ and $\mu_{\delta }$ with previous studies.} 
\label{tab: PM 2d}
\footnotesize
\begin{tabular}{lcc}

\hline
Method & $\mu_{\alpha *}$ & $\mu_{\delta }$ \\
[0.5ex] 
\hline
& [mas yr$^{-1}$] & [mas yr$^{-1}$] \\ [0.5ex] 
\hline \hline

{\sc Gaia} (Helmi)+18
&\ \,$3.291 \pm 0.0026$ & \ \,$-17.591 \pm 0.0025$  \\

\cite{Vasiliev19b} & $3.285 \pm 0.043$  & $-17.621 \pm 0.043$ \\

\cite{Baumgardt+19} & $3.30\ \,\pm 0.01$\ \,&$-17.60\ \,\pm 0.01$\ \,\\
     
this work (hybrid)& $3.306 \pm 0.013$ & $-17.587 \pm 0.024$ \\

\bf Overall average & \bf 3.296 $\pm$ 0.019 & \bf --17.600 $ \, \pm$ 0.024 \\
\hline

\end{tabular}
\normalsize
\end{table}

We only kept stars with GC membership probability higher then 0.9 according to our mixture model. 
This corresponds to a $1.43 \, \rm mas\,yr^{-1}$ cut in the distribution of PMs of the GC, thus a
$3.6 \,\sigma$ cut.

\subsubsection{{\sc Gaia} color-magnitude filtering}

\begin{figure}
\centering
\includegraphics[width=0.9\hsize]{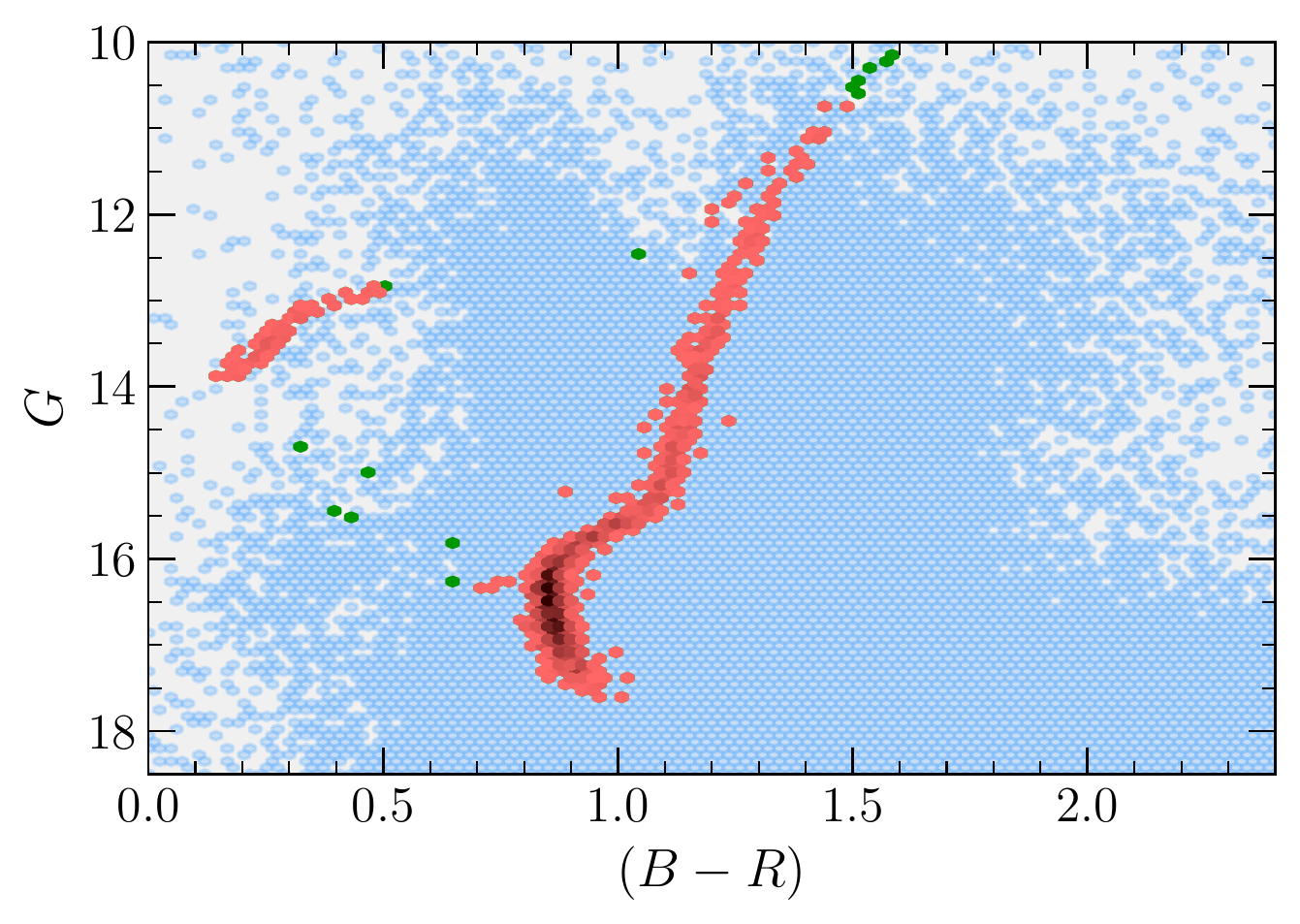}
\caption{
{Color magnitude diagram} (CMD) of NGC~6397 {\sc Gaia DR2} data, after different filtering steps. The \emph{blue points} indicate the stars that failed the PM error, astrometric and photometric flags, maximum projected radius and PM filters. The \emph{green points} the stars that passed these 3 filters but are offset from the CMD. 
The \emph{red points} show the {\sc Gaia} sample of stars after the the previous filters and  subsequent CMD filtering, color-coded from red to dark red, according to increasing star counts.
\label{fig: CMD-Gaia}}
\end{figure}


We  CMD-filtered  the {\sc Gaia} data roughly following  the same KDE method as we used for the HST data.
The only differences were that 1) we used the equivalent filters $G$, $B$ and $R$, and 2) we selected the 3 $\sigma$ region of the KDE, instead of 2 $\sigma$, since the latter appeared to be too conservative a cut for {\sc Gaia}. Figure~\ref{fig: CMD-Gaia} displays the final stars after this filter, along with the previously filtered subsets.

\subsubsection{Removal of X-ray binaries}
\label{sssec: Gaia-XRB}
X-ray binaries are also in the {\sc Gaia} sample. We therefore removed the five X-ray binaries that were matched to {\sc Gaia} stars. 

\subsubsection{{\sc Gaia} final numbers and comparison to HST}

\begin{figure}
\centering
\includegraphics[width=0.9\hsize]{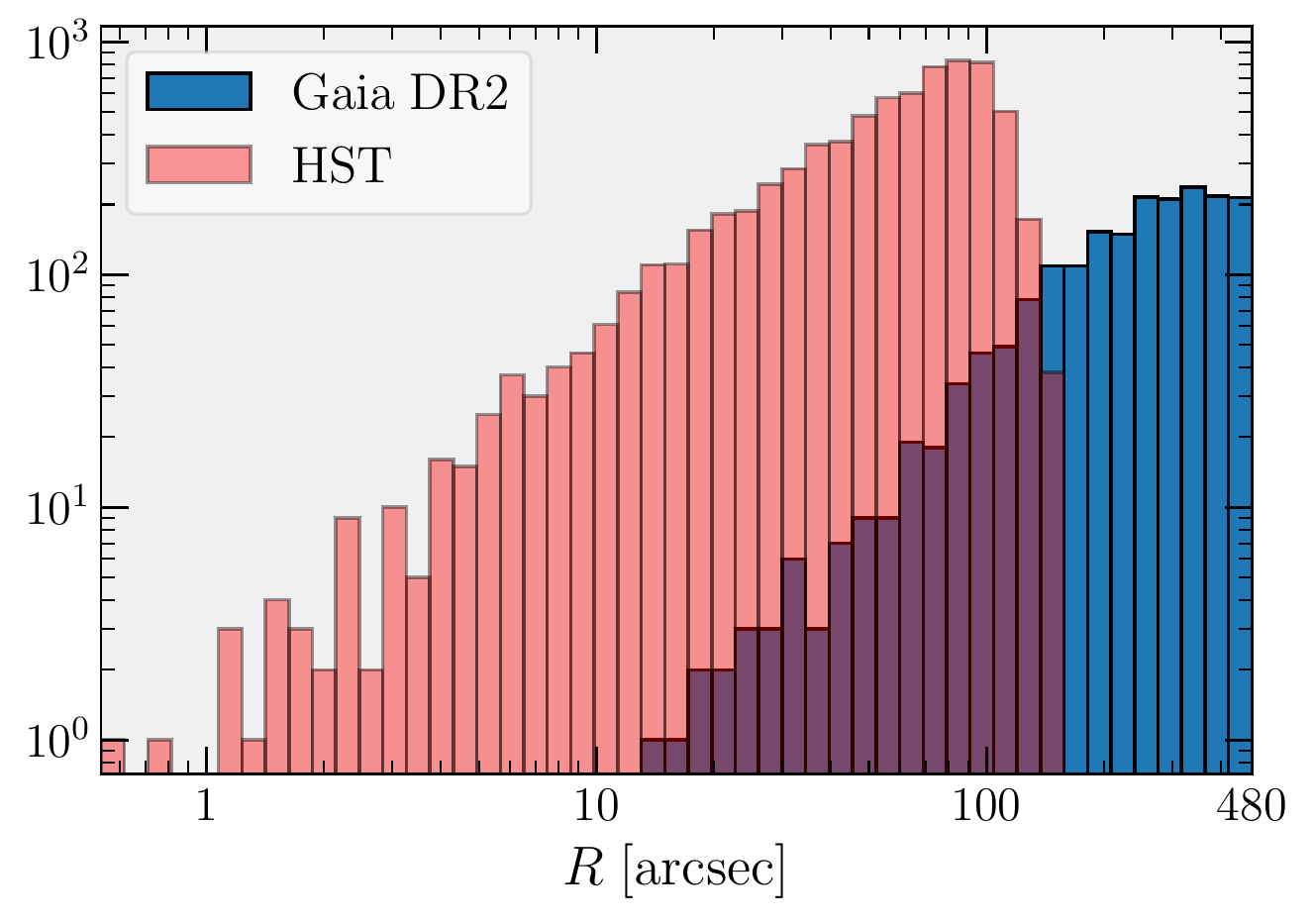}
\caption{
Distribution of projected radii of stars from the GC center of \protect\cite{Goldsbury+10}, with cleaned HST data in red and cleaned {\sc Gaia DR2} data in blue. This plot indicates that HST is much better suited to probe a possible IMBH in the center.
\label{fig: hist-rproj}}
\end{figure}

The cleaned {\sc Gaia} sample contains 1905 stars.
%
Figure~\ref{fig: hist-rproj} shows that the spatial coverage of {\sc Gaia DR2} is indeed poorer than HST when taking into account only stars 
from the clean samples. While the cleaned HST sample contains 221 stars within the $9''$ IMBH radius of influence (for an IMBH with a mass of $600$ M$_\odot$ as measured by \citealt{Kamann+16}, see Sect.~\ref{ssec: hst}), the cleaned {\sc Gaia} presents none, which is clearly insufficient to provide reliable results on the presence of an IMBH.


\subsection{MUSE data cleaning}
\label{ssec: MUSEclean}

We filtered the MUSE sample by removing stars whose probability of being a member of the GC, according to their position in velocity - metallicity space (\citealt{Kamann+16},  kindly provided by S. Kamann), was less than 0.9. This step left us with 6595 stars among the original 7130. We then removed stars with LOS velocity errors greater than $2.232\,\rm km\,s^{-1}$ (half of the GC velocity dispersion), as well as stars that did not match the HST stars in a symmetric $\leq 1$ arcsec match.
We were left with 532 stars, of which 4 were previously identified as X-ray binaries (see section~\ref{sssec: x-ray}), 
yielding a final MUSE sample containing 528 stars, thus adding LOS velocities to $\sim 7\%$ of the HST sample.


Our Gaussian prior on $v_{\rm LOS}$ used the mean 
provided by \cite{Husser+16}: $\langle v_{\rm LOS} \rangle = 17.84 \,\kms$.
For the uncertainty on $\langle v_{\rm LOS} \rangle$, we allowed a much wider dispersion of $2.5\,\kms$ instead of the $0.07\,\kms$ of \citeauthor{Husser+16}, given the relatively wide range of bulk LOS velocities reported in other studies
\citep{Milone+06,Lind+08,Carretta+09b,Lovisi+12}.

\subsection{Merging of the different datasets}
\label{ssec: stack-data}

Although we ran \mpo\ on HST and {\sc Gaia} individually for data homogeneity, we preferred to combine the data from HST, MUSE and {\sc Gaia} to probe the mass and velocity anisotropy parameters with more accuracy.
We started with the HST filtered subset, which accounted for most of the data we would use, and then we merged it to the other datasets following the steps below:

\begin{enumerate}
    \item We restricted the {\sc Gaia} stars to those with $G$ magnitudes within the limits of F606W magnitudes from the HST subset, (i.e., 16.11 and 22.14), given the quasi equivalence between these two filters. This step ensured that mass segregation effects  would be the same for all data subsets.
    \item We removed {\sc Gaia} stars that were symmetrically matched to HST stars to better than 1~arcsec (see Sect.~\ref{sssec: posacc}) since HST PMs presented smaller errors (Fig.~\ref{fig: PMerrs}).
    \item We incorporated LOS velocities from MUSE, according to the approach described in~\ref{ssec: MUSEclean}.
\end{enumerate}

After these steps, we were left with 8255 stars: 7209 of which had PMs from HST, with  583 of those presenting LOS velocities from MUSE, as well as 1046 additional stars with PMs from {\sc Gaia DR2}.


\section{Streaming motions in NGC~6397}
\label{ssec: target}


\subsection{Rotation}
\label{ssec: rotation}
The presence of rotation in quasi-spherical systems makes the kinematical modeling more difficult. Any rotation will be interpreted by codes neglecting it, such as  {\sc MAMPOSSt-PM}, as disordered motions, and should lead to different mass profiles and deduce velocity anisotropies that are more tangential than in reality. 
    
Fortunately, NGC~6397 does not appear to have significant amount of rotation.
Indeed, while \cite{Gebhardt+95} found that the integrated light of the core of NGC~6397 showed rotation with a projected amplitude of $2\,\rm km\,s^{-1}$,
\cite{Kamann+16} concluded with individual stars from VLT/MUSE observations (see~Sect.~\ref{ssec: muse} below) that rotation contributes negligibly to the second velocity moment in this GC.
\cite{Vasiliev19c} reported a typical systematic uncertainty in {\sc Gaia DR2} PMs of up to $\sim 0.02$ mas yr$^{-1}$ at any radius
and considered rotation to be confirmed only when its peak amplitude exceeded $\sim 0.05$ mas yr$^{-1}$, which was not the case for his NGC~6397 measurements.
On the other hand, \cite{Bianchini+18} found POS rotation in \ngc\ with {\sc Gaia DR2} at a $2\,\sigma$ level for this GC, but with $v/\sigma = 0.03$ only, the smallest value of the 51 GCs they analyzed. This is consistent with the quasi-null mean POSt velocity profile seen in Figure~\ref{fig: pseudo-anis}. 
Finally,     \cite*{Sollima+19} detected a rotation of $0.48\,\rm km \,s^{-1}$ from a 3D analysis, which is less than 10 per cent of its velocity dispersion.
We therefore neglect rotation in NGC~6397.

\subsection{Radial motions}

Similarly to rotation, any radial streaming motions will be interpreted by codes such as \mpo\ as radial dispersion.
However, as seen in the top panel of Figure~\ref{fig: pseudo-anis}, the {\sc Gaia DR2} PMs do not show strong signs of radial motions.

\section{Practical considerations}

We now discuss the practical implementation of \mpo\ to our cleaned sample of the stellar motions inside \ngc.

\subsection{Surface density}
\label{ssec: surf-dens}
\subsubsection{Basic approach}
\label{sssec: radcomp}
Eq.~(\ref{eq: pilop}) requires the knowledge of the GC surface density (SD) profile, $\Sigma_{\rm GC}(R)$.
More importantly, Eq.~(\ref{eq: gsystem}) requires the knowledge of the 3D number density $\nu(r)$ when integrating the local VDF along the LOS. 
While {\sc MAMPOSSt-PM} has a mode where it  jointly fits the parameters of $\nu(r)$, $M(r)$ and $\beta(r)$ to the distribution of stars in projected phase space (see eq. [11] of \citealt{Mamon+13}, which is different from our eq.~[\ref{eq: pilop}]), this requires the data to have constant completeness as a function of projected radius.
If the
astrometric measures of PMs are not independent of projected radius, one first needs to estimate (and deproject) the SD profile based on a wider dataset than that with PM values to obtain priors on the parameters of $\nu(r)$. Such an analysis had been  performed for mass-orbit modeling of galaxy clusters with spectroscopic information whose completeness depended on projected radius (e.g., \citealt{Mamon+19}).

\begin{figure}
    \includegraphics[width=\hsize]{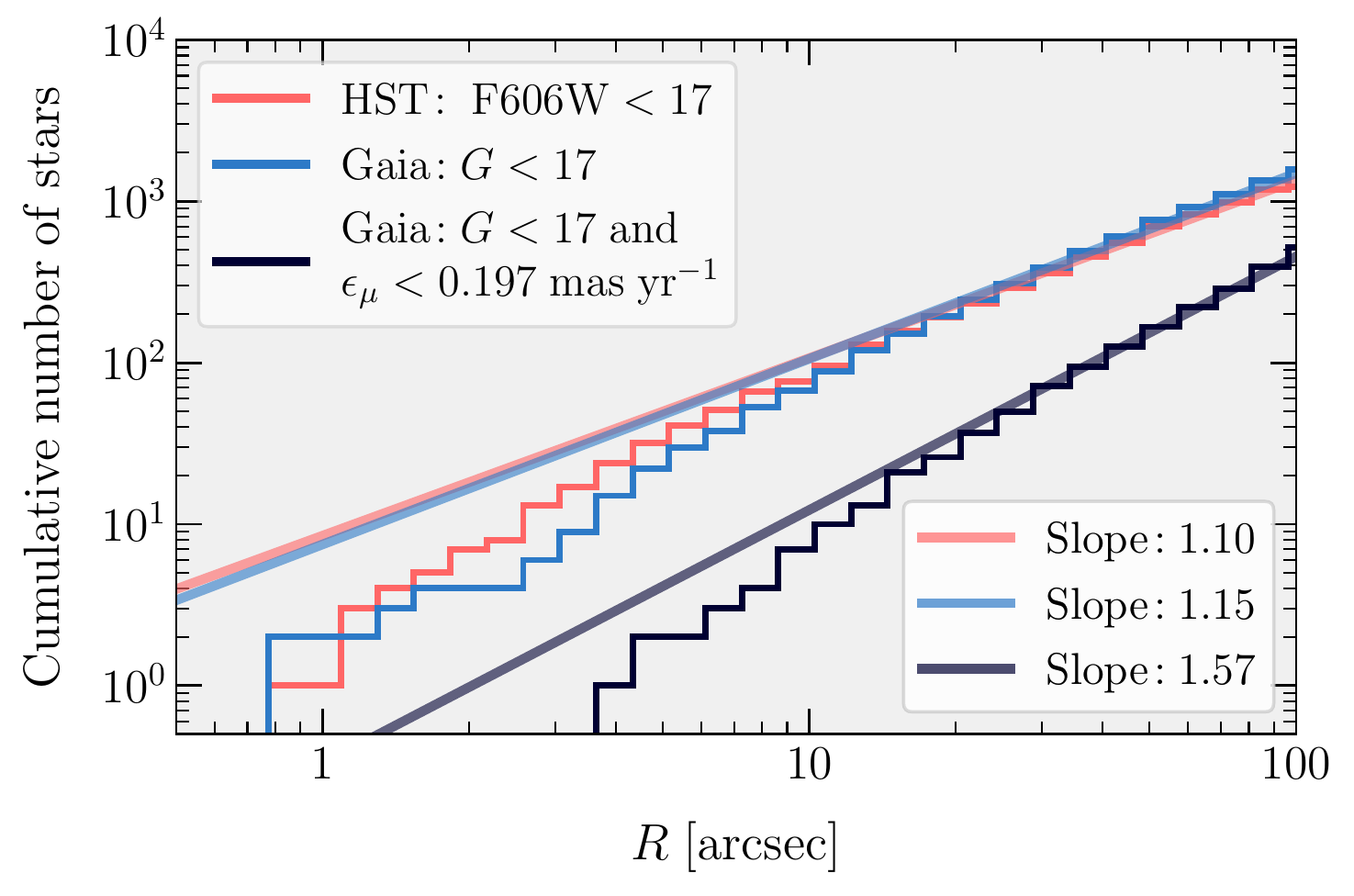}
    \caption{Cumulative distribution functions of projected radii for $G < 17$ {\sc Gaia} stars (\emph{blue}), for the subset of $G < 17$ {\sc Gaia} stars with precise PMs from eq.~(\ref{eq: err_lind18}) (\emph{black}), and for the subset of HST stars (\emph{red}). The \emph{straight lines} are power laws to guide the eye. 
    }

    \label{fig: RCDF_Gaia}
\end{figure}

The PM data are likely to be incomplete at small projected radii because of the increased crowding of stars as one moves inward, especially for {\sc Gaia}, whose mirrors are smaller  than that of HST.
We  noticed that {\sc Gaia} data with PMs traced less well the inner cusp of the SD than the full {\sc Gaia} data (requesting only positions and $G$ magnitude). 
Figure~\ref{fig: RCDF_Gaia}  shows that the {\sc Gaia} PM subsample with $G < 17$ is indeed incomplete
in the inner regions relative to the full $G < 17$ {\sc Gaia} sample since the distribution of projected radii of {\sc Gaia} stars from the GC center  is  shallower for stars with precise PMs than for all stars (including those without PM measurements).
A Kolmogorov-Smirnov test indicates that there is only $10^{-5}$  probability that the distributions of projected radii of the subsample with precise PMs and that of the full sample (including stars without PM measurements and those with imprecise ones) arise from the same parent population. 
Another difficulty is to measure the outer envelope given confusion with field stars. This is important because \mpo\ computes  outward integrals.

We therefore needed to provide \mpo\ with a  precomputed  SD profile.
In \mpo, this profile had to be a simple analytical one- or two-parameter model (where free parameters are scale and possibly shape). More precisely, \mpo\ assumes  Gaussian priors on the  precomputed SD profile parameters. 

\subsubsection{Choice of model and main parameters}
\label{sssec: SD-sersic}

The choice of a good model for SD is crucial because the mass of a possible IMBH is linked to the inner slope of the SD profile \citep{vanderMarel&Anderson&Anderson10}. 
\ngc\ has long been known to have a cuspy (steep) inner SD profile
 \citep{Auriere82,Lauzeral+92,Lugger+95,Noyola&Gebhardt&Gebhardt06,Kamann+16}, 
which prompted \cite{Djorgovski&King86} to classify it as core-collapsed. 
Unfortunately, no simple analytic model was ever fit to the SD profile of NGC~6397 extending to large projected radii.\footnote{\cite{Martinazzi+14} and \cite{Kamann+16} fitted the SD profile of \ngc\ to large projected radii with Chebyshev polynomials (of $\log \Sigma$ vs. $\log\,R$) and multiple Gaussians, respectively. The former  have no analytical deprojections, and while Gaussians are easily deprojected, the multiple Gaussians involve too many parameters for \mpo.}
%
%
\cite{Trager+95} estimated 
an effective (projected half-light) radius $R_{\rm e}=2\farcm90$ for NGC~6397 using ground-based data and analyzing the SD profile in different apertures.
Using {\sc Gaia}~DR2 data,
\cite{deBoer+19} recently inferred $R_{\rm e}=3\farcm35$ (which we infer from their 3D half-luminosity radius $r_{\rm h}$, coupled with their assumed distance and with the ratio $R_{\rm e}/r_{\rm h} \simeq 0.732$ kindly estimated for us by M.~Gieles). But their analysis was based on a fit to two dynamical models that both assumed a cored inner density profile, which clearly does not represent the inner SD profile of NGC~6397.

We thus had to estimate the SD profile ourselves.
%
The cuspy profile led us to fit a S\'ersic model (\citealt{Sersic63}, \citealt{Sersic68}) to the distribution of projected radii since the S\'ersic model has a shape parameter (S\'ersic index) that makes it flexible to handle both cuspy and cored density profiles (e.g., Fig.~A.1 of \citealt{Vitral&Mamon20}).
Furthermore, there are simple analytical forms for the precise (but not exact) deprojection of the S\'ersic model.
In  appendix~\ref{app: surf-dens}, we extend the analysis of \cite{Vitral&Mamon20} of the precision of the different analytical approximations as a function of 
$R_{\rm e}$ and $n$, to even smaller values of $R/R_{\rm e}$ representative of our data.
\mpo\ will then use the best-fit values of $\log R_{\rm e}$ and $n$ and their uncertainties as the respective mean and standard deviation of the Gaussian priors.

We required HST data to fit correctly the inner cusp, and we also needed {\sc Gaia} data to fit correctly the outer envelope. 
We therefore fitted the GC surface density profile $\Sigma_{\rm GC}(R)$ together with the field star surface density $\Sigma_{\rm FS}$ (assumed constant) from the {\sc Gaia}~DR2 photometric data, supplemented by the HST data to better probe the very inner regions. 
The two datasets are not cleaned in any way, except for cuts in magnitude and maximum (minimum) projected radius.
The HST data may not be radially complete (we did not have a parent sample with all stars including those without PM measurements), but this is the best we can do.
Our approach does not remove the Milky Way field stars, but instead incorporates them in our SD fit.
It allows us to keep {\sc Gaia} stars with no PM information, which would be filtered out in a PM cut, even though their contribution to the surface density shape is relevant.


\subsubsection{Stitching HST and {\sc Gaia}}
\label{sssec: SD-fit}

\begin{figure*}
\includegraphics[width=0.47\hsize]{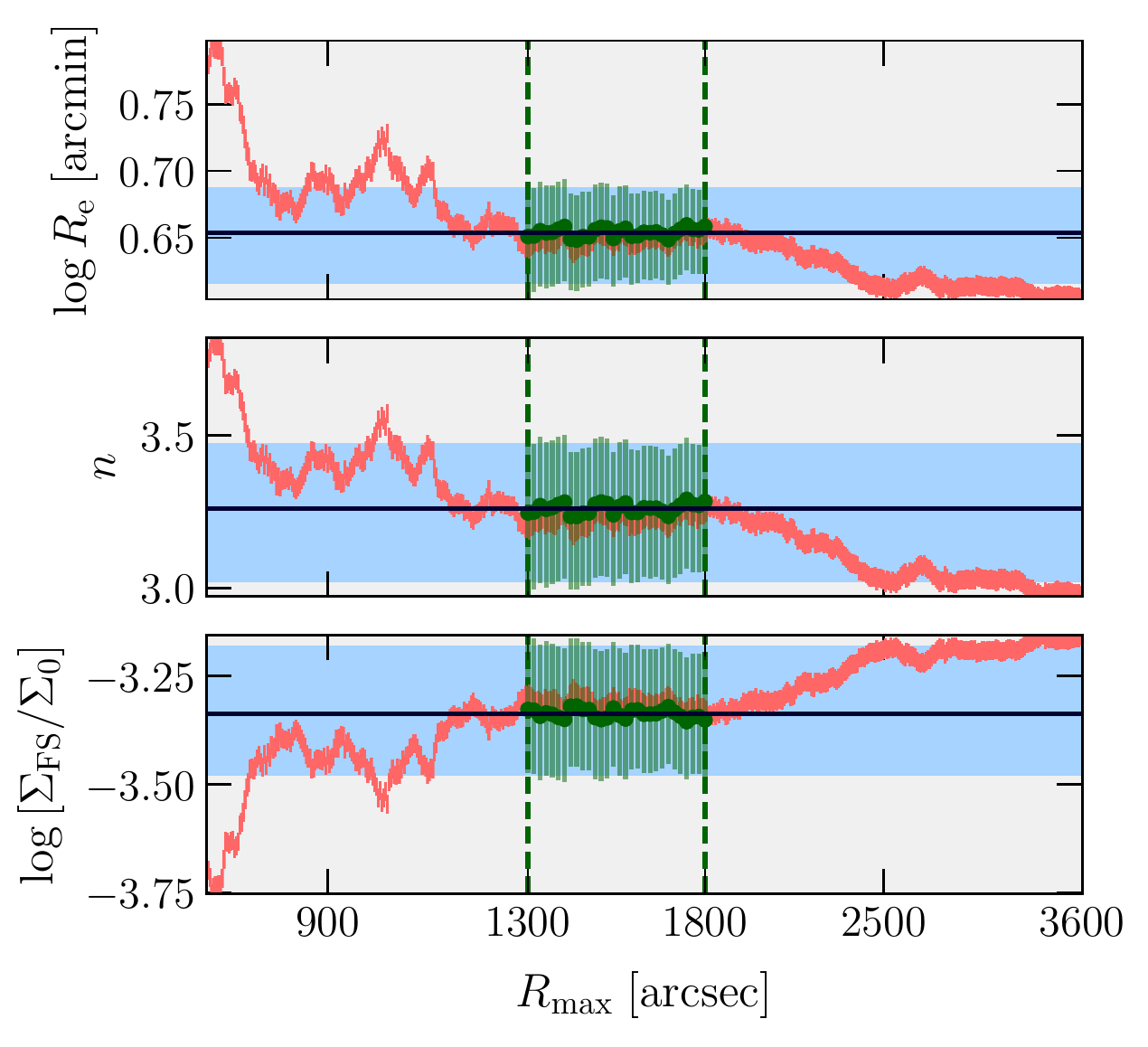}
\includegraphics[width=0.47\hsize]{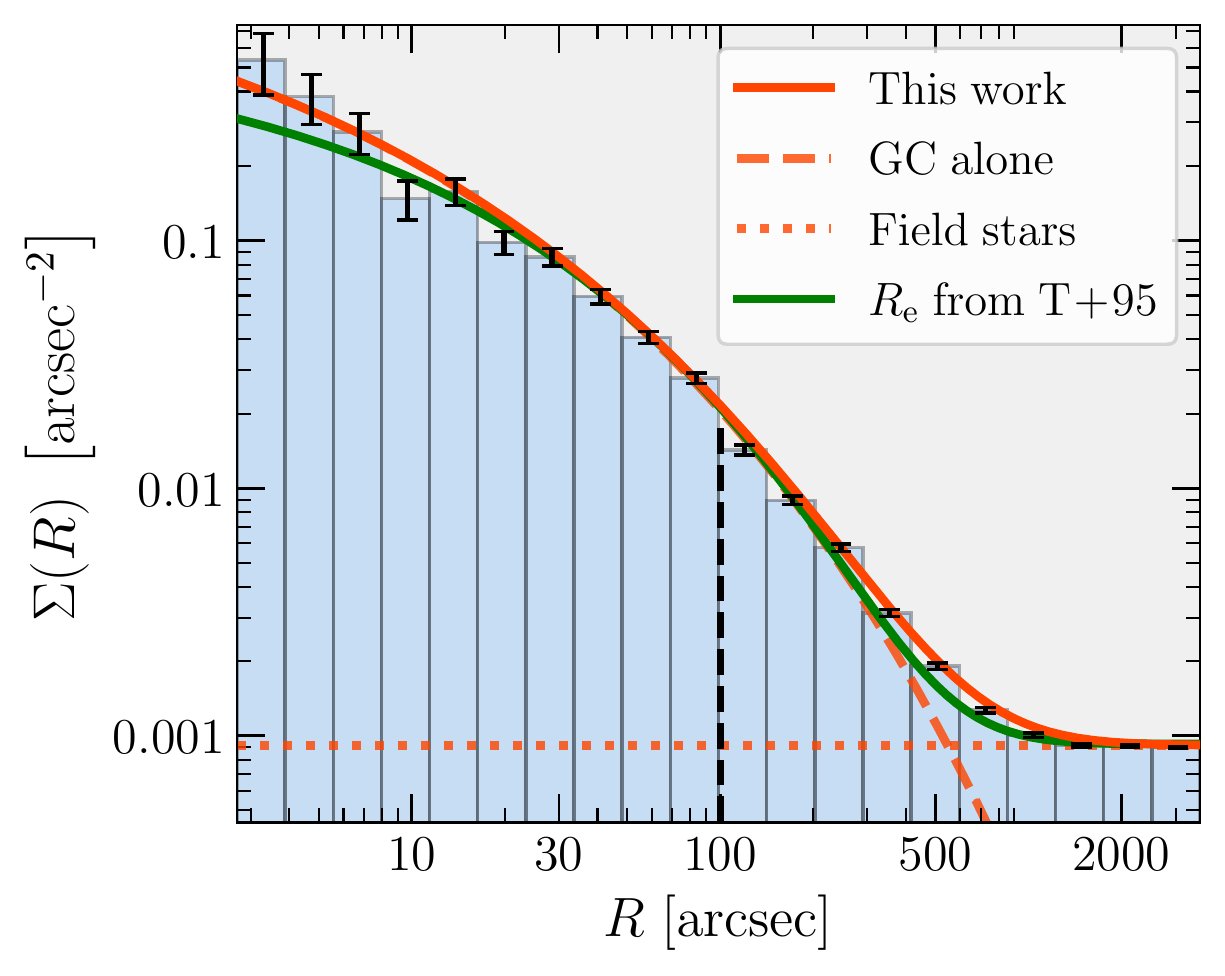}
\caption{Fits to the stellar surface density profile of \ngc.
\textbf{Left}: S\'ersic plus constant field star surface density fits, 
as a function of the maximum allowed radius, $R_{\rm max}$. 
The values of $R_{\rm e}$, $n$, and $\Sigma_{\rm FS}$ obtained by MLE for 500 values of $R_{\rm max}$ are shown in \emph{red}. 
The \emph{black horizontal line} and \emph{light blue shaded region} respectively show  the mean marginal values of 30 MCMC fits and their uncertainties,  performed in the range of $R_{\rm max}$ (delimited by the \emph{green vertical lines}) where $R_{\rm e}$, $n$ and $\Sigma_{\rm FS}$ are roughly independent of $R_{\rm max}$.
%
\textbf{Right}: Goodness of fit of the surface density profile. The \emph{histogram} shows the empirical profile, using 20 logarithmic radial bins extending from the innermost data point to 1 degree. 
The \emph{curves} show different models: our MCMC fit (\emph{red}) 
of a S\'ersic model (\emph{dashed}) plus constant field stars SD (\emph{dotted}), as well as the total (\emph{solid}) to compare with the data. We also show the S\'ersic plus constant background prediction from MLE fits, but with the effective radius fixed to the value found by \citeauthor{Trager+95}  (\emph{green}). 
The \emph{vertical black line} highlights the separation between HST and {\sc Gaia} data used for the fits.
%
\label{fig: SD-fit}}
\end{figure*}


%

We stitched the HST and {\sc Gaia} datasets as follows:
\begin{enumerate}
    \item We cut the  {\sc Gaia} stars (GC stars plus interlopers) at $G=17$, where we expect negligible mass segregation. We explored different magnitude limits, but moderate changes affected little our fits. However, using very faint magnitude limits ($G = 19$ or 20.5), we noticed a lack of continuity in the transition from the HST to the {\sc Gaia} radial range, which appears to be a consequence of overcrowding in the cluster's center \citep{Arenou+18}, where fainter magnitude stars are more difficultly detected.
    
    \item We counted the number $\mathcal{N}_{\rm G}$ of {\sc Gaia} stars inside the annulus
    $2\farcs7 < R < 100''$,
    corresponding to the region where our HST data seems complete (see Figure~\ref{fig: hist-rproj}). 
    
    \item We removed the {\sc Gaia} stars in that annulus and added the $\mathcal{N}_{\rm G}$ HST brightest stars (a tiny fraction of which may be interlopers) in that annulus, so we could correctly probe the inner surface density profile.
\end{enumerate}


\subsubsection{Surface density results}
\label{sssec: sdens_results}

We performed {maximum likelihood estimation} (MLE) to fit $\log R_{\rm e}$ and $n$ to the distribution of projected radii, using a model SD profile with a S\'ersic GC plus a uniform (constant SD) field star contribution, considering the data up to a maximum projected radius, $R_{\rm max}$.\footnote{We performed MLE using 
{\sc scipy.optimize.differential\_evolution} in {\sc Python}.}
This maximum radius must be chosen large enough to capture the behavior of the SD profile in the outer envelope of the GC as well as the contribution of the field stars, but not too large to be overwhelmed by the field stars in the MLE fit.

We performed such MLE fits of the SD profile for 500 values of log-spaced $R_{\rm max}$ between $0\fdg2$ and $1^\circ$.
Figure~\ref{fig: SD-fit} (left) shows a plateau for $R_{\rm max}$ in the range  $\sim 1300'' - 1800''$, where  $R_{\rm e}$, $n$ and the field star SD are each indeed constant.
At lower (resp. higher) $R_{\rm max}$ the field star SD fluctuates at lower (higher) values with corresponding higher (lower) and fluctuating $R_{\rm e}$ and $n$.


For better precision, we then performed 30 MCMC runs for log-spaced $R_{\rm max}$ between 1300 and 1800 arcsec and assigned the mean of best-fitted values and their uncertainties (green crosses and error bars of Figure~\ref{fig: SD-fit}, respectively) as the global S\'ersic radius and index means and uncertainties. 
The results are displayed in
Figure~\ref{fig: SD-fit} (bottom).
We found $R_{\rm e}=4\farcm51\pm0\farcm36$ and $n=3.26\pm0.23$.
Our effective radius is significantly higher than the estimates of \citeauthor{Trager+95} and \cite{deBoer+19}, which both fail to properly capture the cusp. Indeed, the S\'ersic SD profile with the same index ($n=3.26$) but with the \citeauthor{Trager+95} effective radius  (green curve in the right panel of Fig.~\ref{fig: SD-fit}) fails to reproduce the steep inner SD profile.

\subsection{Multiple populations}
\label{sec: multi-pops}


Mass-orbit modeling codes such as \mpo\ assume a given population with a given SD profile, as well as given kinematics (i.e., $p({\bf v}|R)$ for \mpo). Systems with multiple populations, each with their SD profiles and kinematics, should be analyzed  jointly. \mpo\ can handle such multiple populations. In practice, we  predetermine priors for the SD profile of each population, and then run \mpo\ with several tracer components. 


There are two mechanisms that bring multiple populations in
GCs. 
First, the formation of a GC may occur in several episodes, with different populations with their  own chemistry and kinematics.
Second, the short two-body relaxation time of GCs in general and especially of core-collapsed GCs such as \ngc, allows for energy exchanges between stars that are sufficient to drive them toward energy equipartition. This leads to mass segregation, where massive stars are confined to smaller radii 
and have lower typical velocities at a given radius.
We examine these two mechanisms in turn.

\subsubsection{Multiple chemical populations}
\label{ssec: evolpath}

Many GCs show signs of different chemical populations (e.g., \citealt{Carretta+09b}), in particular \ngc\   \citep{Carretta+09,Milone+12b}.
Unfortunately, the separation of stars into different chemical populations in \ngc\ was only performed with HST data limited to small projected radii. Such a chemical separation of stars is impossible with the {\sc Gaia} data, for lack of sufficient wavebands to provide a large enough dimensionality of colors to distinguish chemical populations. 
Moreover, this requires cleaning the data for differential reddening. 
Therefore, the splitting of the stars of \ngc\ into chemical populations, while possible, is beyond the scope of the present article.

Besides, 
\cite{Cordoni+20}, who analyzed the spatial distributions and kinematics of seven GCs (other than \ngc) split by their two detected chemical populations, found that only the two GCs with the highest two-body relaxation times showed signs of different kinematics between their two chemical populations.
If the two populations have roughly the same mass functions, then two-body relaxation should wash out any differences in their positions in projected phase space \citep{Vesperini+13}, except in their outer regions where two-body relaxation is incomplete.
The very low relaxation time of \ngc\ (600 Myr at its half-mass radius according to \citealt{Baumgardt+19}) thus suggests that its chemical populations should have similar kinematics, at least within its effective radius of $4\farcm5$ as we found above.
\subsubsection{Mass segregation}
\label{sssec: mass-seg}

We explored the range of masses of the stars in our datasets by comparing synthetic CMDs to our observed ones.
Following \cite{CarballoBello+12},
we generated a CMD with a set of isochrones from the \textsc{PARSEC} (\citealt{Marigo+17}, \citealt{Bressan+12}, \citealt{Pastorelli+19} and references therein) software. We provided $n_{\rm inTPC} = 15$ since \cite{Marigo+17} mention this suffices to recover the main details of thermal pulse cycle variations in the evolutionary tracks. We also assigned $\eta_{\rm Reimers} = 0.477$, which is the median value of the Reimers coefficient (\citealt{Reimers75}) among clusters, provided by \cite{McDonald&Zijlstra15}; in their figure~4, they showed that NGC~6397 presents a value of $\eta_{\rm Reimers}$ corresponding to this value within $\sim 1\,\sigma$. 
We adopted an age of 12.87 Gyr and a metallicity of [Fe/H] = --1.54 
from \cite{MarinFranch+09}, 
while adopting our distance of 2.39 kpc (Sect.~\ref{ssec: distance}) and a Galactic extinction $A_V=0.50$ obtained from $E(B-V)=0.160$ 
\citep{Schlafly&Finkbeiner11}\footnote{We used the Galactic Dust Reddening and Extinction tool, https://irsa.ipac.caltech.edu/applications/DUST/ in the center of \ngc.} and $R_V=3.1$.
We note that this value of extinction is 14\% lower than $E(B-V)=0.188$ found in the older reddening maps of \cite{Schlegel+98}, but consistent with $A_V=0.52\pm0.02$ found by \cite{Valcin+20} despite their larger distance of 2.67 kpc (Table~\ref{tab:dist}).

\begin{figure}
\centering
\includegraphics[width=\hsize]{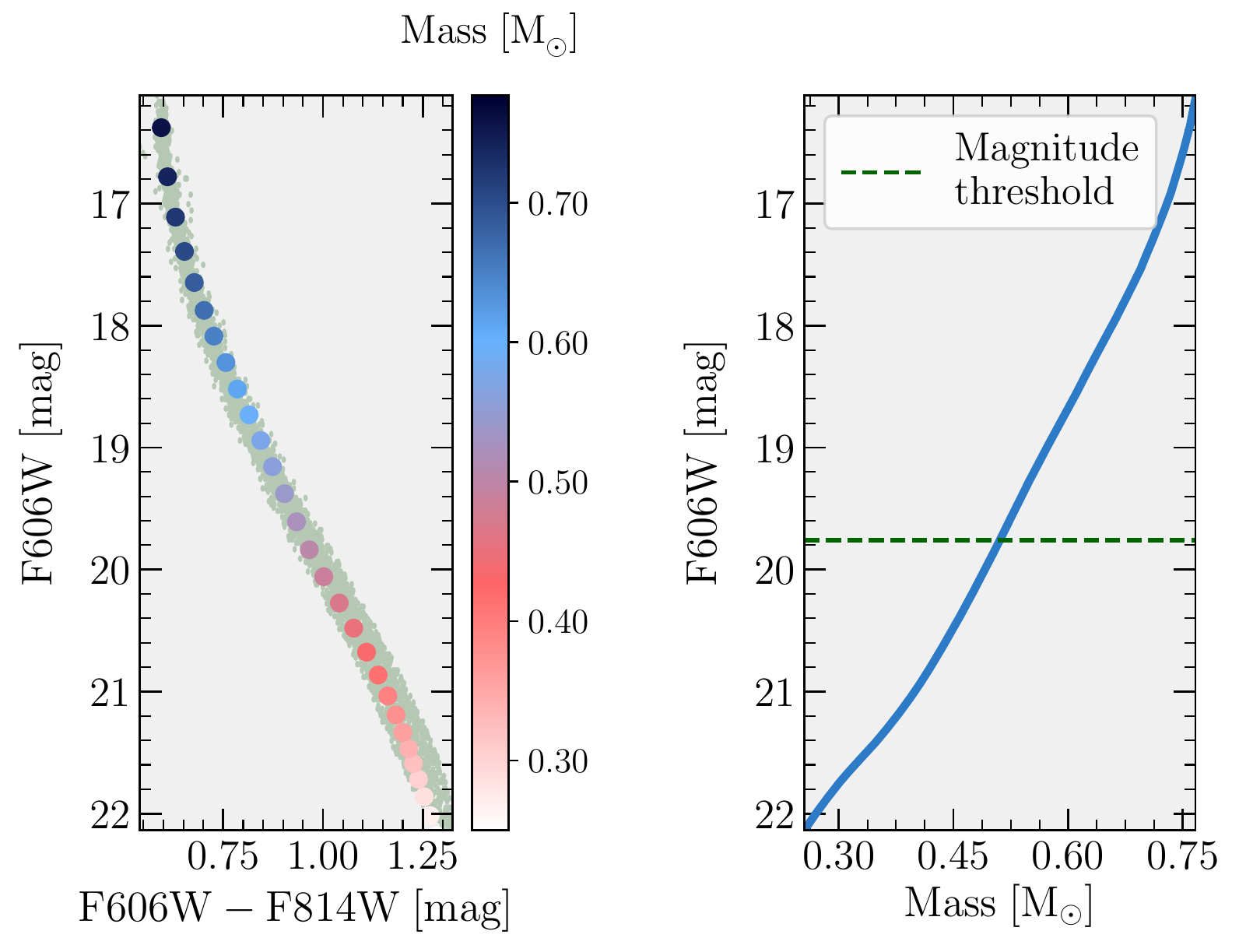}
\centering
\caption{Stellar masses.
{\bf Left}: Comparison of the cleaned HST data and the fitted \textsc{PARSEC} isochrone, displaying the expected stellar mass relative to each main-sequence position.
{\bf Right}: magnitude - mass relation from the {\sc PARSEC} model. The \emph{green dashed horizontal line} shows  the magnitude threshold that we used to split our sample into two mass populations.
}
\label{fig: isochrone-parsec}
\end{figure}

The left panel of Figure~\ref{fig: isochrone-parsec} shows that the {\sc PARSEC} model fits well the CMD of the cleaned subsample of HST stars.
The mass ratio among the heaviest and the
lightest stars in our cleaned HST sample
is 2.57, sufficiently high to worry about the possibility of mass segregation in our cleaned HST sample. 

In fact, mass segregation is visible in NGC~6397 \citep{Heyl+12,Martinazzi+14}.
Analyzing HST stars in the range $3' < R < 7\farcm5$,
\citeauthor{Heyl+12} found mass segregation of main-sequence stars: brighter stars show
4\% lower median projected radius and 12\% lower median PM moduli than fainter stars.
In a subsequent study by the same team, \cite{Goldsbury+13} showed that two characteristic radii scale as a $M^\gamma$, with $\gamma=-1.0\pm0.1$ for \ngc. 
\citeauthor{Martinazzi+14}
found that the mean mass of main-sequence stars drops with \emph{physical} radius (after deprojection) from $>0.7\,\msun$ for $r < 10''$ to $\simeq 0.56\,\msun$ for $r>100''$ (after correcting for  the completeness with magnitude,  estimating it by adding artificial PSF-convolved stars to the images), thus a 20\% effect.


\textsc{MAMPOSSt-PM} is able to treat multiple stellar populations together. We therefore also performed SD fits with two populations of stars, as we explain below.


The magnitude threshold we chose to separate the two population of stars was based on the analysis of \cite{Heyl+12}: Their figures 7 and 12 indicate that NGC~6397 main-sequence stars present a (small) variation of radial distribution and velocity dispersion profiles, respectively, at a magnitude of $\rm F814W\sim 18.75$, which corresponds to $\rm F606W= 19.76$. We thus used this limit to divide our cleaned subset of 8255 stars into two populations, one with brighter magnitudes (6527 heavier stars) and the other with fainter ones (1728 less massive stars). 

Unfortunately, we could not perform a surface density fit as robust as for the single population case because 1) our data was considerably incomplete at higher magnitudes (the fainter subset) and 2) as  mentioned in Sect.~\ref{sssec: SD-fit}, the HST plus {\sc Gaia} stacked subset presented discontinuous trends when allowing stars with fainter magnitudes ($G>17$). 
We therefore let \mpo\ fit the SD profile of each population from the kinematics only, more precisely from the conditional probabilities $p({\bf v}|R)$. We adopted 
Gaussian priors for $\log R_{\rm e}$, with mean equal to that  found by our previous SD fit of the single population (Sect.~\ref{ssec: surf-dens}) and a wide (0.2 dex) uncertainty. 
We also considered Gaussian priors for the S\'ersic indices, but with lower  uncertainties to avoid a degeneracy in its marginal distribution.

We helped \textsc{MAMPOSSt-PM} by providing narrow Gaussian mass priors for each population. We derived mass fractions for each population using the power-law main-sequence stellar mass function of slope $\alpha=-0.52$,\footnote{The slope $\alpha=-0.52$ of the main-sequence stellar mass function of \ngc\ is given in H. Baumgardt's very useful web site on GCs, https://people.smp.uq.edu.au/HolgerBaumgardt/globular.}
together with the mass limits of our subsets. Indeed, a power-law relation $\diff N / \diff m \propto m^{\alpha}$ implies a total stellar mass in the range of stellar masses $(m_1,m_2)$ of
\begin{equation} \label{eq: mass-frac}
    M_{\rm total} \propto \int_{m_{1}}^{m_{2}} m \frac{\diff N}{\diff m} \diff m = \frac{m_{2}^{2+\alpha} - m_{1}^{2+\alpha}}{2+\alpha} \ ,
\end{equation}
and thus derive
\begin{equation} \label{eq: mass-frac-bis}
    \frac{M_{\rm bright}}{M_{\rm faint}} = \frac{m_{\rm bright}^{2+\alpha} - m_{\rm cut}^{2+\alpha}}{m_{\rm cut}^{2+\alpha} - m_{\rm faint}^{2+\alpha}} \ ,
\end{equation}
where $M_{\rm bright}$ and $M_{\rm faint}$ are the masses of the brighter and fainter populations, respectively and $m_{\rm bright}$, $m_{\rm faint}$ and $m_{\rm cut}$ are the respective highest, lowest and two-population threshold masses of the global subset.
With 
$m_{\rm bright} = 0.77\,\msun$,
$m_{\rm cut} = 0.51\,\msun$,
and
$m_{\rm faint} = 0.25\,\msun$ (right panel of Fig.~\ref{fig: isochrone-parsec}), Eq.~(\ref{eq: mass-frac-bis}) yields a bright mass fraction of 0.56.
Since our main mass estimates of NGC~6397's mass with one single population were centered around $10^5$ M$_{\odot}$, we passed logarithmic Gaussian priors to each population, centered at
$\log (M/\msun) = 4.7$,
with standard deviation of $0.05$.







\subsubsection{\mpo\ assumptions and priors}
\label{sssec: assume}
To summarize, \mpo\ assumes that \ngc, lying at a distance of 2.39 kpc, with its center at the position found by \cite{Goldsbury+10}, is a spherical self-gravitating  system with no internal streaming motions (rotation, expansion/collapse, etc.). It also assumes that the \emph{local} velocity distribution function is an
ellipsoidal 3D Gaussian aligned with the spherical coordinates.

By relying on the Jeans equation, \mpo\ assumes that stars are test particles orbiting the the gravitational potential, and that they do not interact with one another. While binary stars have their own motions and violate the test particle  hypothesis, these are in great part filtered out of our data. However, the low two-body relaxation time of \ngc\ means that star-star interactions are not negligible, which violates the test particle assumption.

We now summarize our adopted priors.
The surface density profile of GC stars is assumed to follow the S\'ersic law.  
We adopted Gaussian priors for $\log R_{\rm e}$ and for $n$ (see Sect.~\ref{ssec: surf-dens}) for both single- and two-population analyses. 

Following Sect.~\ref{sssec: pm-filter}, we adopted Gaussian priors on the bulk LOS velocity and PM of the GC  and Pearson~VII priors on the modulus of the bulk PM of the field stars.
The field star LOS distribution is Gaussian with a wide (1.7 dex) uncertainty.
The log ratio of field stars to GC stars is flat between --5 and --2. 

Our standard runs assume isotropic velocities.  We also ran cases with anisotropic velocities,  and following \cite{Mamon+19} we used flat priors on
$\beta_{\rm sym} = \beta/(1-\beta/2)$, which varies from --2 for circular orbits to 0 for isotropic orbits and +2 for radial orbits, and with $\beta_{\rm sym}\to\beta$ for $|\beta|\ll 1$. Our flat priors were limited to $-1.9 < \beta_{\rm sym} < 1.9$, where the two extremes correspond to $\beta = -38$ and $\beta=0.97$, respectively (see eq.~[\ref{eq: anisotropy}]).

We also adopted flat priors on the GC stellar log mass (3 to 6) for the single-population runs and Gaussian priors with equal means and wide dispersions for the two-population runs (Sect.~\ref{sssec: mass-seg}).
The (sub-)cluster of unseen objects is modeled with different density models with a wide Gaussian (1~dex) prior on the scale radius uncertainty, assuming isotropic velocities.

\section{Analysis}
\label{sec: Analysis}

\subsection{Marginal distributions and covariances}

We explored parameter space to determine marginal distributions and parameter covariances using the same Markov Chain Monte Carlo (MCMC) approach as \cite{Mamon+13,Mamon+19}, see section 4.7 of \cite{Mamon+19}. This makes use of the public {\tt Fortran} {\sc CosmoMC} code \citep{Lewis&Bridle02}.\footnote{https://cosmologist.info/cosmomc/.}
In particular, we used 6 MCMC chains run in parallel and stopped the exploration of parameter space after one of the chains reached a number of steps $N_{\rm steps} = 10\,000 \,N_{\rm free}$, where $N_{\rm free}$ is the number of free parameters of the model.
The only difference is that we now discard the 3000\,$N_{\rm free}$ first steps of each MCMC chain (burn-in phase) instead of the first 2000\,$N_{\rm free}$ as \citeauthor{Mamon+19} previously did.

\subsection{Bayesian information}
\label{ssec: BIC}

There are several ways to compare the different results of \textsc{MAMPOSSt-PM} using different priors. The simplest would be to compare the log-likelihoods. But there is then the risk of over-fitting (under-fitting) the data when using too many (few) free parameters.
We considered two model selection criteria to distinguish our different models and priors.
We first considered the
{corrected Akaike Information Criterion}
\citep{Sugiyara78}
\begin{equation}
      \mathrm{AICc} =\mathrm{AIC} + 2 \, \frac{N_{\mathrm{free}} \,  (1 + N_{\mathrm{free}})}{N_{\mathrm{data}} - N_{\mathrm{free}} - 1} \ ,
\end{equation}
where AIC is the original 
{Akaike Information Criterion}
\citep{akaike1973information}
\begin{equation}
 \label{eq: AIC}
    \mathrm{AIC} = - 2 \, \ln \mathcal{L_{\mathrm{MLE}}} + 2 \, N_{\mathrm{free}} \ ,
\end{equation}
and where ${\cal L}_{\rm MLE}$ is  highest likelihood found when exploring the parameter space, $N_{\rm free}$ is the number of free parameters, and $N_{\rm data}$ the number of data points.\footnote{The main results of this paper use $N_{\mathrm{data}} = 8255$ and $N_{\mathrm{free}} \sim 10$.}
We also considered the {Bayes Information Criterion} (BIC, \citealt{Schwarz78}):
\begin{equation} \label{eq: BIC}
    \mathrm{BIC} = - 2 \, \ln \mathcal{L_{\mathrm{MLE}}} + N_{\mathrm{free}} \ln N_{\mathrm{data}} \ .
\end{equation}

Given that the relative likelihood (given the data) of one model relative to a reference one is 

\begin{equation}
\rm \exp\left(-{AIC-AIC_{\rm ref}\over 2}\right)
\label{eq: pAIC}
\end{equation}
\citep{Akaike83},
a model with a higher AICc (or AIC) relative to a reference model can be ruled out with 95\% confidence if $\rm AICc > AICc_{\rm ref}+6$.
BIC follows the analogous to Eq.~(\ref{eq: pAIC}) for the relative likelihood \citep{Kass&Rafferty95}:
\begin{equation}
\rm \exp\left(-{BIC-BIC_{\rm ref}\over 2}\right) \ ,
\label{eq: pBIC}
\end{equation}
so a model with a higher BIC relative to a reference model can be ruled out with 95\% confidence if $\rm BIC > BIC_{\rm ref}+6$.


Since BIC penalizes extra parameters more (factor $\ln N_{\rm data} \simeq 9.0$)  than AICc (i.e., by a factor two for our dataset), BIC effectively prefers simpler models than does AICc.
BIC is well suited  for situations where the true model is among the tested ones, while AIC(c) is more robust in the opposite case \citep{Burnham&Anderson02}.
In practice, we do not expect any of our models to be true: for example, there is no expectation that the surface density of stars in a GC should precisely follow a S\'ersic model.
To paraphrase the statistician George Box: ``All models are wrong, but some are useful'' \citep{Box79}.\footnote{The angular fluctuation spectrum of the cosmic microwave background appears to be a counter-example.}

We therefore present both AICc and BIC in our results. However, we give preference to AICc whenever having to decide between different models.

\section{Mass-orbit modeling results}
\label{sec: results}

\begin{table*}
\caption{Main results and priors for \textsc{MAMPOSSt-PM} runs, using the merged sample from the cleaned HST, {\sc Gaia} and MUSE datasets}
\label{tab: mpo-NGC6397-imbh}
\centering
\renewcommand{\arraystretch}{1.2}
\tabcolsep=2.5pt
\footnotesize
\begin{tabular}{r@{\hspace{3mm}}cccr@{\hspace{3mm}}cccccrrrccc}
\hline\hline             
\multicolumn{1}{c}{Model} &
\multicolumn{1}{c}{$r_{\beta}$} &
\multicolumn{1}{c}{$\beta(r)$} & 
\multicolumn{1}{c}{$\Sigma$} &
\multicolumn{1}{c}{$N_{\rm free}$} & 
\multicolumn{1}{c}{$R^{-1}$} & 
\multicolumn{1}{c}{$\beta_{0}$} &
\multicolumn{1}{c}{$\beta_{\rm out}$} &
\multicolumn{1}{c}{$r_{\rm scale}$} &
\multicolumn{1}{c}{$n$} & \multicolumn{1}{c}{$M_{\rm BH}$} &
\multicolumn{1}{c}{$M_{\rm CUO}$} &
\multicolumn{1}{c}{$M_{\rm GC}$} & 
\multicolumn{1}{c}{$- \Delta \ln{\mathcal{L}_{\rm max}}$} &
\multicolumn{1}{c}{$\Delta \rm AICc$} &
\multicolumn{1}{c}{$\Delta \rm BIC$} \\
\multicolumn{1}{c}{} &
\multicolumn{1}{c}{[pc]} &
\multicolumn{1}{c}{} &
\multicolumn{1}{c}{} &
\multicolumn{1}{c}{} &
\multicolumn{1}{c}{} &
\multicolumn{1}{c}{} &
\multicolumn{1}{c}{} &
\multicolumn{1}{c}{[arcmin]} & \multicolumn{1}{c}{} & \multicolumn{1}{c}{[M$_\odot$]} & \multicolumn{1}{c}{[$10^{3} $ M$_\odot$]} & \multicolumn{1}{c}{[$10^{4} $ M$_\odot$]} & 
\multicolumn{1}{c}{} &
\multicolumn{1}{c}{} &
\multicolumn{1}{c}{} \\ 
\multicolumn{1}{c}{(1)} &
\multicolumn{1}{c}{(2)} &
\multicolumn{1}{c}{(3)} &
\multicolumn{1}{c}{(4)} &
\multicolumn{1}{c}{(5)} &
\multicolumn{1}{c}{(6)} &
\multicolumn{1}{c}{(7)} &
\multicolumn{1}{c}{(8)} &
\multicolumn{1}{c}{(9)} &
\multicolumn{1}{c}{(10)} &
\multicolumn{1}{c}{(11)} & 
\multicolumn{1}{c}{(12)} & 
\multicolumn{1}{c}{(13)} & 
\multicolumn{1}{c}{(14)} &
\multicolumn{1}{c}{(15)} &
\multicolumn{1}{c}{(16)} \\ 
\hline
\IMBHisoSingle & \multicolumn{1}{c}{--} & iso & S & 8 & $  0.003 $ & \multicolumn{1}{c}{\darkg{\bf 0 }} & \multicolumn{1}{c}{\darkg{\bf 0 }} & $  4.38^{+ 0.41}_{- 0.48} $ & $  3.29^{+ 0.21}_{- 0.06} $ & $  658^{+ 70}_{- 338} $ & \multicolumn{1}{c}{--} & $  9.75^{+ 0.66}_{- 0.70} $ &  18.82 &  29.62 &  8.90 \\
\IMBHfreebetarbetaSingle & $  1.06^{+ 52.66}_{- 0.99} $ & gOM & S & 11 & $ \darkr{\bf 0.070} $ & $  0.06^{+ 0.04}_{- 0.13} $ & $ -0.05^{+ 0.66}_{- 0.50} $ & $  4.15^{+ 0.68}_{- 0.26} $ & $  3.28^{+ 0.16}_{- 0.06} $ & $  506^{+ 229}_{- 200} $ & \multicolumn{1}{c}{--} & $  9.50^{+ 0.98}_{- 0.43} $ &  18.25 &  34.49 &  34.81 \\
\IMBHfreebetaSingle & TAND & gOM & S & 10 & $  0.007 $ & $  0.05^{+ 0.03}_{- 0.06} $ & $ -0.17^{+ 0.17}_{- 0.08} $ & $  4.24^{+ 0.51}_{- 0.41} $ & $  3.29^{+ 0.18}_{- 0.06} $ & $  453^{+ 221}_{- 207} $ & \multicolumn{1}{c}{--} & $  9.76^{+ 0.68}_{- 0.72} $ &  18.45 &  32.89 &  26.20 \\
\IMBHfreebetainSingle & TAND & gOM & S & 9 & $  0.005 $ & \multicolumn{1}{c}{\darkg{\bf 0 }} & $ -0.08^{+ 0.10}_{- 0.12} $ & $  4.21^{+ 0.58}_{- 0.34} $ & $  3.27^{+ 0.22}_{- 0.04} $ & $  530^{+ 190}_{- 224} $ & \multicolumn{1}{c}{--} & $  9.61^{+ 0.85}_{- 0.56} $ &  18.50 &  30.98 &  17.28 \\
\IMBHfreebetaoutSingle & TAND & gOM & S & 9 & $  0.003 $ & $ -0.01^{+ 0.06}_{- 0.02} $ & \multicolumn{1}{c}{\darkg{\bf 0 }} & $  4.24^{+ 0.56}_{- 0.34} $ & $  3.30^{+ 0.19}_{- 0.08} $ & $  576^{+ 143}_{- 274} $ & \multicolumn{1}{c}{--} & $  9.57^{+ 0.86}_{- 0.52} $ &  18.85 &  31.68 &  17.98 \\
\IMBHisoDouble & \multicolumn{1}{c}{--} & iso & S & 11 & $  0.017 $ & \multicolumn{1}{c}{\darkg{\bf 0 }} & \multicolumn{1}{c}{\darkg{\bf 0 }} & $  3.27^{+ 0.39}_{- 0.42} $ & $  3.31^{+ 0.08}_{- 0.09} $ & $  511^{+ 158}_{- 207} $ & \multicolumn{1}{c}{--} & $  5.45^{+ 0.74}_{- 0.95} $ &  7.09 &  12.17 &  12.49 \\ & \multicolumn{1}{c}{--} & iso & S & \multicolumn{1}{c}{--} & \multicolumn{1}{c}{--} & \multicolumn{1}{c}{\darkg{\bf 0 }} & \multicolumn{1}{c}{\darkg{\bf 0 }} & $  8.35^{+ 0.94}_{- 2.07} $ & $  3.23^{+ 0.09}_{- 0.07} $ & \multicolumn{1}{c}{--} & \multicolumn{1}{c}{--} & $  5.61^{+ 0.34}_{- 0.87} $ & \multicolumn{1}{c}{--} & \multicolumn{1}{c}{--} & \multicolumn{1}{c}{--} \\
\NoneisoSingle & \multicolumn{1}{c}{--} & iso & S & 7 & $  0.002 $ & \multicolumn{1}{c}{\darkg{\bf 0 }} & \multicolumn{1}{c}{\darkg{\bf 0 }} & $  4.02^{+ 0.36}_{- 0.34} $ & $  3.79^{+ 0.11}_{- 0.24} $ & \multicolumn{1}{c}{--} & \multicolumn{1}{c}{--} & $  9.18^{+ 0.58}_{- 0.44} $ &  25.98 &  41.93 &  14.20 \\
\NoneisoDouble & \multicolumn{1}{c}{--} & iso & S & 10 & $  0.009 $ & \multicolumn{1}{c}{\darkg{\bf 0 }} & \multicolumn{1}{c}{\darkg{\bf 0 }} & $  2.46^{+ 0.34}_{- 0.14} $ & $  3.33^{+ 0.19}_{- 0.06} $ & \multicolumn{1}{c}{--} & \multicolumn{1}{c}{--} & $  4.32^{+ 0.83}_{- 0.41} $ &  13.68 &  23.35 &  16.66 \\ & \multicolumn{1}{c}{--} & iso & S & \multicolumn{1}{c}{--} & \multicolumn{1}{c}{--} & \multicolumn{1}{c}{\darkg{\bf 0 }} & \multicolumn{1}{c}{\darkg{\bf 0 }} & $  6.46^{+ 1.61}_{- 0.77} $ & $  3.28^{+ 0.07}_{- 0.10} $ & \multicolumn{1}{c}{--} & \multicolumn{1}{c}{--} & $  5.39^{+ 0.51}_{- 0.64} $ & \multicolumn{1}{c}{--} & \multicolumn{1}{c}{--} & \multicolumn{1}{c}{--} \\
\IMBHCUOisoSingle & \multicolumn{1}{c}{--} & iso & S & 10 & $  0.007 $ & \multicolumn{1}{c}{\darkg{\bf 0 }} & \multicolumn{1}{c}{\darkg{\bf 0 }} & $  5.58^{+ 0.08}_{- 0.79} $ & $  3.27^{+ 0.05}_{- 0.09} $ & $ \darkor{\it 42^{+ 92}_{- 26}} $ & \multicolumn{1}{c}{--} & $  10.94^{+ 0.25}_{- 0.93} $ &  10.09 &  16.17 &  9.48 \\ & \multicolumn{1}{c}{--} & iso & P & \multicolumn{1}{c}{--} & \multicolumn{1}{c}{--} & \multicolumn{1}{c}{\darkg{\bf 0 }} & \multicolumn{1}{c}{\darkg{\bf 0 }} & $  0.13^{+ 0.03}_{- 0.05} $ & \multicolumn{1}{c}{--} & \multicolumn{1}{c}{--} & $  2.08^{+ 0.08}_{- 0.90} $ & \multicolumn{1}{c}{--} & \multicolumn{1}{c}{--} & \multicolumn{1}{c}{--} & \multicolumn{1}{c}{--} \\
\CUOisoSingle & \multicolumn{1}{c}{--} & iso & S & 9 & $  0.002 $ & \multicolumn{1}{c}{\darkg{\bf 0 }} & \multicolumn{1}{c}{\darkg{\bf 0 }} & $ \darkor{\it 5.31^{+ 0.36}_{- 0.52}} $ & $  3.27^{+ 0.05}_{- 0.09} $ & \multicolumn{1}{c}{--} & \multicolumn{1}{c}{--} & $ \darkor{\it 10.68^{+ 0.52}_{- 0.68}} $ &  9.86 &  13.70 & $ \darkbl{\bf  0.00 }$ \\ & \multicolumn{1}{c}{--} & iso & P & \multicolumn{1}{c}{--} & \multicolumn{1}{c}{--} & \multicolumn{1}{c}{\darkg{\bf 0 }} & \multicolumn{1}{c}{\darkg{\bf 0 }} & $  0.12^{+ 0.02}_{- 0.04} $ & \multicolumn{1}{c}{--} & \multicolumn{1}{c}{--} & $  2.01^{+ 0.16}_{- 0.76} $ & \multicolumn{1}{c}{--} & \multicolumn{1}{c}{--} & \multicolumn{1}{c}{--} & \multicolumn{1}{c}{--} \\
\CUOHernquistisoSingle 
& \multicolumn{1}{c}{--} & iso & S & 9 & $  0.004 $ & \multicolumn{1}{c}{\darkg{\bf 0 }} & \multicolumn{1}{c}{\darkg{\bf 0 }} & $ \darkor{\it 5.28^{+ 0.38}_{- 0.55}} $ & $  3.25^{+ 0.08}_{- 0.06} $ & \multicolumn{1}{c}{--} & \multicolumn{1}{c}{--} & $ \darkor{\it 10.41^{+ 0.56}_{- 0.64}} $ &  11.14 &  16.26 &  2.56 \\ & \multicolumn{1}{c}{--} & iso & H & \multicolumn{1}{c}{--} & \multicolumn{1}{c}{--} & \multicolumn{1}{c}{\darkg{\bf 0 }} & \multicolumn{1}{c}{\darkg{\bf 0 }} & $  0.08^{+ 0.02}_{- 0.04} $ & \multicolumn{1}{c}{--} & \multicolumn{1}{c}{--} & $  3.09^{+ 0.26}_{- 1.57} $ & \multicolumn{1}{c}{--} & \multicolumn{1}{c}{--} & \multicolumn{1}{c}{--} & \multicolumn{1}{c}{--} \\
\CUOSersicisoSingle & \multicolumn{1}{c}{--} & iso & S & 10 & $  0.012 $ & \multicolumn{1}{c}{\darkg{\bf 0 }} & \multicolumn{1}{c}{\darkg{\bf 0 }} & $ \darkor{\it 5.30^{+ 0.36}_{- 0.51}} $ & $  3.24^{+ 0.08}_{- 0.06} $ & \multicolumn{1}{c}{--} & \multicolumn{1}{c}{--} & $ \darkor{\it 10.62^{+ 0.54}_{- 0.66}} $ &  9.69 &  15.37 &  8.68 \\ & \multicolumn{1}{c}{--} & iso & S & \multicolumn{1}{c}{--} & \multicolumn{1}{c}{--} & \multicolumn{1}{c}{\darkg{\bf 0 }} & \multicolumn{1}{c}{\darkg{\bf 0 }} & $  0.16^{+ 0.05}_{- 0.06} $ & $  0.92^{+ 1.68}_{- 0.11} $ & \multicolumn{1}{c}{--} & $  1.92^{+ 0.35}_{- 0.69} $ & \multicolumn{1}{c}{--} & \multicolumn{1}{c}{--} & \multicolumn{1}{c}{--} & \multicolumn{1}{c}{--} \\
\CUOfreebetaSingle & TAND & gOM & S & 11 & $ \darkr{\bf 0.031} $ & $ -0.02^{+ 0.06}_{- 0.04} $ & $ -0.03^{+ 0.11}_{- 0.14} $ & $ \darkor{\it 5.31^{+ 0.36}_{- 0.52}} $ & $  3.27^{+ 0.05}_{- 0.08} $ & \multicolumn{1}{c}{--} & \multicolumn{1}{c}{--} & $ \darkor{\it 10.69^{+ 0.53}_{- 0.67}} $ &  9.92 &  17.83 &  18.15 \\ & \multicolumn{1}{c}{--} & gOM & P & \multicolumn{1}{c}{--} & \multicolumn{1}{c}{--} & \multicolumn{1}{c}{\darkg{\bf 0 }} & \multicolumn{1}{c}{\darkg{\bf 0 }} & $  0.12^{+ 0.03}_{- 0.04} $ & \multicolumn{1}{c}{--} & \multicolumn{1}{c}{--} & $  1.91^{+ 0.30}_{- 0.67} $ & \multicolumn{1}{c}{--} & \multicolumn{1}{c}{--} & \multicolumn{1}{c}{--} & \multicolumn{1}{c}{--} \\
\CUOisoDouble
& \multicolumn{1}{c}{--} & iso & S & 12 & $  0.014 $ & \multicolumn{1}{c}{\darkg{\bf 0 }} & \multicolumn{1}{c}{\darkg{\bf 0 }} & $  4.42^{+ 0.16}_{- 0.85} $ & $  3.27^{+ 0.09}_{- 0.07} $ & \multicolumn{1}{c}{--} & \multicolumn{1}{c}{--} & $  6.69^{+ 0.08}_{- 1.43} $ & $ \darkbl{\bf  0.00 }$ & $ \darkbl{\bf  0.00 }$ &  7.33 \\ & \multicolumn{1}{c}{--} & iso & S & \multicolumn{1}{c}{--} & \multicolumn{1}{c}{--} & \multicolumn{1}{c}{\darkg{\bf 0 }} & \multicolumn{1}{c}{\darkg{\bf 0 }} & $  9.38^{+ 1.02}_{- 2.07} $ & $  3.25^{+ 0.05}_{- 0.11} $ & \multicolumn{1}{c}{--} & \multicolumn{1}{c}{--} & $  4.96^{+ 1.12}_{- 0.10} $ & \multicolumn{1}{c}{--} & \multicolumn{1}{c}{--} & \multicolumn{1}{c}{--} \\ & \multicolumn{1}{c}{--} & iso & P & \multicolumn{1}{c}{--} & \multicolumn{1}{c}{--} & \multicolumn{1}{c}{\darkg{\bf 0 }} & \multicolumn{1}{c}{\darkg{\bf 0 }} & $  0.11^{+ 0.02}_{- 0.04} $ & \multicolumn{1}{c}{--} & \multicolumn{1}{c}{--} & $  1.72^{+ 0.28}_{- 0.67} $ & \multicolumn{1}{c}{--} & \multicolumn{1}{c}{--} & \multicolumn{1}{c}{--} & \multicolumn{1}{c}{--} \\
\CUOfreebetaDouble
& TAND & gOM & S & 16 & $  0.009 $ & $  0.03^{+ 0.06}_{- 0.05} $ & $  0.00^{+ 0.08}_{- 0.15} $ & $ \darkor{\it 3.99^{+ 0.57}_{- 0.56}} $ & $  3.27^{+ 0.10}_{- 0.07} $ & \multicolumn{1}{c}{--} & \multicolumn{1}{c}{--} & $ \darkor{\it 6.16^{+ 0.63}_{- 0.89}} $ & -0.56 &  6.91 &  42.29 \\ & TAND & gOM & S & \multicolumn{1}{c}{--} & \multicolumn{1}{c}{--} & $ -0.11^{+ 0.10}_{- 0.08} $ & $  0.13^{+ 0.15}_{- 0.68} $ & $  9.85^{+ 1.00}_{- 2.47} $ & $  3.24^{+ 0.07}_{- 0.10} $ & \multicolumn{1}{c}{--} & \multicolumn{1}{c}{--} & $  5.42^{+ 0.66}_{- 0.55} $ & \multicolumn{1}{c}{--} & \multicolumn{1}{c}{--} & \multicolumn{1}{c}{--} \\ & \multicolumn{1}{c}{--} & iso & P & \multicolumn{1}{c}{--} & \multicolumn{1}{c}{--} & \multicolumn{1}{c}{\darkg{\bf 0 }} & \multicolumn{1}{c}{\darkg{\bf 0 }} & $  0.11^{+ 0.03}_{- 0.04} $ & \multicolumn{1}{c}{--} & \multicolumn{1}{c}{--} & $  1.84^{+ 0.19}_{- 0.90} $ & \multicolumn{1}{c}{--} & \multicolumn{1}{c}{--} & \multicolumn{1}{c}{--} & \multicolumn{1}{c}{--} \\
\hline
\end{tabular}

\parbox{\hsize}{\textit{Notes}: Columns are (1): run number; 
(2): anisotropy radius (where TAND is tied anisotropy number density, $r_\beta = r_{\rm scale}$); 
(3): velocity anisotropy model (``iso'' for isotropic); 
(4): surface density model (``S'' for S\'ersic, ``P'' for Plummer, ``H'' for Hernquist); 
(5):  number of free parameters;
(6): MCMC convergence criterion ($R^{-1} \leq 0.02$ is considered as properly converged, worse convergence runs are shown in bold red); 
(7): anisotropy value at $r=0$;
(8): anisotropy value at $r_{\rm out} = 8'$; 
(9): 
scale radius (effective --- half projected number --- radius $R_{\rm e}$ for S\'ersic, otherwise
radius of 3D density slope --2, with $r_{-2}=1.63\,R_{\rm e}$ for Plummer and
$r_{-2} = 0.28\,R_{\rm e}$ for Hernquist); 
(10): 
S\'ersic index; 
(11): 
black hole mass; 
(12): 
mass of inner subcluster of unresolved objects (CUO);
(13): mass
of the stellar population considered;
(14): difference in minus natural logarithm of the maximum likelihood relative to model~\CUOisoDouble;
(15): difference in AIC (eq.~[\ref{eq: AIC}]) relative to best value;
(16): difference in BIC (eq.~[\ref{eq: BIC}]) relative to best value.
Blue bold zeros for $- \Delta \ln{\mathcal{L}_{\rm max}}$, $\Delta \rm AICc$ and $\Delta \rm BIC$ represent the reference values among all runs, for each column, respectively. Values in bold gray were fixed parameters.
The values of columns (9) to (13) are at maximum likelihood (\emph{black}) or medians (\emph{orange} and italics, when the MLE is outside the 16-84 percentiles). The uncertainties are from the 16th and 84th percentiles of the marginal distributions.
The number of stars in each subset is 8255, the maximum allowed PM error (eq.~[\ref{eq: err_lind18}]) for those runs is 0.197 mas yr$^{-1}$, the distance was considered as 2.39 kpc, and the GC center was that of \cite{Goldsbury+10}. For models considering two main-sequence populations, the parameters of the brighter (fainter) population is displayed first (next).
}
\end{table*}

\begin{figure*}
\centering
\includegraphics[width=0.45\hsize]{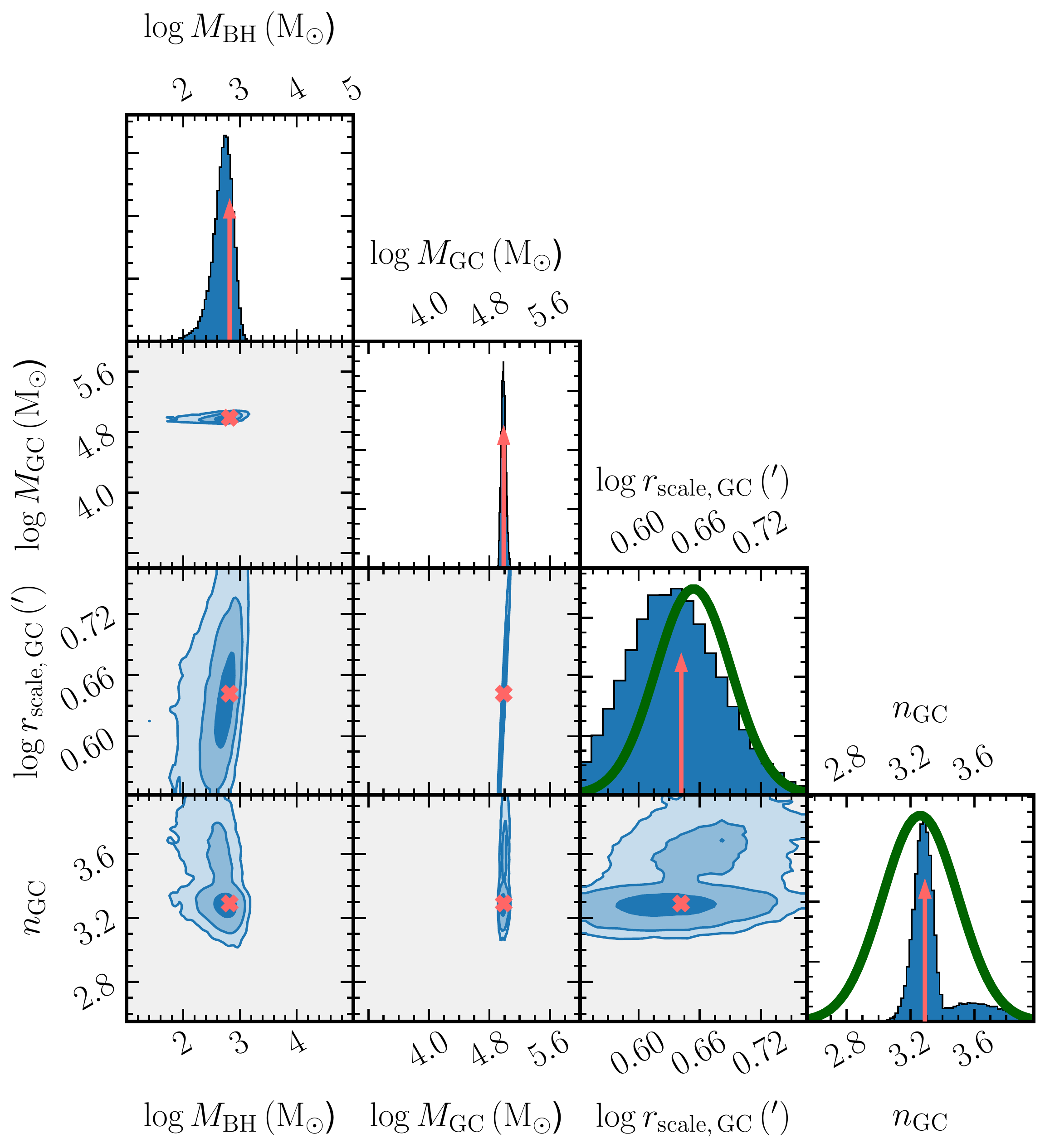}
\includegraphics[width=0.545\hsize]{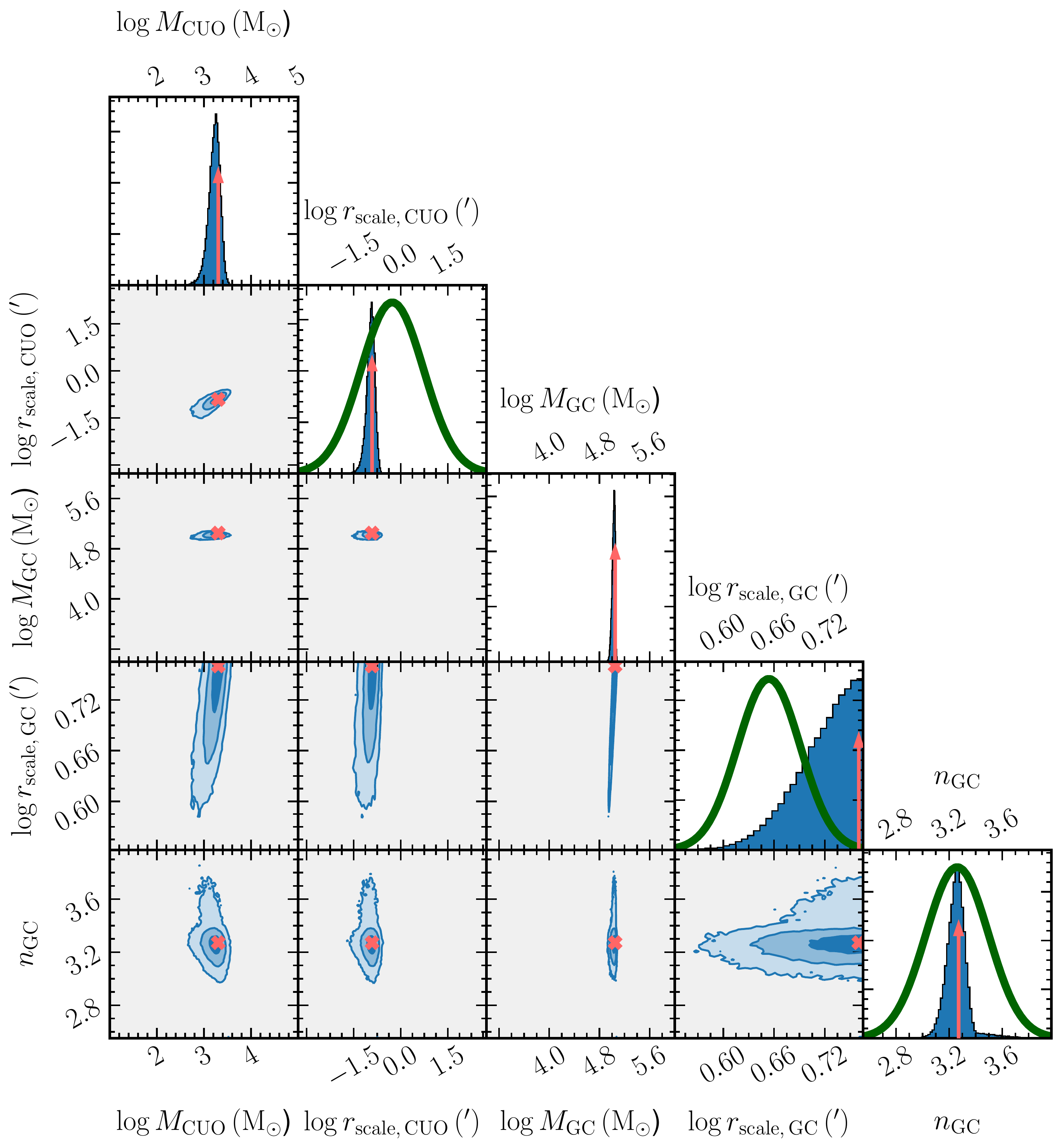}

\caption{Selected marginal distributions and covariances for models~\IMBHisoSingle\ (IMBH, \emph{left}) and \CUOisoSingle\ (subcluster of unresolved objects, CUO, \emph{right}), both for single main-sequence population and isotropic velocities.
Only the elements of the MCMC chains past the burn-in phase are considered.
The \emph{green curve} shows the Gaussian prior, while its absence indicates flat priors in the provided range. The \emph{red arrows and crosses} indicate maximum likelihoods. The \emph{contours} are the equivalent of 1, 2, and $3\,\sigma$.
The two panels only show four out of eight (left) and five out of nine (right) of the free parameters. 
\label{fig: run-IMBH}}
\end{figure*}

\subsection{\textsc{MAMPOSSt-PM} runs}
\label{ssec: res-runs}

Table~\ref{tab: mpo-NGC6397-imbh} displays our standard runs, along with some robustness tests on velocity isotropy, mass segregation and presence of central components.
Our runs have come with 4 variations for the inner mass distribution: 1) no IMBH nor (sub)-cluster of unresolved objects (CUO, see Sect.~\ref{sssec: dark-matter}), 2) IMBH with no CUO, 3) CUO without an IMBH, and 4) both IMBH and CUO. 

The left panel of Figure~\ref{fig: run-IMBH} shows the marginal distributions and covariances from the MCMC output of model~\IMBHisoSingle (IMBH without CUO and with isotropic stars), for a selection of missing parameters (missing are the bulk GC motion and the log ratio of field stars to GC members). The MCMC ran well in the sense that the MLE values (red arrows) are close to the peaks of the marginal distributions (blue shaded histograms). One notices that higher IMBH masses are obtained with higher GC scale radii, independently of the S\'ersic index. The reason is that, at given total stellar GC mass, more concentrated stars will lead to more mass in the inner regions, meaning less IMBH mass. This is the opposite of what \cite{vanderMarel&Anderson&Anderson10} proposed, but consistent with what \cite{Baumgardt17} found for \ngc.

The right panel of Figure~\ref{fig: run-IMBH} shows the same for model~\CUOisoSingle\ (CUO without IMBH and with isotropic stars). Again, the MLE values are coincident with the peaks of the marginal distributions. But surprisingly, the MCMC favors a GC scale radius at the upper edge of the $3\,\sigma$ prior obtained from the SD profile fit of Sect.~\ref{sssec: sdens_results}. We will discuss this below. The mass of the CUO increases with the GC and CUO scale radii, and decreases with the GC S\'ersic index (because low index leads to flatter SD profiles, which are somewhat analog to wider SD profiles). Interestingly, the scale radius of the CUO population is small ($7''$), as we will discuss in Sect.~\ref{sssec: surf-NSC}.

\subsection{Velocity anisotropy}
\label{ssec: vel-anis-MPOPM}

The combination of HST data properly probing the inner regions of \ngc\ with the {\sc Gaia} data probing the outer regions allows us to estimate the velocity anisotropy across the cluster.
We ran \mpo\ using different priors on the anisotropy. 
Our standard prior has isotropic velocities throughout the GC. Our other priors assume the gOM anisotropy model (we found that the softer varying \citealt{Tiret+07} model performs almost as well, but not better). 


The highest likelihoods (lowest $-\Delta \ln {\cal L}_{\rm max}$) between the first five models are achieved by model~\IMBHfreebetarbetaSingle\ (everything free), but it did not converge ($R^{-1} = 0.07 > 0.02$). 
%
In particular, model~\IMBHfreebetarbetaSingle\ leads to quasi-isotropic inner and outer velocities, with a narrow constraint for $\beta_0$ and a quite wide one for $\beta_\infty$. 
The transition radius, $r_\beta$, is very poorly constrained.
Compared to model~\IMBHfreebetarbetaSingle, the isotropic model~\IMBHisoSingle\ is very strongly preferred by BIC (according to Eq.~[\ref{eq: pBIC}] for $\rm \Delta BIC = 26$) and marginally so (92\% confidence according to Eq.~[\ref{eq: pAIC}] for $\Delta \rm AIC = 4.87$) by AICc.
Model~\IMBHfreebetaSingle\ with the anisotropy radius tied to the effective radius of the SD profile (TAND) is strongly preferred by BIC  Bayesian evidence (with respect to model~\IMBHfreebetarbetaSingle), but not with AICc.
Interestingly, model~\IMBHfreebetaSingle\ leads to negligible anisotropy in the center with narrow uncertainty. But it points to mildly tangential anisotropy at $r = r_{\rm out} = 8'$, which corresponds to our maximum projected radius $R_{\rm max}$.
Forcing central isotropy with TAND (model~\IMBHfreebetainSingle) leads to less tangential outer velocities. Conversely, forcing outer isotropy with TAND (model~\IMBHfreebetaoutSingle) leads to strongly well constrained isotropic inner velocities. 

We also ran a CUO case, where the stars had free inner and outer anisotropy, but with TAND. Compared to its isotropic analog (model~\CUOisoSingle), model~\CUOfreebetaSingle\ with a single population has a slightly worse likelihood (but did not fully converge), and the inner and outer anisotropies are consistent with isotropic orbits ($\beta_0 = -0.02\pm0.05$, $\beta_{\rm out}=-0.03\pm0.05$).
An analogous run for the two-population CUO model 
also yields close to isotropic orbits for the bright population ($\beta_0 = 0.03\pm0.05$, $\beta_{\rm out} = 0.00\pm0.12$). The faint population has inner velocities consistent with isotropy ($\beta_0 = -0.1\pm0.1$). On the other hand, its outer anisotropy is poorly constrained 
($\beta_{\rm out}=0.1^{+0.2}_{-0.7}$) because of the near complete lack of faint {\sc Gaia} stars with decent PM errors, hence we are limited to the inner regions. Still, the outer velocities are again consistent with isotropy.

In summary, there is strong evidence for isotropy with BIC, but weaker evidence with AICc. However, even the anisotropic runs produce anisotropy profiles that are very close to isotropic throughout. We thus conclude that the visible stars in \ngc\ have quasi-isotropic orbits, at least for the stars brighter than $\rm F606W = 19.76$ and up to $R = 8'$. We therefore adopted isotropic orbits as our standard when investigating other quantities.

\subsection{Intermediate mass black hole}
\label{ssec: no-BH}

We tested the scenario with an IMBH and no CUO (models~\IMBHisoSingle\ to \IMBHisoDouble).
Among models~\IMBHisoSingle\ to \IMBHisoDouble, the most likely one and very strongly favored by AICc (model~\IMBHisoDouble, which is the analog of model~\IMBHisoSingle, but with 2 populations for the GC stars) yields an IMBH mass of $511_{-207}^{+158}\,\msun$, while model~\IMBHisoSingle, which is weakly favored by BIC over model~\IMBHisoDouble, yields an IMBH mass of $658_{-338}^{+70}\,\msun$. Both models~\IMBHisoSingle\ and \IMBHisoDouble\ indicate an IMBH mass above $200\,\msun$ at 95\% confidence.
%
%
Furthermore, AICc (resp. BIC) indicates very strong (resp. quite strong) evidence for  the presence of an IMBH in the absence of a central diffuse component (i.e., comparing IMBH models~\IMBHisoSingle\ and \IMBHisoDouble\ to models~\NoneisoSingle\ and \NoneisoDouble, respectively).


However, both AICc and BIC indicate strong evidence against the IMBH hypothesis in comparison with the presence of a CUO, with differences of 15.9 in AICc and 8.9 in BIC between isotropic, single-population models~\IMBHisoSingle\ and \CUOisoSingle. With two-population GC stars, we have differences of 12.2 in AICc between isotropic models~\IMBHisoDouble\ and \CUOisoDouble. But with BIC, the difference is only 5.2, leading to only moderately strong evidence (92\% confidence) of favoring the CUO over the IMBH.
But given our preference for AICc for the complex physics of GC kinematics with imperfect models for the surface density profile for example (see Sect.~\ref{sec: Analysis}), we conclude that the evidence is strong in favor of a dark component that is diffuse for the two-population model.
Hence, the unseen inner matter of \ngc\ is very likely to  be diffuse. We therefore now investigate the CUO model in more detail.

\subsection{Inner subcluster of unresolved objects (CUO)}
\label{ssec: inner-cluster}

\begin{figure}
\centering
\includegraphics[width=0.65\hsize]{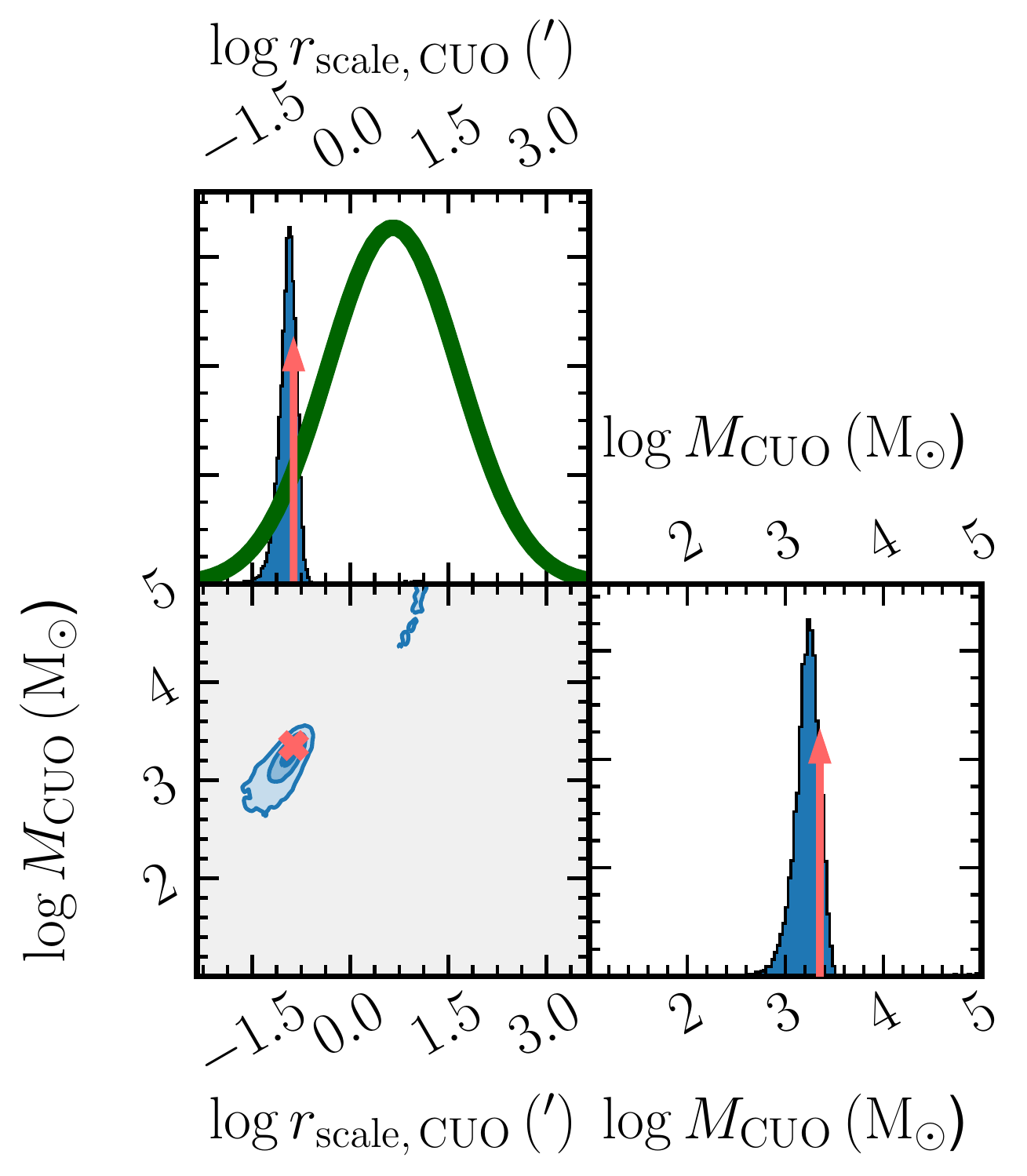}
\caption{Selected marginal distributions of the CUO effective radius and mass, and their covariance, 
for a preliminary \mpo\ run for an isotropic, single-population plus Plummer CUO SD profile, similar to model~\CUOisoSingle, but with a prior on the log CUO scale radius centered at $r_{-2, \rm CUO} = R_{\rm e,GC} = 4\farcm51$. 
The notation is the same as in Figure~\ref{fig: run-IMBH}.
\label{fig: CUO_bad_R}
}
\end{figure}

\begin{figure*}
\centering
\includegraphics[width=\hsize]{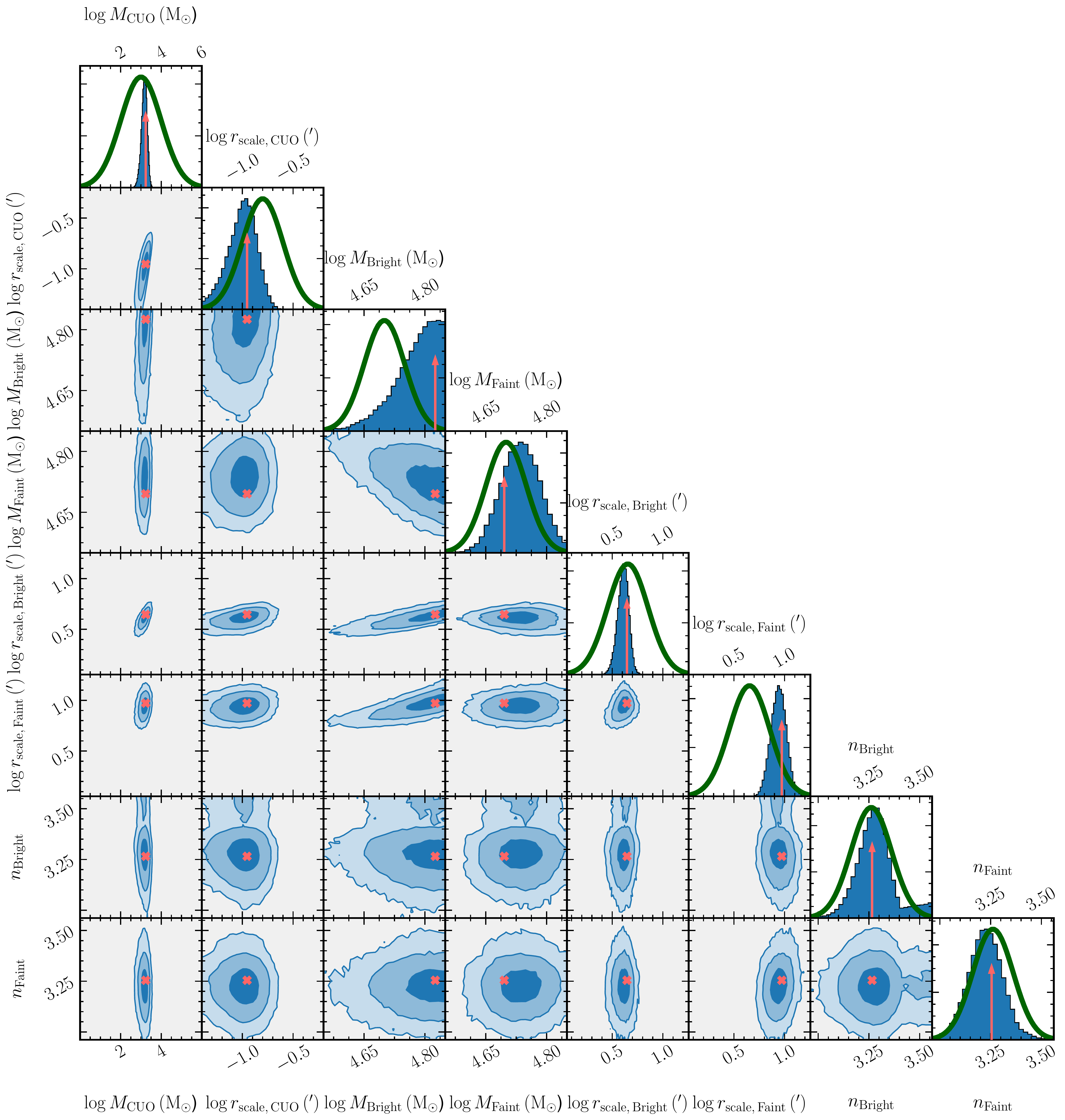}

\caption{Same as Figure~\ref{fig: run-IMBH}, but for  model~\CUOisoDouble\ with a inner subcluster of unresolved objects (CUO) with two populations of stars, a brighter (heavier) and a fainter (lighter) one, all with isotropic velocities. 
Only eight of twelve free parameters are shown.
\label{fig: NSC}}
\end{figure*}

\begin{figure*}
\centering
\includegraphics[width=0.95\hsize]{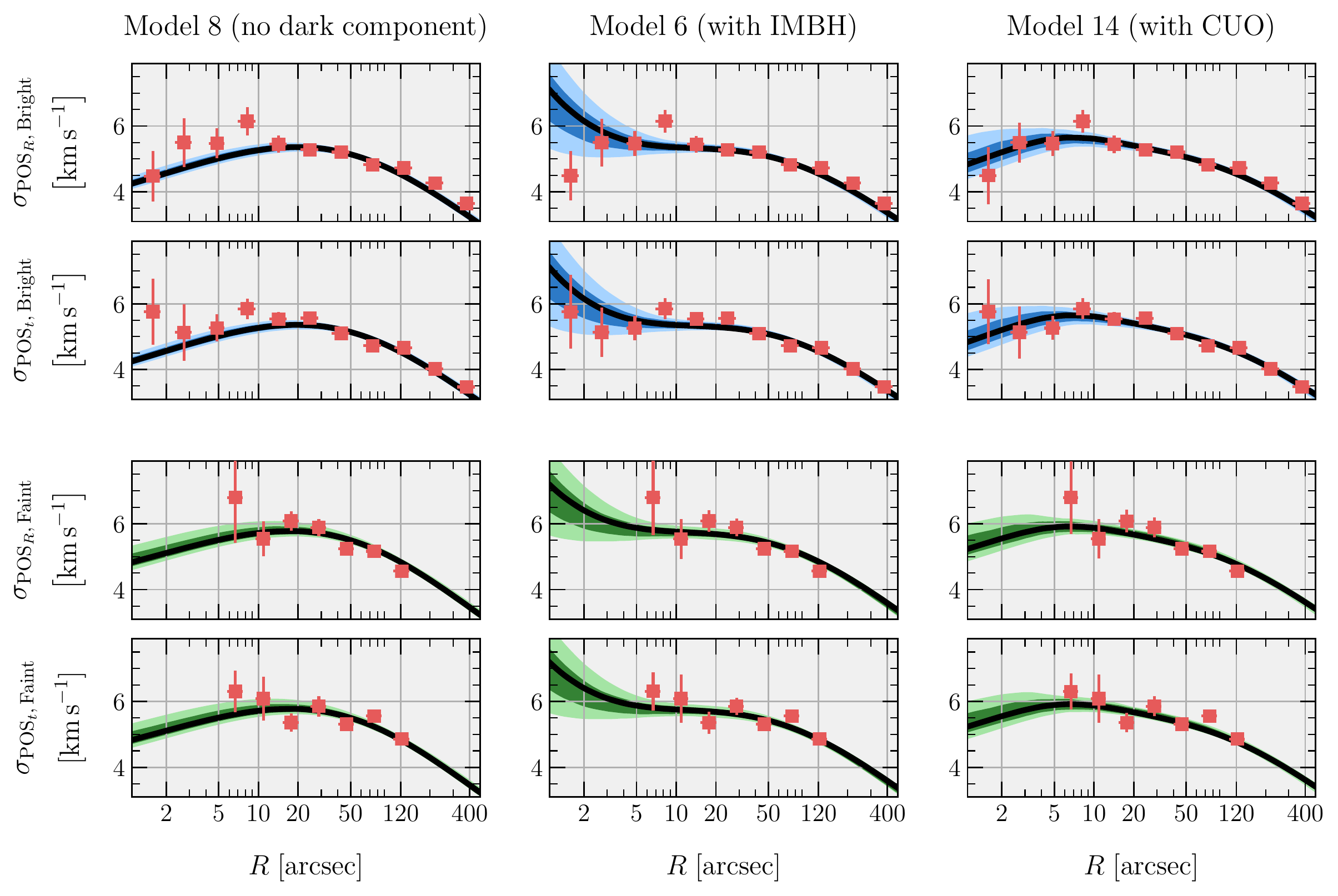}

\caption{Goodness of fit plots of plane of sky (POS) velocity dispersions as a function of the projected radius 
for models~\NoneisoDouble\ (no dark component, {\bf left}), \IMBHisoDouble\ (IMBH, {\bf middle}) and
\CUOisoDouble\ (central unresolved objects, {\bf right}), which all consider two stellar populations, one of brighter (\emph{blue}) stars and another of fainter (\emph{green}) ones. 
The \emph{black curves display} the maximum likelihood solutions, while the \emph{darker and lighter shaded regions} show the [16, 84] and [2.5, 97.5] percentiles, respectively. The \emph{red squares} show the data in logarithmic spaced bins. The vertical error bars were calculated using a bootstrap method, while the horizontal error bars considered the radial quantization noise.
\label{fig: good-fit}
}
\end{figure*}




\subsubsection{CUO density profile}
\label{sssec: surf-NSC}
 
%
If the dark component is diffuse as a CUO instead of a singular IMBH, we first need to measure its extent.
\mpo\ can provide constraints on the shape and scale radius of the density profile of the unseen population, through the sole use of the conditional probabilities of velocity at given projected radius, $p({\bf v}|R)$, without directly fitting the distribution of projected radii.

We first assumed a \cite{Plummer1911} model for the CUO with the same effective radius as that of the GC stars ($4\farcm 5$), but with a wide standard deviation (1~dex) for the Gaussian prior on log scale radius.
Interestingly, as seen in Figure~\ref{fig: CUO_bad_R},
\mpo\ converged to $r_{-2}=7''$, thus an effective (half-projected number) radius of only $4\farcs3$ for the CUO,
\footnote{We used the Plummer model relations $\nu(r) \propto (r^2+a^2)^{-5/2}$, $\Sigma(R) \propto (R^2+a^2)^{-2}$, $M(r)/M_\infty = r^3/(r^2+a^2)^{3/2}$, with $R_{\rm e}/a=1/2$ and $r_{-2}/a=\sqrt{2/3}$.}
which is 60 times lower than the median of the prior, thus confirming our suspicion that the CUO might indeed be significantly more concentrated than the main-sequence stars.

We then ran \mpo\ for three models with a much smaller scale radius prior for the CUO ($30''$), to allow for more accurate fits.
We used three different CUO density models: a S\'ersic model, a \citeauthor{Plummer1911} model, and a \cite{Hernquist90} model.
%
%
%
%
All three led to very small scale radii, which convert to effective (half projected number) radii $R_{\rm e}=9\farcs6, 17\farcs4$ and $4\farcs4$ for the S\'ersic, Hernquist and Plummer models, respectively.\footnote{We used the Hernquist model relations $\nu(r) \propto a^4/[r\,(r+a)^3]$, $M(r)/M_\infty = r^2/(r+a)^2$, with $R_{\rm e}/a=1.8153$ \citep{Hernquist90}, with his expression for   $\Sigma(R)$, and also $r_{-2}/a=1/2$.}

The S\'ersic model (12) is the most likely one, closely followed by the Plummer model (10).  AICc has a weak preference for the Plummer model, while BIC prefers the Plummer model, with strong evidence against the more complex S\'ersic model, but weak evidence against the Hernquist model (11). 
Model~\CUOSersicisoSingle\ with the S\'ersic CUO density profile produced a low S\'ersic index of  $n = 0.92$, which leads to a shallow inner slope
that is not too different from the zero slope of the inner  Plummer density profile (see Figure A.1 from \citealt{Vitral&Mamon20}). The steeper inner profile of the Hernquist density profile makes it less similar to the $n=0.92$ S\'ersic model than is the Plummer profile, as confirmed by the Hernquist model~\CUOHernquistisoSingle\ producing the lowest likelihood (highest $-\ln {\cal L}$) among the three models.

In summary, it is hard to distinguish which is the best density model for the CUO scenario. There is weak evidence for a shallow slope.
We adopted the Plummer model given that it is the preferred of the three density models for both AICc and BIC, albeit with weak evidence for both.



\subsubsection{Presence of an IMBH in addition to the CUO}
\label{sssec: NSC-IMBH}

One may ask whether the center of \ngc\ can host both an IMBH and a diffuse dark component.
Model~\IMBHCUOisoSingle\ contains both, but it is somewhat less likely than model~\CUOisoSingle, and in comparison it  is strongly disfavored by BIC, although only weakly disfavored by AICc. We note that the Plummer model used for the CUO is the one that best distinguishes the CUO from a possible additional IMBH. Moreover, the recovered mass of the additional IMBH is so small $42^{+92}_{-26}\,\msun$ that it can no longer be called an IMBH. 
%


\subsection{Two-mass populations}
\label{ssec: two-mass}

Table~\ref{tab: mpo-NGC6397-imbh} also displays the \mpo\ results of 4 models with two populations, split by apparent magnitude, hence by mass (Sect.~\ref{sssec: mass-seg}).
Model~\CUOisoDouble, with CUO but no IMBH, has the highest likelihood of the three models considering isotropy. It is very strongly favored with AICc over models~\IMBHisoDouble\ (with IMBH but no CUO) and \NoneisoDouble\ (no IMBH nor CUO). It is also preferred by BIC, which strongly favors it over model~\NoneisoDouble, but only marginally favors it over model~\IMBHisoDouble\ ($\Delta\rm BIC = 5.16$, leading to 92\% confidence in preferring model~\CUOisoDouble\ over model~\IMBHisoDouble, according to Eq.~[\ref{eq: pBIC}]).
This preference of the CUO model over the IMBH model resembles that found for the single population (Sect.~\ref{ssec: no-BH}).

But there are differences in the \mpo\ results between single-population and two-population models, both in their Bayesian evidence and in their best-fit parameters.
Indeed, the two-population model~\CUOisoDouble\ shows the highest likelihood and AICc evidence of the first fourteen models listed in Table~\ref{tab: mpo-NGC6397-imbh}. In particular, comparing two-population models to their single-population equivalents, there is strong AICc evidence ($\Delta \rm AICc = 13.7$) favoring two-population model~\CUOisoDouble\ than 1-population model~\CUOisoSingle. But there is strong BIC evidence ($\Delta \rm BIC = 7.3$) the other way, with model~\CUOisoSingle\ displaying the best BIC evidence of all fourteen models listed in Table~\ref{tab: mpo-NGC6397-imbh}.

The reader may note that 
the GC stellar masses summed over the one or two populations are 9\% greater in the two-population runs for CUO without IMBH scenarios (model~\CUOisoDouble\ versus model~\CUOisoSingle) and 13\% higher for IMBH without CUO scenarios (model~\IMBHisoDouble\ versus model~\IMBHisoSingle).
%
Moreover, the IMBH mass is 22\% lower in the two-population 
model~\IMBHisoDouble\ relative to the single-population model~\IMBHisoSingle.
Given that the standard deviations in the log GC mass (single or sum of Bright and Faint) are less than 0.03 dex,  the difference in the means of $\log M_{\rm GC}$  of 0.086 dex is highly significant for samples of order of $10^5$ points, as we checked with a Student~t test.
Similarly, the standard deviations in $\log M_{\rm IMBH}$ are of order 0.2 dex; thus the difference of 0.037 dex in their means is again highly significant, as we also checked with the Student~t test. 

\mpo\ yields interesting results on the differences between the bright and faint populations.
First, the two-population runs can be tested for the respective masses in each. Despite our prior of equal masses for each, \mpo\ returns best-fit bright population mass fractions of 0.49, 0.44, and 0.57 for models~\IMBHisoDouble\ (IMBH without CUO), \NoneisoDouble\ (no IMBH nor CUO), and \CUOisoDouble\ (CUO without IMBH).  Only model~\CUOisoDouble\ has a bright fraction close to the expected value of 0.56 (Sect.~\ref{sssec: mass-seg}).

Secondly, in all three two-population models,
the brighter population has a much lower scale radius than its fainter counterpart, by factors of 2.5, 2.6 and 2.1 for models~\IMBHisoDouble, \NoneisoDouble, and \CUOisoDouble, respectively. These lower scale radii for the bright population are highly statistically significant. Indeed, the fractions of MCMC chain elements leading to higher scale radius of the brighter population are less than 0.04\% for models~\IMBHisoDouble, \NoneisoDouble, and \CUOisoDouble.
Therefore, \mpo\ is able to find very strong kinematic signatures of luminosity (hence mass) segregation, by fitting $p({\bf v}|R)$ with the same priors on the scale radii of the two populations, without directly fitting the distribution of projected radii. 

Recall that we used the same Gaussian priors on scale radius for both populations. As can be seen in Figure~\ref{fig: NSC}, these priors may have been too narrow, and \mpo\ may have thus under-estimated the differences in the scale radii of the bright versus faint populations. Figure~\ref{fig: good-fit} illustrates the quality of the 3 classes of models in reproducing the observed velocity dispersion profiles. Model~\NoneisoDouble, with no additional dark component, clearly underestimates the velocity dispersions below $10''$. Model~\IMBHisoDouble, with an IMBH, overestimates it below $4''$. Finally, model~\CUOisoDouble, with a CUO, does best (the non-perfect match appears to be caused by sampling noise).

\section{Conclusions and discussion}
\label{sec: Discussion}

\subsection{Main results}
We ran the \mpo\ code on a combined set of PMs from HST and {\sc Gaia}, as well as with LOS velocities from MUSE. 
Our results indicate that the GC star velocities are close to isotropic out to 2 effective radii.
Our models with inner central components (IMBH or diffuse subcluster of unresolved objects, CUO) are strongly preferred over models without any.
Models where the central mass is in the form of an IMBH find favored masses of $500-650\,\msun$, with a 5th lower percentile of $200\,\msun$.
But we found better likelihoods and Bayesian evidence for a diffuse central unseen mass component (CUO), with effective (projected) radius of order of 2.5 to $5''$ and mass of order 1000 to $2000\,\msun$.

\subsection{Robustness}
\label{ssec: robust}

We tested the robustness of our results for several variations in our assumptions and our cuts to the data. These are: Single versus two stellar populations, the adopted SD profile, the minimum allowed PM error, and the HST quality flags.

First, the preference for the CUO component is relatively robust to our analysis with a single bright population or two populations. It is also quite robust to our assumptions on velocity anisotropy from free inner and outer values to isotropic.
The differences in AICc and BIC of IMBH relative to CUO are
+16 and +9 for single-population isotropic, 
+15 and +8 for single-population, free inner and outer anisotropy but fixing the anisotropy transition radius to the density scale radius  (TAND), 
and
+12 and +5 for double-population isotropic.
In other words, CUO is very strongly preferred to IMBH by both AICc and BIC. 
We also ran an isotropic model restricting the sample to stars with F606W and $G$ magnitudes brighter than 17.5, to limit the effects of possible mass segregation of these bright stars with the fainter, lower-mass stars. We found, again, that the CUO scenario is strongly preferred over the IMBH one  ($\Delta \rm AICc = 7.0$). 

The preference for CUO versus IMBH is also robust to the choice of the density profile. Indeed, adopting a S\'ersic profile, but fixing the effective radius to $2\farcm9$ as found by \cite{Trager+95}, produces a worse match to the kinematic data. However, it strongly prefers the CUO model to the IMBH one, but provides much weaker constraints on the central masses. Similarly, forcing an inner cored profile to the luminous mass, with a Plummer model,
keeping the same 
effective radius, the kinematics indicate much worse AICc compared to S\'ersic. The AICc then strongly prefers anisotropic models, for which it still  strongly prefers the CUO option to the IMBH one.

In addition, the preference for a diffuse central mass (CUO) versus an IMBH is also robust to the choice of maximum allowed PM error. We obtain the same $\Delta \rm AICc = 12$ 
for both $\epsilon_\mu < 0.276\,\rm mas\,yr^{-1}$ 
and our standard choice of  $\epsilon_\mu < 0.197,\rm mas\,yr^{-1}$, as well as the same $\Delta \rm BIC = 5$. On the other hand, 
to $\epsilon_\mu < 0.141\,\rm mas\,yr^{-1}$, yields $\Delta \rm AICc = 6.5$ instead, presumably because of the small dataset satisfying this low maximum allowed PM error.

Finally, we also tested the robustness of our conclusions when also applying HST PM quality flags, such as the reduced $\chi^2$ flag. Limiting our sample to  stars satisfying $(\chi^{2}_{\mu_{\alpha} \, \cos{\delta}} + \chi^{2}_{\mu_{\delta}})/2 < 1.25$,
which is the threshold chosen by \cite{Bellini+14} for NGC 7078, we noticed that the fitted parameters agreed within $1\,\sigma$ to the ones in Table~\ref{tab: mpo-NGC6397-imbh}. Furthermore, the CUO scenario was again favored over the IMBH one with moderately strong AICc evidence ($\Delta \rm AICc = 4.91$, i.e., 91 per cent confidence) instead of the strong AICc evidence observed for the subset without this $\chi^{2}$ filter.
Moreover, \cite{Bellini+14} argue that no quality parameter filters are required for nearby GCs with multiple observations, such as \ngc, when the PM dispersion depends little on them, as we indeed checked for both the reduced $\chi^2$ and for their QFIT parameter. Our conclusions are thus robust with respect to the HST data quality flags.


%





\subsection{Effects of the dataset}
\label{ssec: discus-dataset}
\begin{table}
    \centering
    \caption{Masses and CUO scale radius obtained using different datasets}
    \tabcolsep=2.5pt
    \renewcommand{\arraystretch}{1.5}
    \begin{tabular}{lrrrrr}
Dataset   & \multicolumn{1}{c}{$N_{\rm data}$} & \multicolumn{1}{c}{$M_{\rm GC}$ }& \multicolumn{1}{c}{$M_{\rm IMBH}$} & \multicolumn{1}{c}{$M_{\rm CUO}$} & \multicolumn{1}{c}{$r_{\rm CUO}$}\\
    &     & \multicolumn{1}{c}{[$10^4\,\msun$]} & \multicolumn{1}{c}{[$\msun$]}  & \multicolumn{1}{c}{[$10^3\,\msun$]} & \multicolumn{1}{c}{[arcmin]} \\
\hline
MUSE & 528 &  $8.81^{+1.66}_{-1.15}$ & $22^{+398}_{-2}$ & \multicolumn{1}{c}{--} & \multicolumn{1}{c}{--} \\
MUSE & 528 &  $8.77^{+1.69}_{-1.16}$ & \multicolumn{1}{c}{--} & $2.73^{+3.29}_{-2.70}$ & $0.32^{+98.86}_{-0.30}$ \\
HST & 7209 & $8.80^{+0.74}_{-0.40}$ & $472^{+172}_{-238}$ & \multicolumn{1}{c}{--} &  \multicolumn{1}{c}{--} \\
HST & 7209 & $10.20^{+0.89}_{-1.09}$ & \multicolumn{1}{c}{--} & $1.87^{+0.25}_{-0.79}$& $0.11^{+0.03}_{-0.04}$ \\
{\sc Gaia} & 1905 & $10.79^{+0.08}_{-1.32}$ & $165^{+912}_{-141}$ & \multicolumn{1}{c}{--} &  \multicolumn{1}{c}{--} \\
{\sc Gaia} & 1905 & $10.11^{+0.74}_{-0.76}$ & \multicolumn{1}{c}{--} & $0.35^{+5.25}_{-0.32}$& $15.02^{+76.26}_{-13.01}$ \\
all 3 (mod.\,\IMBHisoSingle) & 8255 &  $9.75^{0.66}_{-0.70}$ & $658^{+70}_{-338}$ & \multicolumn{1}{c}{--} &  \multicolumn{1}{c}{--}  \\
all 3 (mod.\,\CUOisoSingle) & 8255 & $10.68^{+0.52}_{-0.68}$ & \multicolumn{1}{c}{--} & $2.01^{+0.16}_{-0.76}$ & $0.12^{+0.02}_{-0.04}$\\
\hline
    \end{tabular}
\parbox{\hsize}{Notes: The models are all isotropic single-component. The last  two refer to our models~\IMBHisoSingle\ and \CUOisoSingle, respectively.
}
    \label{tab: massvsdata}
\end{table}

It is useful to check how efficient are our different datasets in reaching our conclusions. Such a comparison will be useful for future mass-orbit modeling of other GCs, for example in the absence of HST and LOS data.

Table~\ref{tab: massvsdata} displays a summary of the masses and CUO size from single-component isotropic runs using different datasets.
We used the same SD profile for all samples.

First, with MUSE data alone, we cannot constrain the mass of the IMBH.
In contrast, \cite{Kamann+16} found $M=600\pm200\,\msun$ for the IMBH.
The difference can be explained by the much more liberal cut in velocity errors used by \citeauthor{Kamann+16}, $5\,\kms$, compared to our cut at $2.2\,\kms$, which led them to have 4608 stars compared to our 528 stars.
Our small MUSE sample prevents us from 
constraining the CUO mass and size.

This is also true using {\sc Gaia} data alone.
It is not surprising, given that we have only 152 {\sc Gaia} stars within the inner $100''$ and none within the inner $10''$. This low number is caused by our cut on PM errors. Later {\sc Gaia} releases will have lower PM errors, and will thus have larger numbers of inner stars to use for mass-orbit modeling. 

Moreover, a subset with just HST and MUSE, extending up to $2\farcm5$ is not able to constrain the outer isotropy indicated by \textsc{Gaia DR2}. Indeed, a run similar to model 13 with this dataset yielded $\beta(r = 8') = -1.0^{+ 0.5}_{- 0.3}$, which is a significantly tangential anisotropy, in contrast with $\beta(8') = -0.03_{-0.14}^{+0.11}$ when Gaia data is also considered. 


In summary, {\sc Gaia} offers much improvement to HST(+MUSE) data, by 1) allowing a better estimation of the surface density profile, hence of the 3D density profile, and 2) allow measuring velocity anisotropy well beyond the effective radius.
With the prior knowledge of the surface density profile and velocity anisotropy profile, the addition of {\sc Gaia} provides somewhat better constrained masses for the inner component (IMBH or CUO). On the other hand, {\sc Gaia} data by itself, has much too few stars with sufficiently accurate PMs at low projected radii to probe, the IMBH or CUO mass, and even less the CUO size.   This highlights the requirement of combining HST and Gaia for proper mass-orbit modeling of the PM of stars in GCs.

\subsection{Velocity anisotropy}
\label{discus-beta}

\begin{figure}
    \centering
    \includegraphics[width=\hsize]{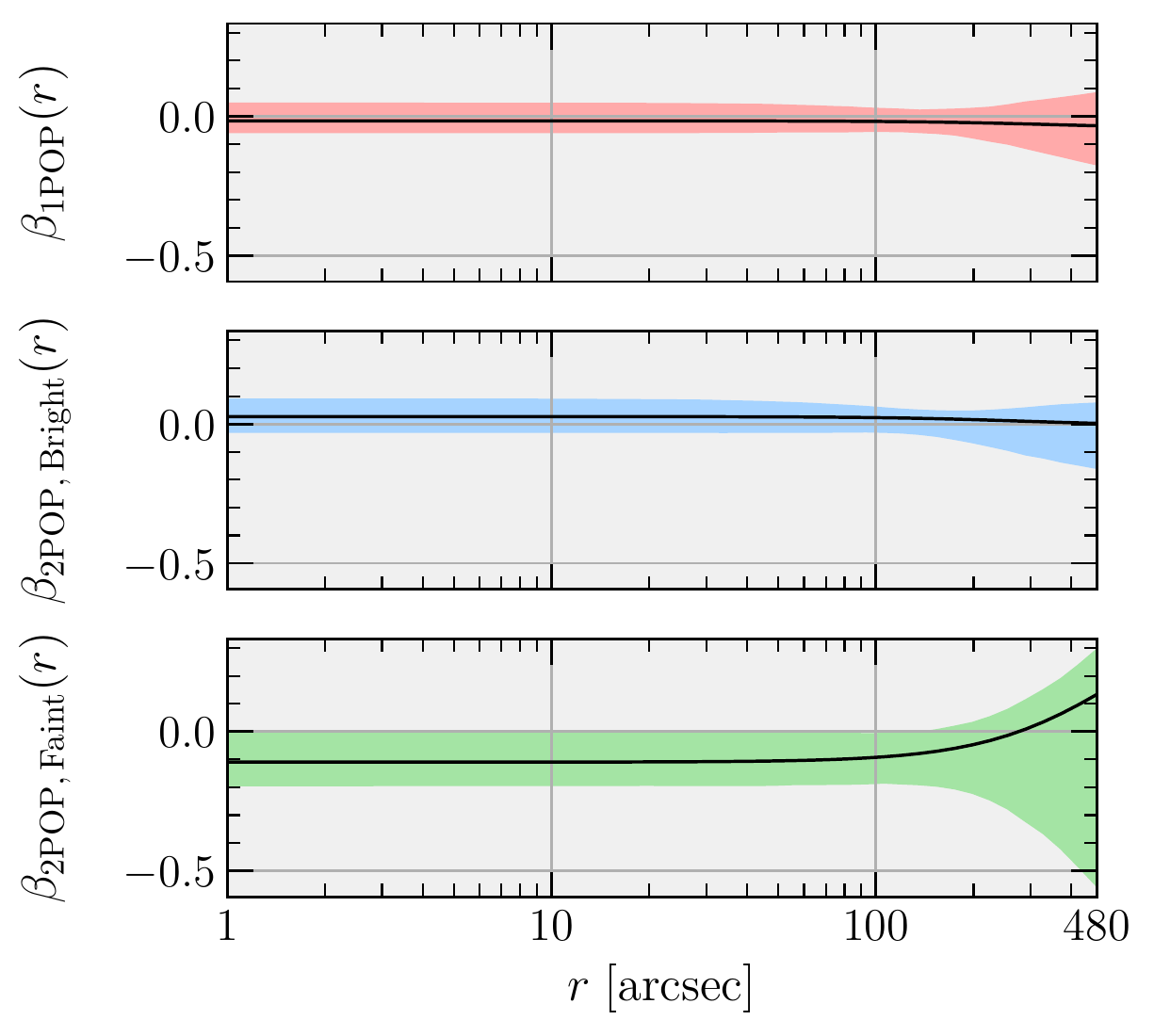}
    \caption{Velocity anisotropy profiles of \ngc\ for models~\CUOfreebetaSingle\ (\emph{top}) and \CUOfreebetaDouble\ (bright in \emph{middle} and faint at \emph{bottom}).
    }
    \label{fig: betaofr}
\end{figure}

Our combination of HST data probing the inner regions of \ngc\ and {\sc Gaia} data probing the outer regions, allowed us to obtain a wide-range view of the variation of the orbital anisotropy with radius.
Bayesian evidence favors isotropy throughout the cluster.
This confirms the \emph{projected} isotropy previously determined by 
\cite{Heyl+12} from $3\farcm5$ to $7'$ using HST (with 10\% individual errors), as well as 
\cite{Watkins+15a}: $\sigma_{\rm POSt} =  0.98\, \sigma_{\rm POSr}$, up to $R_{\rm e}/10$ using HST, and by \cite{Jindal+19} at larger radii
 up to $12' \simeq 3\,R_{\rm e}$ using {\sc Gaia}, with signs of radial motions at $R>8'$.

But a constant projected isotropy can hide radial variations of the 3D anisotropy $\beta(r)$.
Figure~\ref{fig: betaofr} displays the first known constraints on the velocity anisotropy profile of \ngc, which we obtained from our \mpo\ analyses from $1''$ to $8'$.
Indeed, it is highly consistent with isotropic orbits throughout, as seen in Figure~\ref{fig: betaofr}.
Our parametric representation of $\beta(r)$ prevents us to see a possible sharp up- or down-turn in this radial range. Nevertheless, \mpo, like other mass-orbit modeling algorithms, probes physical radii beyond the maximum projected radius. This suggests that the orbits of stars at two effective radii are not much disturbed by the Milky Way, despite the differences in the two components of the mean POS velocity beyond $R=8'$. 

%

Only two previous studies have measured velocity anisotropy profiles in GCs. Cluster $\omega$~Cen was modeled by 
\cite{vanderMarel&Anderson&Anderson10}, who fit the PM dispersions measured in bins of projected radius. They found a gOM anisotropy profile with $\beta_0 = 0.13\pm0.02$ and $\beta_\infty=-0.52\pm0.22$, thus slightly radial in the center and slightly tangential in the outer regions.
Messier~15 was analyzed by \cite{denBrok+14}, who used an updated version of JAM \citep{Cappellari08}. They found a gOM anisotropy profile with $\beta_0=-0.21\pm0.30$ and $\beta_\infty=0.015\pm0.12$, hence consistent with isotropic at all radii.

In comparison, the present study provides the first Bayesian mass-orbit modeling of discrete GC data. Our model~\CUOfreebetaSingle, as well as the bright population of model~\CUOfreebetaDouble, both yield inner anisotropies consistent with zero with $\approx \pm0.05$ uncertainty (Table~\ref{tab: mpo-NGC6397-imbh}). Both yield anisotropies at $8'$ of $|\beta|\leq 0.03$ with uncertainties $\approx 0.14$ (Table~\ref{tab: mpo-NGC6397-imbh}). 

The velocity isotropy of \ngc\ at fairly large radii appears to contradict models where core-collapsed GCs have isotropic inner motions and very radial outer motions (e.g., \citealt{Takahashi95}).
But such models are for isolated GCs, whereas \ngc\ has crossed the Galactic disk many times and as recently as 4~Myr ago (Sect.~\ref{sssec: dark-matter}). 
GC stars are perturbed by tidal forces from the Milky Way (disk, bulge and halo, which sum up to a total gravitational potential not far from spherical, see  fig.~2.19 of \citealt{Binney&Tremaine08}).
Indeed, we noticed a difference in the mean POS velocities between the radial and tangential components beyond $8'$ (top panel of Fig.~\ref{fig: pseudo-anis}), suggesting that the outer parts of \ngc\ are out of equilibrium. 

While the tidal field on an embedded GC inside a galaxy should be compressive and radial \citep{Dekel+03}, it is now understood that this does not lead to radial orbits in the outer envelopes of GCs, where the stars are most susceptible to tidal perturbations. 
Indeed, comparing stars of same binding energy, those that are on radial orbits  spend more time at larger radii and feel stronger tidal forces, while those that are on tangential orbits are less tidally perturbed (see \citealt{Giersz&Heggie97}). This reduces the radial outer anisotropy caused by two-body relaxation.
This is confirmed by the isotropy at the half-mass radius of GCs with short half-mass relaxation times determined with mass-orbit modeling by \cite{Watkins+15a}, who also found increasingly radial orbits at $r_{\rm h}$ for GCs with increasingly longer half-mass relaxation times.

This is also consistent 
with $N$-body simulations of the internal motions of a GC subject to the tidal field of a point-mass galaxy \citep{Tiongco+16,Zocchi+16} or of a realistic galactic potential by itself or within an infalling dwarf galaxy \citep{Bianchini+17}. These simulations indicate that, after the establishment of radial outer anisotropy at core collapse, the outer velocity anisotropy  becomes less radial thanks to the tidal field of the Milky Way, especially when the tidal field is strong, as measured by small values of the ratio of Jacobi to half-mass radii, $r_{\rm J}/r_{\rm h}$, also called the {filling factor}.
One can compare these simulations in detail to \ngc. We estimate the filling factor for \ngc\  using its estimated pericentric and apocentric radii of 2.9 and 6.6 kpc \citep{GaiaHelmi+18} and the Milky Way mass profile of \cite{Cautun+20}, which produces $r_{\rm J}=31$ to 51~pc, hence $r_{\rm h}/r_{\rm J} = 0.08$ to 0.14, assuming $R_{\rm e}/r_{\rm h}=0.74$.\footnote{We expect this $R_{\rm e}/r_{\rm h}$ ratio from the deprojected S\'ersic of index 3.3 (\citealt{Simonneau&Prada04} approximation).}
The single simulation of \cite{Zocchi+16} reaches this range of $r_{\rm h}/r_{\rm J}$ for snapshots 3 to 11, for which $\beta(r_{\rm h})$ decreases from 0.13 to $-0.02$, as the tides preferentially remove the radial orbits, leading to $\beta(r_{\rm h})= 0.00$ to 0.08 for $r_{\rm h}/r_{\rm J}=0.12$, decreasing to more tangential values for higher filling factors.
Similarly, among the the MW-only simulations of \cite{Bianchini+17}, who  also included a realistic initial mass function and stellar evolution, only one of them has a filling factor at 10 Gyr that matches the one deduced above for \ngc, and this simulation produced $\beta(r_{\rm h}) = 0.075$.
So both studies would predict that a core-collapsed cluster like \ngc, subject to the tidal field of the Milky Way, should recover quasi-isotropic orbits at $r_{\rm h}$, consistent with what we found.
This needs to be checked with more refined $N$-body simulations that consider elongated orbits of a GC around its galaxy and possibly the incorporation of binaries at the start of the simulation.

There are several additional possible mechanisms for producing outer isotropic orbits. 
First, 
the passage through the differentially rotating Galactic disk could lead to exchanges of angular momentum leading to tangential orbits, in particular for the outer less bound stars. But on the scale of a GC, the differences in disk rotation velocity must be minute. 
Second, the stars in the outer parts of the GC must suffer violent relaxation during passages through the Galactic disk, but the passages are so short (a few Myr) that there may not be enough time for violent relaxation to effectively perturb the GC stars.
Third, solenoidal modes of turbulence on Galactic gas clouds generate angular momentum within these clouds and within the stars formed therein, which in turn can transfer this angular momentum to the passing GC. But again, the short duration of these passages limits the amount of angular momentum that the GC stars can acquire.
Therefore, the outer isotropy appears to be produced by the tidal field of the Milky Way preferentially removing those stars that were on radial orbits.

\subsection{Intermediate mass black hole}
\label{ssec: discus-imbh}

Our IMBH mass estimates between 500 and $600\,\msun$, 
with uncertainties of $\approx 200\,\msun$, are consistent with those of \cite{Kamann+16} ($M_{\rm IMBH} = 600\pm200\,\msun$) and \cite{Tremou+18} ($M_{\rm IMBH} < 610\,\msun$).
Our 95\% confidence lower limits (i.e., 5th percentiles) are $201\,\msun$ for both models~\IMBHisoSingle\ (single population) and \IMBHisoDouble\ (two-population). 

But if IMBHs grow by BH mergers, one expects that the gravitational radiation emitted during these mergers are anisotropic and lead to substantial recoil of the BH remnant of the merger \citep{Peres62}.
Such recoils are strong enough  for 70\% of IMBHs of mass $< 1000\,\msun$ to escape a GC of escape velocity $50\,\kms$
over time \citep{HolleyBockelmann+08}. 
We can estimate the escape velocity from the GC center,
$v_{\rm esc} = \sqrt{-2\,\Phi(0)}$,
for our best IMBH model~\IMBHisoDouble,
given that the central potential is
$\Phi_0 = -(2/\pi)\,b^n\,\Gamma(n)/\Gamma(2n)\,G\,M/R_{\rm e}$ for the S\'ersic model \citep{Ciotti91}, where $b(n)$ follows the relation given in \cite{Ciotti&Bertin99}.
Model~\IMBHisoDouble\ then has $v_{\rm esc}=20.9\,\kms$ and $13.1\,\kms$ for the Bright and Faint 
components, respectively. Summing these 
two
terms quadratically  leads to a global escape velocity of 
$v_{\rm esc}=24.7\,\kms$ for model~\IMBHisoDouble.
This escape velocity is lower than the $50\,\kms$ assumed by \citeauthor{HolleyBockelmann+08}, which should lead to an even much higher fraction of escaping IMBHs of mass $1000\,\msun$. 
And our 95\% confidence upper limits on the IMBH mass (without a CUO) are lower, again implying that a yet higher fraction of IMBHs escape their GC.
A simpler model by \citeauthor{HolleyBockelmann+08} indicates that 25 IMBH+BH mergers should lead to 85\% of $500\,\msun$ IMBHs escaping their GC of escape velocity $17\,\kms$.
This weakens our confidence in the central IMBH scenario for \ngc.

\subsection{Cluster of unresolved objects}
\label{ssec: discus-CUO}

\begin{figure}
    \centering
    \includegraphics[width=0.9\hsize]{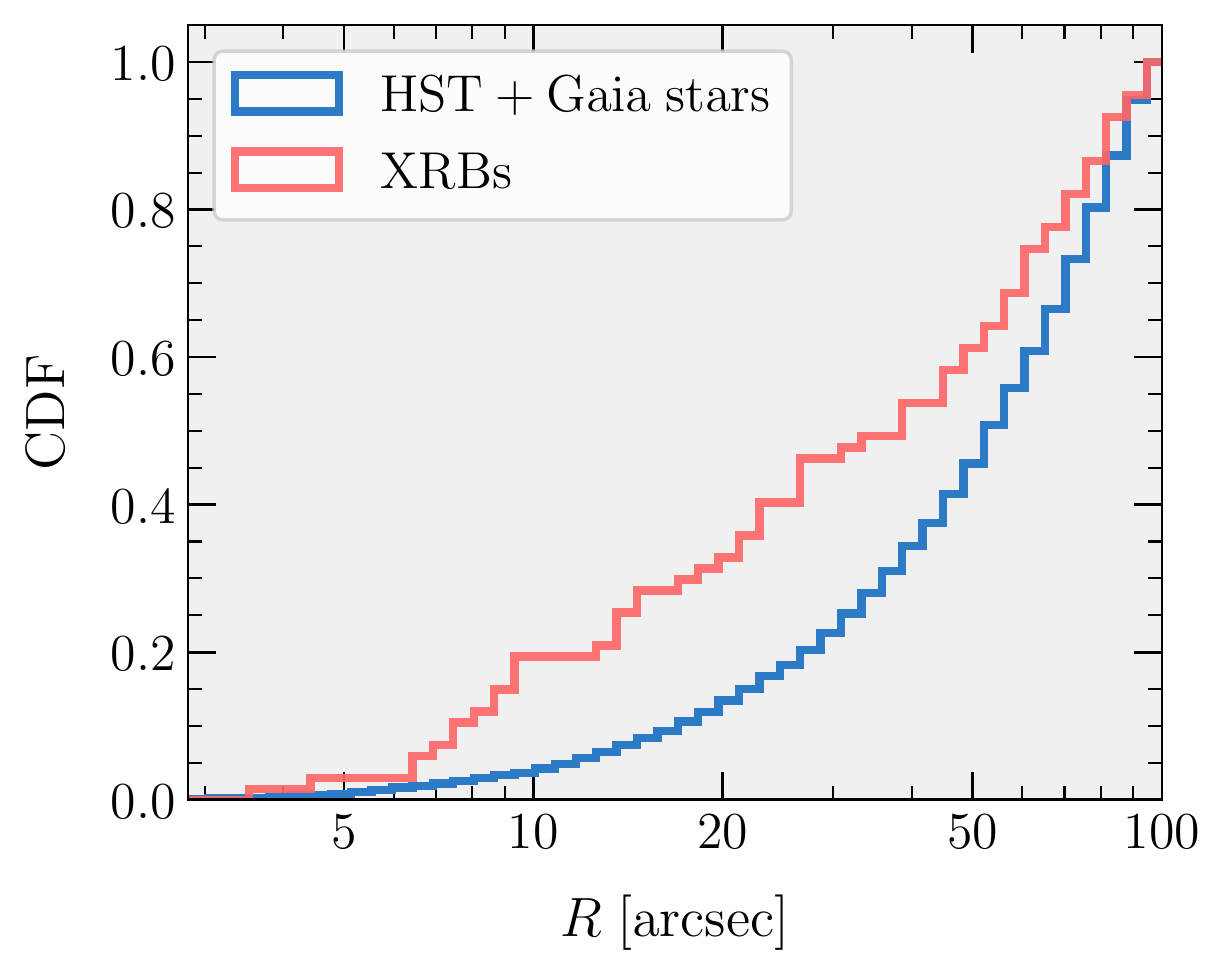}
    \caption{
    Cumulative distribution functions of projected radii for our HST$+${\sc Gaia} subset in blue and for the X-ray binaries from \cite{Bahramian+20} in red. We considered the subsets in the range of $2.7'' < R_{\rm proj} < 100''$, where the HST sample seemed complete, according to Figure~\ref{fig: hist-rproj}. 
    }
    \label{fig: XRBs}
\end{figure}


The scale radius of the CUO in \ngc\ is consistent with a feature in
the SD profile of NGC~6397. Indeed, as seen in the right panel of Figure~\ref{fig: SD-fit}, the SD profile suggests a separation between two populations at $R=10''$, consistent with the CUO effective radius of a $2\farcs5$ to $5''$.

We now discuss the nature of the subcluster of unresolved objects that we found in the center of \ngc.
The much smaller scale radius for the CUO indicates that the objects must be more massive than the stars that we studied,  thus more massive than $m_{\rm bright}=0.77\,\msun$ (Sect.~\ref{sssec: mass-seg}).
Indeed, such unresolved massive objects would sink to the center by dynamical friction \cite{Chandrasekhar43}.
The core-collapsed state of \ngc\ is a strong argument in favor of a population of heavier stars in the cluster's inner regions, as it suggests a short two-body relaxation time, hence mass segregation. 
These heavier objects constituting the CUO could be
white dwarfs, neutron stars, stellar black holes, or unresolved binaries.


We compared the radial distribution from our cleaned merged sample and from X-ray binaries from \cite{Bahramian+20}. 
Figure~\ref{fig: XRBs} shows the {cumulative distribution function} (CDF) of these two datasets in the range of $2.7-100''$, where our HST data was complete. It is clear that these two populations do not follow the same radial distribution (we find a KS $p$-value of $1.7 \times 10^{-4}$).
Furthermore, the bulk of the X-ray binaries seems to be located within $6''-50''$ arcsec (Figure~\ref{fig: XRBs}).
This is consistent with the CUO effective radius of 2.5 to $5''$  (Table~\ref{tab: mpo-NGC6397-imbh}).

One may ask 
which among white dwarfs, neutron stars, BHs and massive binaries dominates the mass of the CUO. 
We can first discard unresolved binaries of main-sequence stars. Indeed, if they are unobserved, their total magnitude must be fainter than $\rm F606W = 22$, which is the rough magnitude limit of our HST sample ({\sc Gaia} is not relevant given the very low effective radius of the CUO). There is no way that their total mass can exceed that of the bulk of our sample.
We could then have unresolved binaries of a main-sequence star with a compact star (white dwarf or neutron star) or possibly a BH. But the main-sequence star will have a mass of $m_{\rm faint}=0.25\,\msun$ (Sect.~\ref{sssec: mass-seg}), thus at least three times lower than our cut at $m_{\rm bright} = 0.77\,\msun$ and at least 20 times lower than that of a BH. Therefore, the main-sequence mass can be neglected and we are left with the compact star or BH.

\begin{figure}
    \centering
    \includegraphics[width=0.9\hsize]{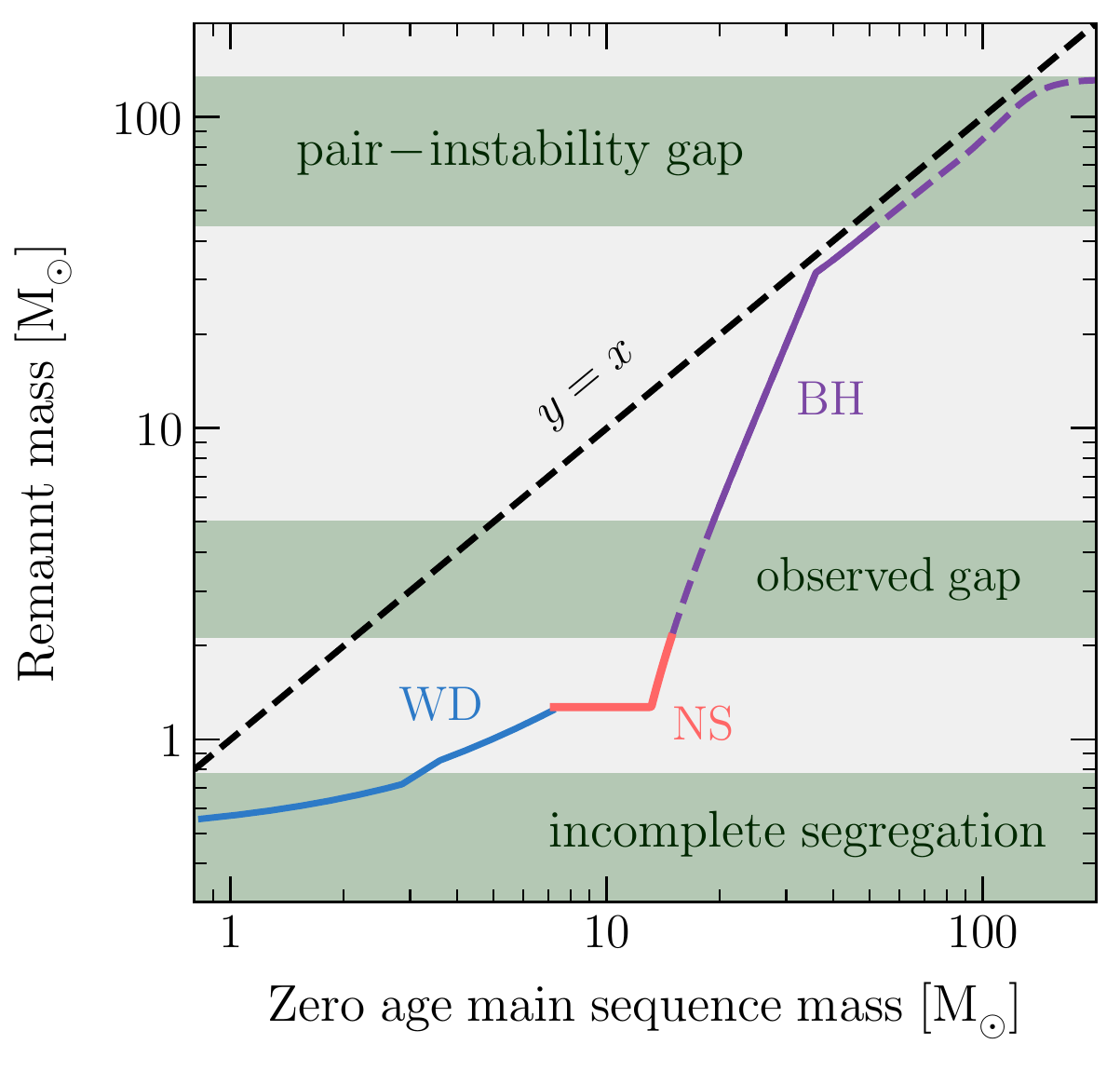}
    \caption{Initial -- final mass relation of white dwarfs (WD, MIST model from \citealt{Cummings+18}, \emph{blue}), neutron stars (NS, \textit{red}) and black holes (BH, \textit{purple})  (PARSEC/SEVN model with $Z=0.0002$ from \citealt{Spera+15}). Factors of two changes in $Z$ are barely visible in the red (purple) line and do not affect the blue line. The \emph{lower green band} indicates incomplete mass segregation because some of the main-sequence stars are more massive than the white dwarfs. The \emph{middle green band} indicates the gap where no BHs have (yet) been detected. The \emph{upper green bands} highlight the gap where pair-instability supernovae fully explode the progenitor star without leaving a black hole. The \emph{black line} shows equality as a reference.
    }
    \label{fig: IFMR}
\end{figure}

We can  compare the total mass in white dwarfs, neutron stars and BHs, by integrating over the {zero-age} (stellar mass function
of the) {main sequence} (ZAMS):
\begin{equation}
\int_{m_{\rm min}}^{m_{\rm max}} m_{\rm remnant}(m)\,n(m)\,{\rm d}m \ ,
\label{eq: massfracs}
\end{equation}
where $n(m)$ is the ZAMS.
We adopted the initial - final (remnant) mass relations from equations (4)--(6) of \cite{Cummings+18} for white dwarfs and from equations (C1), (C2), (C11), and (C15) of
\cite{Spera+15} for neutron stars and BHs.
Figure~\ref{fig: IFMR} displays these initial - final mass relations.
We took a minimum remnant mass of $0.77\,\msun$, corresponding to the maximum main-sequence mass (Sect.~\ref{sssec: mass-seg}), below which mass segregation is not effective against the most massive main-sequence stars.
The maximum possible neutron star mass is $M\simeq2.15\,\msun$ \citep{Rezzolla+18}.
We conservatively assumed a mass gap between 2.15 and $5\,\msun$ from the lack of LIGO detections in this mass range. We also considered a maximum stellar BH mass above which pair-instability supernovae fully explode the progenitor star leaving no remnant, adopting $M_{\rm BH,max}=45\,\msun$ \citep{Farmer+19} or $52\,\msun$ \citep{Woosley17}. We ignored the formation of very massive BHs above the pair-instability gap (i.e., $M_{\rm BH} > 133\,\msun$ \citealt{Woosley17}).

Summing the masses of each component by integrating Eq.~(\ref{eq: massfracs}) over the ZAMS mass function (i.e., the initial mass function, which in this mass range always has the \citealt{Salpeter55} slope of --2.3) leads to BHs accounting for  $\approx 58\%$ of the CUO, with only $\approx\,30\%$ from white dwarfs and $\approx\,12\%$ from neutron stars.
These fractions vary little with the maximum allowed BH mass:
with 55\% of the CUO mass in BHs with $M_{\rm BH} < 45\,\msun$ \citep{Farmer+19} or 60\% with $M_{\rm BH}<52\,\msun$ \citep{Woosley17}.
These fractions are insensitive  to the metallicity of \ngc\ from $Z = 0.00013$ to 0.0004 (i.e., from 1\% to 3\% solar, consistent with the metallicities derived for \ngc\ by \citealt{MarinFranch+09} and \citealt{Jain+20}, respectively).

However, we must take this CUO mass dominance by BHs with caution because BHs may merge, losing of order of 5\% of their mass to gravitational waves (e.g., \citealt{Abbott+16}), leading to kicks, some of which are strong enough to drive them out of the GC (Sect.~\ref{ssec: discus-imbh}).
Still, one part of the BH population may merge and end up escaping the GC, while another part has not merged, but would nevertheless be located at low radii thanks to orbital decay by dynamical friction. 
If the CUO mass fraction in BHs were at least $f_0 = 0.55$ or $0.6$ before merging and escapes, and if a mass fraction $f_{\rm d}$ of this BH component disappeared through mergers or escapes, then BHs would still dominate the CUO if $f_0 \, (1-f_{\rm d}) > 1-f_0$, i.e. $f_{\rm d} < 2-1/f_0 = 0.19$ or 0.33 considering the maximum BH mass of $45\,\msun$ \citep{Farmer+19} or $52\,\msun$ \citep{Woosley17}.
Put another way, BHs could contribute to half the CUO mass, if the maximum surviving BH mass is $40\,\msun$ according to our integrations of Eq.~(\ref{eq: massfracs}).
The more massive BHs would sink faster to the center by dynamical friction and preferentially merge, leaving these lower mass ones.
This discussion is a simplification because orbital decay by dynamical friction is stochastic, and one needs to test this with $N$-body simulations. 

Our preference for the CUO model agrees with the suggestions by \cite{Zocchi+19} and \cite{Mann+19} that such a CUO can mimic an IMBH.
Both studies obtained constraints on the mass of the stellar BH population in $\omega$~Cen and 47~Tuc, respectively. While those GCs are not core-collapsed, contrary to  \ngc, it is interesting to compare our results with theirs. We determine that the CUO accounts for 1 to 2\% of the GC mass, and if BHs do not merge or escape, the bulk of the CUO should be in BHs. \citeauthor{Zocchi+19} tried different BH mass fractions, and their best fits were obtained with  1 to 5\% of the GC mass in BHs. \citeauthor{Mann+19} used as input a  BH component whose mass is 1.4\% of the GC mass, but they show that only 8.5\% of this mass can be retained to avoid negative IMBH mass.  
\citeauthor{Zocchi+19} did not provide a value for the scale radius of their stellar BHs, while \citeauthor{Mann+19} provided a radius as an input parameter, which amounts to 6\% of their fitted GC scale radius.
In comparison, we are able to strongly constrain the CUO scale effective to 1.7$\pm$0.5\% of the GC effective radius, making our BH system much more concentrated.

A more robust conclusion of our integration of Eq.~(\ref{eq: massfracs}) is that  white dwarfs always dominate the neutron stars, by a factor of $\approx 4$,  regardless of the possible dominance of BHs in the mass of the CUO component. Moreover, neutron stars can also merge together (or with black holes), and such an event has been detected by LIGO \citep{LIGO17_kilonova}. Presumably, the gravitational waves emitted will be much weaker and the lower momentum vector of the waves should lead to smaller kicks on the remnant (despite the lower masses of neutron stars compared to BHs). In summary, the CUO mass should be dominated by BHs, but must also contain white dwarfs, contributing four times more mass to the CUO than neutron stars.

Interestingly, stellar black holes in a dense inner subcluster, such as our CUO, could represent a major channel for
the gravitational wave  detections by LIGO/VIRGO \citep{PortegiesZwart&McMillan00}.
These detections involve black holes of mass $\gtrsim 20\,\msun$
(see \citealt{Abbott+19}) where electromagnetic wave detections are still lacking (e.g., \citealt{Casares07}). Black hole mergers should be even more frequent, presently,  among the core-collapsed GCs such as \ngc, whose high inner densities allow for faster dynamical processes, such as dynamical friction, dynamical creation of black hole binaries \citep{Heggie&Hut03}, and subsequent hardening in three-body encounters \citep{Heggie75}.
Our dynamical analysis provides a promising route to determine the locations of these stellar mass black holes.

\subsection{Final thoughts}
The future of GC and IMBH science is very exciting, thanks to {\sc Gaia} now supplementing HST PMs at large projected radii.
Our discovery of a diffuse central mass, composed in large part of stellar mass black holes, enrichens the physics of the inner regions of GCs, and 
renders the search of IMBHs in Milky Way GCs even more delicate.
Continued pointings of GCs with HST and soon \textit{James Webb Space Telescope} will lead to longer baselines and more accurate PMs.
The third data release of the \textit{{\sc Gaia}} mission will  double the PM precision, thus enabling more accurate mass-orbit modeling, not only of nearby GCs such as NGC~6397, but also of more distant ones, 
in conjunction with HST data.
Future gravitational wave missions such as the
\textit{Laser Interferometer Space Antenna} (LISA) will probe IMBHs as well as stellar BHs, including in our Milky Way \citep{Sesana+20}.
The physics of the inner parts of GCs with possible IMBHs as well as subclusters of stellar-mass black holes is truly enticing.

\begin{acknowledgements}

We are most grateful to Andrea Bellini and Sebastian Kamann for respectively providing PM and LOS velocity data for NGC~6397, without which this paper would not have been possible.
We also thank Thomas de Boer and Mark Gieles for providing their unpublished estimate of the projected half-light radius of NGC~6397.
We acknowledge the anonymous referee who provided useful comments and references. We also thank Gwena\"el Bou\'e and Radek Wojtak for useful comments and suggestions in our mass modeling approach, Lukas Furtak for his help with the PARSEC code, Marta Volonteri for insightful discussions, and Pierre Boldrini, Laura Watkins and Natalie Webb for useful references.

\\
Eduardo Vitral was funded  by a grant from the Centre National d'\'Etudes Spatiales (CNES) and an AMX doctoral grant from \'Ecole Polytechnique.
\\
This work greatly benefited from public software
packages 
{\sc Scipy}, {\sc Astropy} \citep{AstropyCollaboration+13},
{\sc Numpy} \citep{vanderWalt11},
{\sc CosmoMC} \citep{Lewis&Bridle02},
{\sc emcee} \citep{Goodman&Weare10},
{\sc Matplotlib} \citep{Hunter07}, as well as Pierre Raybaut for developing
the {\sc Spyder} Integrated Development Environment.
We also intensively used TOPCAT \citep{Taylor05}.

\end{acknowledgements}

\bibliography{src}

\begin{thebibliography}{164}
\expandafter\ifx\csname natexlab\endcsname\relax\def\natexlab#1{#1}\fi

\bibitem[{{Abbott} {et~al.}(2016){Abbott}, {Abbott}, {Abbott}, {Abernathy},
  {Acernese}, {Ackley}, {Adams}, {Adams}, {Addesso}, {Adhikari}, \&
  et~al.}]{Abbott+16}
{Abbott}, B.~P., {Abbott}, R., {Abbott}, T.~D., {et~al.} 2016, \prl, 116,
  061102

\bibitem[{{Abbott} {et~al.}(2019){Abbott}, {Abbott}, {Abbott}, {Abraham},
  {Acernese}, {Ackley}, {Adams}, {Adhikari}, {Adya}, {Affeldt}, \&
  et~al.}]{Abbott+19}
{Abbott}, B.~P., {Abbott}, R., {Abbott}, T.~D., {et~al.} 2019, \apjl, 882, L24

\bibitem[{{Abbott} {et~al.}(2017){Abbott}, {Abbott}, {Abbott}, {Acernese},
  {Ackley}, {Adams}, {Adams}, {Addesso}, {Adhikari}, {Adya}, \&
  et~al.}]{LIGO17_kilonova}
{Abbott}, B.~P., {Abbott}, R., {Abbott}, T.~D., {et~al.} 2017, \prl, 119,
  161101

\bibitem[{Akaike(1973)}]{akaike1973information}
Akaike, H. 1973, Information Theory and an Extension of the Maximum Likelihood
  Principle (New York, NY: Springer New York), 199--213

\bibitem[{{Akaike}(1983)}]{Akaike83}
{Akaike}, H. 1983, Internaltional Statistical Institute, 44, 277

\bibitem[{{Anderson} {et~al.}(2008){Anderson}, {Sarajedini}, {Bedin}, {King},
  {Piotto}, {Reid}, {Siegel}, {Majewski}, {Paust}, {Aparicio}, {Milone},
  {Chaboyer}, \& {Rosenberg}}]{Anderson+08}
{Anderson}, J., {Sarajedini}, A., {Bedin}, L.~R., {et~al.} 2008, \aj, 135, 2055

\bibitem[{{Arenou} {et~al.}(2018){Arenou}, {Luri}, {Babusiaux}, {Fabricius},
  {Helmi}, {Muraveva}, {Robin}, {Spoto}, {Vallenari}, {Antoja},
  {Cantat-Gaudin}, {Jordi}, {Leclerc}, {Reyl{\'e}}, {Romero-G{\'o}mez}, {Shih},
  {Soria}, {Barache}, {Bossini}, {Bragaglia}, {Breddels}, {Fabrizio},
  {Lambert}, {Marrese}, {Massari}, {Moitinho}, {Robichon}, {Ruiz-Dern},
  {Sordo}, {Veljanoski}, {Eyer}, {Jasniewicz}, {Pancino}, {Soubiran}, {Spagna},
  {Tanga}, {Turon}, \& {Zurbach}}]{Arenou+18}
{Arenou}, F., {Luri}, X., {Babusiaux}, C., {et~al.} 2018, \aap, 616, A17

\bibitem[{{Aros} {et~al.}(2020){Aros}, {Sippel}, {Mastrobuono-Battisti},
  {Askar}, {Bianchini}, \& {van de Ven}}]{Aros+20}
{Aros}, F.~I., {Sippel}, A.~C., {Mastrobuono-Battisti}, A.~r., {et~al.} 2020,
  \mnras, 499, 4646

\bibitem[{{Astropy Collaboration} {et~al.}(2013){Astropy Collaboration},
  {Robitaille}, {Tollerud}, {Greenfield}, {Droettboom}, {Bray}, {Aldcroft},
  {Davis}, {Ginsburg}, {Price-Whelan}, {Kerzendorf}, {Conley}, {Crighton},
  {Barbary}, {Muna}, {Ferguson}, {Grollier}, {Parikh}, {Nair}, {Unther},
  {Deil}, {Woillez}, {Conseil}, {Kramer}, {Turner}, {Singer}, {Fox}, {Weaver},
  {Zabalza}, {Edwards}, {Azalee Bostroem}, {Burke}, {Casey}, {Crawford},
  {Dencheva}, {Ely}, {Jenness}, {Labrie}, {Lim}, {Pierfederici}, {Pontzen},
  {Ptak}, {Refsdal}, {Servillat}, \& {Streicher}}]{AstropyCollaboration+13}
{Astropy Collaboration}, {Robitaille}, T.~P., {Tollerud}, E.~J., {et~al.} 2013,
  \aap, 558, A33

\bibitem[{{Auriere}(1982)}]{Auriere82}
{Auriere}, M. 1982, \aap, 109, 301

\bibitem[{{Bahramian} {et~al.}(2020){Bahramian}, {Strader}, {Miller-Jones},
  {Chomiuk}, {Heinke}, {Maccarone}, {Pooley}, {Shishkovsky}, {Tudor}, {Zhao},
  {Li}, {Sivakoff}, {Tremou}, \& {Buchner}}]{Bahramian+20}
{Bahramian}, A., {Strader}, J., {Miller-Jones}, J. C.~A., {et~al.} 2020, \apj,
  901, 57

\bibitem[{{Baumgardt}(2017)}]{Baumgardt17}
{Baumgardt}, H. 2017, \mnras, 464, 2174

\bibitem[{{Baumgardt} {et~al.}(2019){Baumgardt}, {Hilker}, {Sollima}, \&
  {Bellini}}]{Baumgardt+19}
{Baumgardt}, H., {Hilker}, M., {Sollima}, A., \& {Bellini}, A. 2019, \mnras,
  482, 5138

\bibitem[{{Baumgardt} \& {Mieske}(2008)}]{Baumgardt&Mieske08}
{Baumgardt}, H. \& {Mieske}, S. 2008, \mnras, 391, 942

\bibitem[{{Bellini} {et~al.}(2014){Bellini}, {Anderson}, {van der Marel},
  {Watkins}, {King}, {Bianchini}, {Chanam{\'e}}, {Chandar}, {Cool}, {Ferraro},
  {Ford}, \& {Massari}}]{Bellini+14}
{Bellini}, A., {Anderson}, J., {van der Marel}, R.~P., {et~al.} 2014, \apj,
  797, 115

\bibitem[{{Bianchini} {et~al.}(2016){Bianchini}, {Norris}, {van de Ven},
  {Schinnerer}, {Bellini}, {van der Marel}, {Watkins}, \&
  {Anderson}}]{Bianchini+16}
{Bianchini}, P., {Norris}, M.~A., {van de Ven}, G., {et~al.} 2016, \apjl, 820,
  L22

\bibitem[{{Bianchini} {et~al.}(2017){Bianchini}, {Sills}, \&
  {Miholics}}]{Bianchini+17}
{Bianchini}, P., {Sills}, A., \& {Miholics}, M. 2017, \mnras, 471, 1181

\bibitem[{{Bianchini} {et~al.}(2018){Bianchini}, {van der Marel}, {del Pino},
  {Watkins}, {Bellini}, {Fardal}, {Libralato}, \& {Sills}}]{Bianchini+18}
{Bianchini}, P., {van der Marel}, R.~P., {del Pino}, A., {et~al.} 2018, \mnras,
  481, 2125

\bibitem[{{Binney} \& {Mamon}(1982)}]{Binney&Mamon82}
{Binney}, J. \& {Mamon}, G.~A. 1982, \mnras, 200, 361

\bibitem[{{Binney} \& {Tremaine}(2008)}]{Binney&Tremaine08}
{Binney}, J. \& {Tremaine}, S. 2008, Galactic Dynamics: Second Edition
  (Princeton University Press, Princeton, NJ, USA)

\bibitem[{{Box}(1979)}]{Box79}
{Box}, G. E.~P. 1979, in Robustness in Statistics, ed. R.~L. {Launer} \& G.~N.
  {Wilkinson} (Academic Press), 201--236

\bibitem[{{Bressan} {et~al.}(2012){Bressan}, {Marigo}, {Girardi}, {Salasnich},
  {Dal Cero}, {Rubele}, \& {Nanni}}]{Bressan+12}
{Bressan}, A., {Marigo}, P., {Girardi}, L., {et~al.} 2012, \mnras, 427, 127

\bibitem[{{Brown} {et~al.}(2018){Brown}, {Casertano}, {Strader}, {Riess},
  {VandenBerg}, {Soderblom}, {Kalirai}, \& {Salinas}}]{Brown+18}
{Brown}, T.~M., {Casertano}, S., {Strader}, J., {et~al.} 2018, \apjl, 856, L6

\bibitem[{{Burnham} \& {Anderson}(2002)}]{Burnham&Anderson02}
{Burnham}, K.~P. \& {Anderson}, D.~R. 2002, A Practival Information-Theoretic
  Approach, 2nd edn. (New York: Springer)

\bibitem[{{Cappellari}(2008)}]{Cappellari08}
{Cappellari}, M. 2008, \mnras, 390, 71

\bibitem[{{Carballo-Bello} {et~al.}(2012){Carballo-Bello}, {Gieles}, {Sollima},
  {Koposov}, {Mart{\'\i}nez-Delgado}, \& {Pe{\~n}arrubia}}]{CarballoBello+12}
{Carballo-Bello}, J.~A., {Gieles}, M., {Sollima}, A., {et~al.} 2012, \mnras,
  419, 14

\bibitem[{{Carretta} {et~al.}(2009{\natexlab{a}}){Carretta}, {Bragaglia},
  {Gratton}, \& {Lucatello}}]{Carretta+09b}
{Carretta}, E., {Bragaglia}, A., {Gratton}, R., \& {Lucatello}, S.
  2009{\natexlab{a}}, \aap, 505, 139

\bibitem[{{Carretta} {et~al.}(2009{\natexlab{b}}){Carretta}, {Bragaglia},
  {Gratton}, {Lucatello}, {Catanzaro}, {Leone}, {Bellazzini}, {Claudi},
  {D'Orazi}, {Momany}, {Ortolani}, {Pancino}, {Piotto}, {Recio-Blanco}, \&
  {Sabbi}}]{Carretta+09}
{Carretta}, E., {Bragaglia}, A., {Gratton}, R.~G., {et~al.} 2009{\natexlab{b}},
  \aap, 505, 117

\bibitem[{{Casares}(2007)}]{Casares07}
{Casares}, J. 2007, in Black Holes from Stars to Galaxies -- Across the Range
  of Masses, ed. V.~{Karas} \& G.~{Matt}, Vol. 238, 3--12

\bibitem[{{Cautun} {et~al.}(2020){Cautun}, {Ben{\'\i}tez-Llambay}, {Deason},
  {Frenk}, {Fattahi}, {G{\'o}mez}, {Grand }, {Oman}, {Navarro}, \&
  {Simpson}}]{Cautun+20}
{Cautun}, M., {Ben{\'\i}tez-Llambay}, A.~r., {Deason}, A.~J., {et~al.} 2020,
  \mnras, 494, 4291

\bibitem[{{Chandrasekhar}(1942)}]{Chandrasekhar42}
{Chandrasekhar}, S. 1942, Principles of Stellar Dynamics (University of Chicago
  Press)

\bibitem[{{Chandrasekhar}(1943)}]{Chandrasekhar43}
{Chandrasekhar}, S. 1943, \apj, 97, 255

\bibitem[{{Chilingarian} {et~al.}(2018){Chilingarian}, {Katkov}, {Zolotukhin},
  {Grishin}, {Beletsky}, {Boutsia}, \& {Osip}}]{Chilingarian+18}
{Chilingarian}, I.~V., {Katkov}, I.~Y., {Zolotukhin}, I.~Y., {et~al.} 2018,
  \apj, 863, 1

\bibitem[{{Ciotti}(1991)}]{Ciotti91}
{Ciotti}, L. 1991, \aap, 249, 99

\bibitem[{{Ciotti} \& {Bertin}(1999)}]{Ciotti&Bertin99}
{Ciotti}, L. \& {Bertin}, G. 1999, \aap, 352, 447

\bibitem[{{Cordoni} {et~al.}(2020){Cordoni}, {Milone}, {Mastrobuono-Battisti},
  {Marino}, {Lagioia}, {Tailo}, {Baumgardt}, \& {Hilker}}]{Cordoni+20}
{Cordoni}, G., {Milone}, A.~P., {Mastrobuono-Battisti}, A., {et~al.} 2020,
  \apj, 889, 18

\bibitem[{{Courteau} {et~al.}(2014){Courteau}, {Cappellari}, {de Jong},
  {Dutton}, {Emsellem}, {Hoekstra}, {Koopmans}, {Mamon}, {Maraston}, {Treu}, \&
  {Widrow}}]{Courteau+14}
{Courteau}, S., {Cappellari}, M., {de Jong}, R.~S., {et~al.} 2014, Reviews of
  Modern Physics, 86, 47

\bibitem[{{Croton} {et~al.}(2006){Croton}, {Springel}, {White}, {De Lucia},
  {Frenk}, {Gao}, {Jenkins}, {Kauffmann}, {Navarro}, \& {Yoshida}}]{Croton+06}
{Croton}, D.~J., {Springel}, V., {White}, S.~D.~M., {et~al.} 2006, \mnras, 365,
  11

\bibitem[{{Cummings} {et~al.}(2018){Cummings}, {Kalirai}, {Tremblay},
  {Ramirez-Ruiz}, \& {Choi}}]{Cummings+18}
{Cummings}, J.~D., {Kalirai}, J.~S., {Tremblay}, P.~E., {Ramirez-Ruiz}, E., \&
  {Choi}, J. 2018, \apj, 866, 21

\bibitem[{{Davis} {et~al.}(2008){Davis}, {Richer}, {Anderson}, {Brewer},
  {Hurley}, {Kalirai}, {Rich}, \& {Stetson}}]{Davis+08}
{Davis}, D.~S., {Richer}, H.~B., {Anderson}, J., {et~al.} 2008, \aj, 135, 2155

\bibitem[{{de Boer} {et~al.}(2019){de Boer}, {Gieles}, {Balbinot},
  {H{\'e}nault-Brunet}, {Sollima}, {Watkins}, \& {Claydon}}]{deBoer+19}
{de Boer}, T.~J.~L., {Gieles}, M., {Balbinot}, E., {et~al.} 2019, \mnras, 485,
  4906

\bibitem[{{Dekel} {et~al.}(2003){Dekel}, {Devor}, \& {Hetzroni}}]{Dekel+03}
{Dekel}, A., {Devor}, J., \& {Hetzroni}, G. 2003, \mnras, 341, 326

\bibitem[{{den Brok} {et~al.}(2014){den Brok}, {van de Ven}, {van den Bosch},
  \& {Watkins}}]{denBrok+14}
{den Brok}, M., {van de Ven}, G., {van den Bosch}, R., \& {Watkins}, L. 2014,
  \mnras, 438, 487

\bibitem[{{Djorgovski} \& {King}(1986)}]{Djorgovski&King86}
{Djorgovski}, S. \& {King}, I.~R. 1986, \apjl, 305, L61

\bibitem[{{Dotter} {et~al.}(2010){Dotter}, {Sarajedini}, {Anderson},
  {Aparicio}, {Bedin}, {Chaboyer}, {Majewski}, {Mar{\'\i}n-Franch}, {Milone},
  {Paust}, {Piotto}, {Reid}, {Rosenberg}, \& {Siegel}}]{Dotter+10}
{Dotter}, A., {Sarajedini}, A., {Anderson}, J., {et~al.} 2010, \apj, 708, 698

\bibitem[{{Drukier} {et~al.}(1998){Drukier}, {Slavin}, {Cohn}, {Lugger},
  {Berrington}, {Murphy}, \& {Seitzer}}]{Drukier+98}
{Drukier}, G.~A., {Slavin}, S.~D., {Cohn}, H.~N., {et~al.} 1998, \aj, 115, 708

\bibitem[{{Event Horizon Telescope Collaboration} {et~al.}(2019){Event Horizon
  Telescope Collaboration}, {Akiyama}, {Alberdi}, {Alef}, {Asada}, {Azulay},
  {Baczko}, {Ball}, {Balokovi{\'c}}, {Barrett}, \&
  et~al.}]{EventHorizonTelescopeCollaboration+19}
{Event Horizon Telescope Collaboration}, {Akiyama}, K., {Alberdi}, A., {et~al.}
  2019, \apjl, 875, L1

\bibitem[{{Farmer} {et~al.}(2019){Farmer}, {Renzo}, {de Mink}, {Marchant}, \&
  {Justham}}]{Farmer+19}
{Farmer}, R., {Renzo}, M., {de Mink}, S.~E., {Marchant}, P., \& {Justham}, S.
  2019, \apj, 887, 53

\bibitem[{{Foreman-Mackey} {et~al.}(2013){Foreman-Mackey}, {Hogg}, {Lang}, \&
  {Goodman}}]{ForemanMackey+13}
{Foreman-Mackey}, D., {Hogg}, D.~W., {Lang}, D., \& {Goodman}, J. 2013, \pasp,
  125, 306

\bibitem[{{Gaia Collaboration} {et~al.}(2018{\natexlab{a}}){Gaia
  Collaboration}, {Helmi}, {van Leeuwen}, {McMillan}, {Massari}, {Antoja},
  {Robin}, {Lindegren}, {Bastian}, {Arenou}, \& et~al.}]{GaiaHelmi+18}
{Gaia Collaboration}, {Helmi}, A., {van Leeuwen}, F., {et~al.}
  2018{\natexlab{a}}, \aap, 616, A12

\bibitem[{{Gaia Collaboration} {et~al.}(2018{\natexlab{b}}){Gaia
  Collaboration}, {Mignard}, {Klioner}, {Lindegren}, {Hern{\'a}ndez},
  {Bastian}, {Bombrun}, {Hobbs}, {Lammers}, {Michalik}, \&
  et~al.}]{GaiaMignard+18}
{Gaia Collaboration}, {Mignard}, F., {Klioner}, S.~A., {et~al.}
  2018{\natexlab{b}}, \aap, 616, A14

\bibitem[{{Gebhardt} {et~al.}(1995){Gebhardt}, {Pryor}, {Williams}, \&
  {Hesser}}]{Gebhardt+95}
{Gebhardt}, K., {Pryor}, C., {Williams}, T.~B., \& {Hesser}, J.~E. 1995, \aj,
  110, 1699

\bibitem[{{Giersz} \& {Heggie}(1997)}]{Giersz&Heggie97}
{Giersz}, M. \& {Heggie}, D.~C. 1997, \mnras, 286, 709

\bibitem[{{Giersz} {et~al.}(2015){Giersz}, {Leigh}, {Hypki}, {L{\"u}tzgendorf},
  \& {Askar}}]{Giersz+15}
{Giersz}, M., {Leigh}, N., {Hypki}, A., {L{\"u}tzgendorf}, N., \& {Askar}, A.
  2015, \mnras, 454, 3150

\bibitem[{{Goldsbury} {et~al.}(2013){Goldsbury}, {Heyl}, \&
  {Richer}}]{Goldsbury+13}
{Goldsbury}, R., {Heyl}, J., \& {Richer}, H. 2013, \apj, 778, 57

\bibitem[{{Goldsbury} {et~al.}(2010){Goldsbury}, {Richer}, {Anderson},
  {Dotter}, {Sarajedini}, \& {Woodley}}]{Goldsbury+10}
{Goldsbury}, R., {Richer}, H.~B., {Anderson}, J., {et~al.} 2010, \aj, 140, 1830

\bibitem[{{Goodman} \& {Weare}(2010)}]{Goodman&Weare10}
{Goodman}, J. \& {Weare}, J. 2010, Communications in Applied Mathematics and
  Computational Science, 5, 65

\bibitem[{{Gratton} {et~al.}(2003){Gratton}, {Bragaglia}, {Carretta},
  {Clementini}, {Desidera}, {Grundahl}, \& {Lucatello}}]{Gratton+03}
{Gratton}, R.~G., {Bragaglia}, A., {Carretta}, E., {et~al.} 2003, \aap, 408,
  529

\bibitem[{{Greene} {et~al.}(2020){Greene}, {Strader}, \& {Ho}}]{Greene+20}
{Greene}, J.~E., {Strader}, J., \& {Ho}, L.~C. 2020, \araa, 58, 257

\bibitem[{{Haehnelt} \& {Rees}(1993)}]{Haehnelt&Rees93}
{Haehnelt}, M.~G. \& {Rees}, M.~J. 1993, \mnras, 263, 168

\bibitem[{{Haiman}(2013)}]{Haiman13}
{Haiman}, Z. 2013, Astrophysics and Space Science Library, Vol. 396, The
  Formation of the First Massive Black Holes, ed. T.~{Wiklind}, B.~{Mobasher},
  \& V.~{Bromm}, 293

\bibitem[{{Hansen} {et~al.}(2007){Hansen}, {Anderson}, {Brewer}, {Dotter},
  {Fahlman}, {Hurley}, {Kalirai}, {King}, {Reitzel}, {Richer}, {Rich}, {Shara},
  \& {Stetson}}]{Hansen+07}
{Hansen}, B. M.~S., {Anderson}, J., {Brewer}, J., {et~al.} 2007, \apj, 671, 380

\bibitem[{{Harris}(1996)}]{Harris96}
{Harris}, W.~E. 1996, \aj, 112, 1487

\bibitem[{{Harris}(2010)}]{Harris10}
{Harris}, W.~E. 2010, arXiv e-prints, arXiv:1012.3224

\bibitem[{{Hawking}(1971)}]{Hawking71}
{Hawking}, S. 1971, \mnras, 152, 75

\bibitem[{{Heggie} \& {Hut}(2003)}]{Heggie&Hut03}
{Heggie}, D. \& {Hut}, P. 2003, The Gravitational Million-Body Problem: A
  Multidisciplinary Approach to Star Cluster Dynamics

\bibitem[{{Heggie}(1975)}]{Heggie75}
{Heggie}, D.~C. 1975, \mnras, 173, 729

\bibitem[{{Heggie} \& {Hut}(1996)}]{Heggie&Hut96}
{Heggie}, D.~C. \& {Hut}, P. 1996, in IAU Symposium, Vol. 174, Dynamical
  Evolution of Star Clusters: Confrontation of Theory and Observations, ed.
  P.~{Hut} \& J.~{Makino}, 303

\bibitem[{{Hernquist}(1990)}]{Hernquist90}
{Hernquist}, L. 1990, \apj, 356, 359

\bibitem[{{Heyl} {et~al.}(2012){Heyl}, {Richer}, {Anderson}, {Fahlman},
  {Dotter}, {Hurley}, {Kalirai}, {Rich}, {Shara}, {Stetson}, {Woodley}, \&
  {Zurek}}]{Heyl+12}
{Heyl}, J.~S., {Richer}, H., {Anderson}, J., {et~al.} 2012, \apj, 761, 51

\bibitem[{{Holley-Bockelmann} {et~al.}(2008){Holley-Bockelmann},
  {G{\"u}ltekin}, {Shoemaker}, \& {Yunes}}]{HolleyBockelmann+08}
{Holley-Bockelmann}, K., {G{\"u}ltekin}, K., {Shoemaker}, D., \& {Yunes}, N.
  2008, \apj, 686, 829

\bibitem[{{Hopkins} {et~al.}(2006){Hopkins}, {Hernquist}, {Cox}, {Di Matteo},
  {Robertson}, \& {Springel}}]{Hopkins+06}
{Hopkins}, P.~F., {Hernquist}, L., {Cox}, T.~J., {et~al.} 2006, \apjs, 163, 1

\bibitem[{{Hoyle} \& {Fowler}(1963)}]{Hoyle&Fowler63}
{Hoyle}, F. \& {Fowler}, W.~A. 1963, \nat, 197, 533

\bibitem[{{Hunter}(2007)}]{Hunter07}
{Hunter}, J.~D. 2007, Computing in Science \& Engineering, 9, 90

\bibitem[{{Husser} {et~al.}(2016){Husser}, {Kamann}, {Dreizler}, {Wendt},
  {Wulff}, {Bacon}, {Wisotzki}, {Brinchmann}, {Weilbacher}, {Roth}, \&
  {Monreal-Ibero}}]{Husser+16}
{Husser}, T.-O., {Kamann}, S., {Dreizler}, S., {et~al.} 2016, \aap, 588, A148

\bibitem[{{Ibata} {et~al.}(2013){Ibata}, {Nipoti}, {Sollima}, {Bellazzini},
  {Chapman}, \& {Dalessandro}}]{Ibata+13}
{Ibata}, R., {Nipoti}, C., {Sollima}, A., {et~al.} 2013, \mnras, 428, 3648

\bibitem[{{Jain} {et~al.}(2020){Jain}, {Prugniel}, {Martins}, \&
  {Lan{\c{c}}on}}]{Jain+20}
{Jain}, R., {Prugniel}, P., {Martins}, L., \& {Lan{\c{c}}on}, A. 2020, \aap,
  635, A161

\bibitem[{{Jindal} {et~al.}(2019){Jindal}, {Webb}, \& {Bovy}}]{Jindal+19}
{Jindal}, A., {Webb}, J.~J., \& {Bovy}, J. 2019, \mnras, 487, 3693

\bibitem[{{Kaaret} {et~al.}(2001){Kaaret}, {Prestwich}, {Zezas}, {Murray},
  {Kim}, {Kilgard}, {Schlegel}, \& {Ward}}]{Kaaret+01}
{Kaaret}, P., {Prestwich}, A.~H., {Zezas}, A., {et~al.} 2001, \mnras, 321, L29

\bibitem[{{Kamann} {et~al.}(2016){Kamann}, {Husser}, {Brinchmann}, {Emsellem},
  {Weilbacher}, {Wisotzki}, {Wendt}, {Krajnovi{\'c}}, {Roth}, {Bacon}, \&
  {Dreizler}}]{Kamann+16}
{Kamann}, S., {Husser}, T.~O., {Brinchmann}, J., {et~al.} 2016, \aap, 588, A149

\bibitem[{{Kass} \& {Rafferty}(1995)}]{Kass&Rafferty95}
{Kass}, R.~E. \& {Rafferty}, A.~E. 1995, J. of Am. Stat. Assoc., 90, 773

\bibitem[{{Lauzeral} {et~al.}(1992){Lauzeral}, {Ortolani}, {Auriere}, \&
  {Melnick}}]{Lauzeral+92}
{Lauzeral}, C., {Ortolani}, S., {Auriere}, M., \& {Melnick}, J. 1992, \aap,
  262, 63

\bibitem[{{Leigh} {et~al.}(2013){Leigh}, {B{\"o}ker}, {Maccarone}, \&
  {Perets}}]{Leigh+13}
{Leigh}, N. W.~C., {B{\"o}ker}, T., {Maccarone}, T.~J., \& {Perets}, H.~B.
  2013, \mnras, 429, 2997

\bibitem[{{Leonard}(1989)}]{Leonard89}
{Leonard}, P. J.~T. 1989, \aj, 98, 217

\bibitem[{{Lewis} \& {Bridle}(2002)}]{Lewis&Bridle02}
{Lewis}, A. \& {Bridle}, S. 2002, \prd, 66, 103511

\bibitem[{{Lima Neto} {et~al.}(1999){Lima Neto}, {Gerbal}, \&
  {M{\'a}rquez}}]{LimaNeto+99}
{Lima Neto}, G.~B., {Gerbal}, D., \& {M{\'a}rquez}, I. 1999, \mnras, 309, 481

\bibitem[{{Lin} {et~al.}(2020){Lin}, {Strader}, {Romanowsky}, {Irwin}, {Godet},
  {Barret}, {Webb}, {Homan}, \& {Remillard}}]{Lin+20}
{Lin}, D., {Strader}, J., {Romanowsky}, A.~J., {et~al.} 2020, \apjl, 892, L25

\bibitem[{{Lind} {et~al.}(2008){Lind}, {Korn}, {Barklem}, \&
  {Grundahl}}]{Lind+08}
{Lind}, K., {Korn}, A.~J., {Barklem}, P.~S., \& {Grundahl}, F. 2008, \aap, 490,
  777

\bibitem[{{Lindegren} {et~al.}(2018){Lindegren}, {Hern{\'a}ndez}, {Bombrun},
  {Klioner}, {Bastian}, {Ramos-Lerate}, {de Torres}, {Steidelm{\"u}ller},
  {Stephenson}, {Hobbs}, {Lammers}, {Biermann}, {Geyer}, {Hilger}, {Michalik},
  {Stampa}, {McMillan}, {Casta{\~n}eda}, {Clotet}, {Comoretto}, {Davidson},
  {Fabricius}, {Gracia}, {Hambly}, {Hutton}, {Mora}, {Portell}, {van Leeuwen},
  {Abbas}, {Abreu}, {Altmann}, {Andrei}, {Anglada}, {Balaguer-N{\'u}{\~n}ez},
  {Barache}, {Becciani}, {Bertone}, {Bianchi}, {Bouquillon}, {Bourda},
  {Br{\"u}semeister}, {Bucciarelli}, {Busonero}, {Buzzi}, {Cancelliere},
  {Carlucci}, {Charlot}, {Cheek}, {Crosta}, {Crowley}, {de Bruijne}, {de
  Felice}, {Drimmel}, {Esquej}, {Fienga}, {Fraile}, {Gai}, {Garralda},
  {Gonz{\'a}lez-Vidal}, {Guerra}, {Hauser}, {Hofmann}, {Holl}, {Jordan},
  {Lattanzi}, {Lenhardt}, {Liao}, {Licata}, {Lister}, {L{\"o}ffler},
  {Marchant}, {Martin-Fleitas}, {Messineo}, {Mignard}, {Morbidelli}, {Poggio},
  {Riva}, {Rowell}, {Salguero}, {Sarasso}, {Sciacca}, {Siddiqui}, {Smart},
  {Spagna}, {Steele}, {Taris}, {Torra}, {van Elteren}, {van Reeven}, \&
  {Vecchiato}}]{Lindegren+18}
{Lindegren}, L., {Hern{\'a}ndez}, J., {Bombrun}, A., {et~al.} 2018, \aap, 616,
  A2

\bibitem[{{Loeb} \& {Rasio}(1994)}]{Loeb&Rasio94}
{Loeb}, A. \& {Rasio}, F.~A. 1994, \apj, 432, 52

\bibitem[{{Lovisi} {et~al.}(2012){Lovisi}, {Mucciarelli}, {Lanzoni}, {Ferraro},
  {Gratton}, {Dalessandro}, \& {Contreras Ramos}}]{Lovisi+12}
{Lovisi}, L., {Mucciarelli}, A., {Lanzoni}, B., {et~al.} 2012, \apj, 754, 91

\bibitem[{{Lugger} {et~al.}(1995){Lugger}, {Cohn}, \& {Grindlay}}]{Lugger+95}
{Lugger}, P.~M., {Cohn}, H.~N., \& {Grindlay}, J.~E. 1995, \apj, 439, 191

\bibitem[{{Madau} \& {Rees}(2001)}]{Madau&Rees01}
{Madau}, P. \& {Rees}, M.~J. 2001, \apjl, 551, L27

\bibitem[{{Mamon} {et~al.}(2013){Mamon}, {Biviano}, \& {Bou{\'e}}}]{Mamon+13}
{Mamon}, G.~A., {Biviano}, A., \& {Bou{\'e}}, G. 2013, \mnras, 429, 3079

\bibitem[{{Mamon} {et~al.}(2019){Mamon}, {Cava}, {Biviano}, {Moretti},
  {Poggianti}, \& {Bettoni}}]{Mamon+19}
{Mamon}, G.~A., {Cava}, A., {Biviano}, A., {et~al.} 2019, \aap, 631, A131

\bibitem[{{Mann} {et~al.}(2019){Mann}, {Richer}, {Heyl}, {Anderson}, {Kalirai},
  {Caiazzo}, {M{\"o}hle}, {Knee}, \& {Baumgardt}}]{Mann+19}
{Mann}, C.~R., {Richer}, H., {Heyl}, J., {et~al.} 2019, \apj, 875, 1

\bibitem[{{Marigo} {et~al.}(2017){Marigo}, {Girardi}, {Bressan}, {Rosenfield},
  {Aringer}, {Chen}, {Dussin}, {Nanni}, {Pastorelli}, {Rodrigues}, {Trabucchi},
  {Bladh}, {Dalcanton}, {Groenewegen}, {Montalb{\'a}n}, \& {Wood}}]{Marigo+17}
{Marigo}, P., {Girardi}, L., {Bressan}, A., {et~al.} 2017, \apj, 835, 77

\bibitem[{{Mar{\'\i}n-Franch} {et~al.}(2009){Mar{\'\i}n-Franch}, {Aparicio},
  {Piotto}, {Rosenberg}, {Chaboyer}, {Sarajedini}, {Siegel}, {Anderson},
  {Bedin}, {Dotter}, {Hempel}, {King}, {Majewski}, {Milone}, {Paust}, \&
  {Reid}}]{MarinFranch+09}
{Mar{\'\i}n-Franch}, A., {Aparicio}, A., {Piotto}, G., {et~al.} 2009, \apj,
  694, 1498

\bibitem[{{Martinazzi} {et~al.}(2014){Martinazzi}, {Pieres}, {Kepler}, {Costa},
  {Bonatto}, \& {Bica}}]{Martinazzi+14}
{Martinazzi}, E., {Pieres}, A., {Kepler}, S.~O., {et~al.} 2014, \mnras, 442,
  3105

\bibitem[{{Mashchenko} \& {Sills}(2005)}]{Mashchenko&Sills&Sills05}
{Mashchenko}, S. \& {Sills}, A. 2005, \apj, 619, 258

\bibitem[{{McDonald} \& {Zijlstra}(2015)}]{McDonald&Zijlstra15}
{McDonald}, I. \& {Zijlstra}, A.~A. 2015, \mnras, 448, 502

\bibitem[{{Merritt}(1985)}]{Merritt85}
{Merritt}, D. 1985, \aj, 90, 1027

\bibitem[{{Merritt}(1987)}]{Merritt87}
{Merritt}, D. 1987, \apj, 313, 121

\bibitem[{{Miller} \& {Hamilton}(2002)}]{Miller&Hamilton02}
{Miller}, M.~C. \& {Hamilton}, D.~P. 2002, \mnras, 330, 232

\bibitem[{{Milone} {et~al.}(2012{\natexlab{a}}){Milone}, {Marino}, {Piotto},
  {Bedin}, {Anderson}, {Aparicio}, {Cassisi}, \& {Rich}}]{Milone+12b}
{Milone}, A.~P., {Marino}, A.~F., {Piotto}, G., {et~al.} 2012{\natexlab{a}},
  \apj, 745, 27

\bibitem[{{Milone} {et~al.}(2012{\natexlab{b}}){Milone}, {Piotto}, {Bedin},
  {Aparicio}, {Anderson}, {Sarajedini}, {Marino}, {Moretti}, {Davies},
  {Chaboyer}, {Dotter}, {Hempel}, {Mar{\'\i}n-Franch}, {Majewski}, {Paust},
  {Reid}, {Rosenberg}, \& {Siegel}}]{Milone+12a}
{Milone}, A.~P., {Piotto}, G., {Bedin}, L.~R., {et~al.} 2012{\natexlab{b}},
  \aap, 540, A16

\bibitem[{{Milone} {et~al.}(2006){Milone}, {Villanova}, {Bedin}, {Piotto},
  {Carraro}, {Anderson}, {King}, \& {Zaggia}}]{Milone+06}
{Milone}, A.~P., {Villanova}, S., {Bedin}, L.~R., {et~al.} 2006, \aap, 456, 517

\bibitem[{{Noyola} \& {Gebhardt}(2006)}]{Noyola&Gebhardt&Gebhardt06}
{Noyola}, E. \& {Gebhardt}, K. 2006, \aj, 132, 447

\bibitem[{{Oppenheimer} \& {Snyder}(1939)}]{Oppenheimer&Snyder39}
{Oppenheimer}, J.~R. \& {Snyder}, H. 1939, Physical Review, 56, 455

\bibitem[{{Osipkov}(1979)}]{Osipkov79}
{Osipkov}, L.~P. 1979, Soviet Astronomy Letters, 5, 42

\bibitem[{{Pastorelli} {et~al.}(2019){Pastorelli}, {Marigo}, {Girardi}, {Chen},
  {Rubele}, {Trabucchi}, {Aringer}, {Bladh}, {Bressan}, {Montalb{\'a}n},
  {Boyer}, {Dalcanton}, {Eriksson}, {Groenewegen}, {H{\"o}fner}, {Lebzelter},
  {Nanni}, {Rosenfield}, {Wood}, \& {Cioni}}]{Pastorelli+19}
{Pastorelli}, G., {Marigo}, P., {Girardi}, L., {et~al.} 2019, \mnras, 485, 5666

\bibitem[{{Pearson}(1916)}]{Pearson16}
{Pearson}, K. 1916, Philosophical Transactions of the Royal Society of London
  Series A, 216, 429

\bibitem[{{Peres}(1962)}]{Peres62}
{Peres}, A. 1962, Phys. Rev., 128, 2471

\bibitem[{{Plummer}(1911)}]{Plummer1911}
{Plummer}, H.~C. 1911, \mnras, 71, 460

\bibitem[{{Portegies Zwart} \& {McMillan}(2000)}]{PortegiesZwart&McMillan00}
{Portegies Zwart}, S.~F. \& {McMillan}, S. L.~W. 2000, \apjl, 528, L17

\bibitem[{{Portegies Zwart} \& {McMillan}(2002)}]{PortegiesZwart&McMillan02}
{Portegies Zwart}, S.~F. \& {McMillan}, S. L.~W. 2002, \apj, 576, 899

\bibitem[{{Read} {et~al.}(2020){Read}, {Mamon}, {Vasiliev}, {Watkins},
  {Walker}, {Pe{\~n}arrubia}, {Wilkinson}, {Dehnen}, \& {Das}}]{Read+20}
{Read}, J., {Mamon}, G.~A., {Vasiliev}, E., {et~al.} 2020, \mnras, in press,
  arXiv:2011.09493

\bibitem[{{Read} \& {Steger}(2017)}]{Read&Steger17}
{Read}, J.~I. \& {Steger}, P. 2017, \mnras, 471, 4541

\bibitem[{{Reid} \& {Gizis}(1998)}]{Reid&Gizis98}
{Reid}, I.~N. \& {Gizis}, J.~E. 1998, \aj, 116, 2929

\bibitem[{{Reimers}(1975)}]{Reimers75}
{Reimers}, D. 1975, Memoires of the Societe Royale des Sciences de Liege, 8,
  369

\bibitem[{{Rezzolla} {et~al.}(2018){Rezzolla}, {Most}, \& {Weih}}]{Rezzolla+18}
{Rezzolla}, L., {Most}, E.~R., \& {Weih}, L.~R. 2018, \apjl, 852, L25

\bibitem[{{Richardson} \& {Fairbairn}(2014)}]{Richardson&Fairbairn14}
{Richardson}, T. \& {Fairbairn}, M. 2014, \mnras, 441, 1584

\bibitem[{{Robin} {et~al.}(2003){Robin}, {Reyl{\'e}}, {Derri{\`e}re}, \&
  {Picaud}}]{Robin+03}
{Robin}, A.~C., {Reyl{\'e}}, C., {Derri{\`e}re}, S., \& {Picaud}, S. 2003,
  \aap, 409, 523

\bibitem[{{Rodrigues}(1840)}]{Rodrigues1840}
{Rodrigues}, O. 1840, Journal de Math\'matiques Pures et Appliqu\'ees, 5, 380

\bibitem[{{Salpeter}(1955)}]{Salpeter55}
{Salpeter}, E.~E. 1955, \apj, 121, 161

\bibitem[{{Schlafly} \& {Finkbeiner}(2011)}]{Schlafly&Finkbeiner11}
{Schlafly}, E.~F. \& {Finkbeiner}, D.~P. 2011, \apj, 737, 103

\bibitem[{{Schlegel} {et~al.}(1998){Schlegel}, {Finkbeiner}, \&
  {Davis}}]{Schlegel+98}
{Schlegel}, D.~J., {Finkbeiner}, D.~P., \& {Davis}, M. 1998, \apj, 500, 525

\bibitem[{{Schmidt}(1963)}]{Schmidt63}
{Schmidt}, M. 1963, \nat, 197, 1040

\bibitem[{{Schwarz}(1978)}]{Schwarz78}
{Schwarz}, G. 1978, Annals of Statistics, 6, 461

\bibitem[{{S\'ersic}(1963)}]{Sersic63}
{S\'ersic}, J.~L. 1963, Bull. Assoc. Argentina de Astron., 6, 41

\bibitem[{{Sersic}(1968)}]{Sersic68}
{Sersic}, J.~L. 1968, Atlas de galaxias australes (Cordoba, Argentina:
  Observatorio Astronomico)

\bibitem[{{Sesana} {et~al.}(2020){Sesana}, {Lamberts}, \&
  {Petiteau}}]{Sesana+20}
{Sesana}, A., {Lamberts}, A., \& {Petiteau}, A. 2020, \mnras, 494, L75

\bibitem[{{Shao} \& {Li}(2019)}]{Shao&Li19}
{Shao}, Z. \& {Li}, L. 2019, \mnras, 489, 3093

\bibitem[{{Shin} {et~al.}(2013){Shin}, {Kim}, \& {Lee}}]{Shin+13}
{Shin}, J., {Kim}, S.~S., \& {Lee}, Y.-W. 2013, Journal of Korean Astronomical
  Society, 46, 173

\bibitem[{{Silverman}(1986)}]{Silverman86}
{Silverman}, B.~W. 1986, Density estimation for statistics and data analysis

\bibitem[{{Simonneau} \& {Prada}(2004)}]{Simonneau&Prada04}
{Simonneau}, E. \& {Prada}, F. 2004, \rmxaa, 40, 69

\bibitem[{{Sollima} {et~al.}(2019){Sollima}, {Baumgardt}, \&
  {Hilker}}]{Sollima+19}
{Sollima}, A., {Baumgardt}, H., \& {Hilker}, M. 2019, \mnras, 485, 1460

\bibitem[{{Spera} {et~al.}(2015){Spera}, {Mapelli}, \& {Bressan}}]{Spera+15}
{Spera}, M., {Mapelli}, M., \& {Bressan}, A.~r. 2015, \mnras, 451, 4086

\bibitem[{{Strigari} {et~al.}(2007){Strigari}, {Bullock}, \&
  {Kaplinghat}}]{Strigari+07}
{Strigari}, L.~E., {Bullock}, J.~S., \& {Kaplinghat}, M. 2007, \apjl, 657, L1

\bibitem[{{Sugiura}(1978)}]{Sugiyara78}
{Sugiura}, N. 1978, Communications in Statistics - Theory and Methods, 7, 13

\bibitem[{{Takahashi}(1995)}]{Takahashi95}
{Takahashi}, K. 1995, \pasj, 47, 561

\bibitem[{{Taylor}(2005)}]{Taylor05}
{Taylor}, M.~B. 2005, Astronomical Society of the Pacific Conference Series,
  Vol. 347, TOPCAT \& STIL: Starlink Table/VOTable Processing Software, ed.
  P.~{Shopbell}, M.~{Britton}, \& R.~{Ebert}, 29

\bibitem[{{The LIGO Scientific Collaboration} {et~al.}(2020){The LIGO
  Scientific Collaboration}, {the Virgo Collaboration}, {Abbott}, {Abbott},
  {Abraham}, {Acernese}, {Ackley}, {Adams}, {Adhikari}, {Adya}, \&
  et~al.}]{TheLIGOScientificCollaboration+20}
{The LIGO Scientific Collaboration}, {the Virgo Collaboration}, {Abbott}, R.,
  {et~al.} 2020, arXiv e-prints, arXiv:2009.01075

\bibitem[{{Thompson} {et~al.}(2020){Thompson}, {Kochanek}, {Stanek}, {Badenes},
  {Jayasinghe}, {Tayar}, {Johnson}, {Holoien}, {Auchettl}, \&
  {Covey}}]{Thompson+20}
{Thompson}, T.~A., {Kochanek}, C.~S., {Stanek}, K.~Z., {et~al.} 2020, Science,
  368, eaba4356

\bibitem[{{Tiongco} {et~al.}(2016){Tiongco}, {Vesperini}, \&
  {Varri}}]{Tiongco+16}
{Tiongco}, M.~A., {Vesperini}, E., \& {Varri}, A.~L. 2016, \mnras, 455, 3693

\bibitem[{{Tiret} {et~al.}(2007){Tiret}, {Combes}, {Angus}, {Famaey}, \&
  {Zhao}}]{Tiret+07}
{Tiret}, O., {Combes}, F., {Angus}, G.~W., {Famaey}, B., \& {Zhao}, H.~S. 2007,
  \aap, 476, L1

\bibitem[{{Trager} {et~al.}(1995){Trager}, {King}, \& {Djorgovski}}]{Trager+95}
{Trager}, S.~C., {King}, I.~R., \& {Djorgovski}, S. 1995, \aj, 109, 218

\bibitem[{{Tremou} {et~al.}(2018){Tremou}, {Strader}, {Chomiuk}, {Shishkovsky},
  {Maccarone}, {Miller-Jones}, {Tudor}, {Heinke}, {Sivakoff}, {Seth}, \&
  {Noyola}}]{Tremou+18}
{Tremou}, E., {Strader}, J., {Chomiuk}, L., {et~al.} 2018, \apj, 862, 16

\bibitem[{{Valcin} {et~al.}(2020){Valcin}, {Bernal}, {Jimenez}, {Verde}, \&
  {Wand elt}}]{Valcin+20}
{Valcin}, D., {Bernal}, J.~L., {Jimenez}, R., {Verde}, L., \& {Wand elt}, B.~D.
  2020, \jcap, 2020, 002

\bibitem[{{van der Marel} \&
  {Anderson}(2010)}]{vanderMarel&Anderson&Anderson10}
{van der Marel}, R.~P. \& {Anderson}, J. 2010, \apj, 710, 1063

\bibitem[{{van der Walt} {et~al.}(2011){van der Walt}, {Colbert}, \&
  {Varoquaux}}]{vanderWalt11}
{van der Walt}, S., {Colbert}, S.~C., \& {Varoquaux}, G. 2011, Computing in
  Science Engineering, 13, 22

\bibitem[{{Vasiliev}(2019{\natexlab{a}})}]{Vasiliev19a}
{Vasiliev}, E. 2019{\natexlab{a}}, \mnras, 482, 1525

\bibitem[{{Vasiliev}(2019{\natexlab{b}})}]{Vasiliev19b}
{Vasiliev}, E. 2019{\natexlab{b}}, \mnras, 484, 2832

\bibitem[{{Vasiliev}(2019{\natexlab{c}})}]{Vasiliev19c}
{Vasiliev}, E. 2019{\natexlab{c}}, \mnras, 489, 623

\bibitem[{{Vesperini} {et~al.}(2013){Vesperini}, {McMillan}, {D'Antona}, \&
  {D'Ercole}}]{Vesperini+13}
{Vesperini}, E., {McMillan}, S. L.~W., {D'Antona}, F., \& {D'Ercole}, A. 2013,
  \mnras, 429, 1913

\bibitem[{{Vitral} \& {Mamon}(2020)}]{Vitral&Mamon20}
{Vitral}, E. \& {Mamon}, G.~A. 2020, \aap, 635, A20

\bibitem[{{Volonteri}(2010)}]{Volonteri10}
{Volonteri}, M. 2010, \aapr, 18, 279

\bibitem[{{Watkins} {et~al.}(2013){Watkins}, {van de Ven}, {den Brok}, \& {van
  den Bosch}}]{Watkins+13}
{Watkins}, L.~L., {van de Ven}, G., {den Brok}, M., \& {van den Bosch}, R.
  C.~E. 2013, \mnras, 436, 2598

\bibitem[{{Watkins} {et~al.}(2015{\natexlab{a}}){Watkins}, {van der Marel},
  {Bellini}, \& {Anderson}}]{Watkins+15a}
{Watkins}, L.~L., {van der Marel}, R.~P., {Bellini}, A., \& {Anderson}, J.
  2015{\natexlab{a}}, \apj, 803, 29

\bibitem[{{Watkins} {et~al.}(2015{\natexlab{b}}){Watkins}, {van der Marel},
  {Bellini}, \& {Anderson}}]{Watkins+15b}
{Watkins}, L.~L., {van der Marel}, R.~P., {Bellini}, A., \& {Anderson}, J.
  2015{\natexlab{b}}, \apj, 812, 149

\bibitem[{{Woosley}(2017)}]{Woosley17}
{Woosley}, S.~E. 2017, \apj, 836, 244

\bibitem[{{Zel'dovich} \& {Novikov}(1966)}]{Zeldovich&Novikov66}
{Zel'dovich}, Y.~B. \& {Novikov}, I.~D. 1966, \azh, 43, 758

\bibitem[{{Zocchi} {et~al.}(2019){Zocchi}, {Gieles}, \&
  {H{\'e}nault-Brunet}}]{Zocchi+19}
{Zocchi}, A., {Gieles}, M., \& {H{\'e}nault-Brunet}, V. 2019, \mnras, 482, 4713

\bibitem[{{Zocchi} {et~al.}(2016){Zocchi}, {Gieles}, {H{\'e}nault-Brunet}, \&
  {Varri}}]{Zocchi+16}
{Zocchi}, A., {Gieles}, M., {H{\'e}nault-Brunet}, V., \& {Varri}, A.~L. 2016,
  \mnras, 462, 696

\end{thebibliography}

\begin{appendix}

\section{Deprojection of the S\'ersic surface density profile}
\label{app: surf-dens}

The S\'ersic surface density model used in this work is deprojected in the same way as \citeauthor{Vitral&Mamon20} (\citeyear{Vitral&Mamon20}, hereafter VM20), but we extend the deprojection to lower radii: $10^{-4}\,R_{\rm e}$ instead of $10^{-3}\,R_{\rm e}$.
We consider again the approximations of \citeauthor{LimaNeto+99} (\citeyear{LimaNeto+99}, hereafter LGM99), \citeauthor{Simonneau&Prada04} (\citeyear{Simonneau&Prada04}, hereafter SP04), as well as the formula of VM20, to find which is better suited for each region of the [$n \times r/R_{\rm e}$] domain. 

\subsection{New coefficients for VM20 approximation extending to very low radii}

The coefficients of the VM20 approximation were originally calculated for a logarithmic grid of [$n \times r/R_{\rm e}$] with $-3 \leq \log(r/R_{\rm e}) \leq 3$ (100 steps) and $0.5 \leq n \leq 10$ (50 steps). However, our HST data could extend to even lower radii than $0.001\,R_{\rm e}$, if $R_e > 7'$ (the lowest projected radius is $0\farcs42$). Indeed, such large effective radii were found for the faint stars of the two-population models~\CUOisoDouble\ and \CUOfreebetaDouble.
We therefore recomputed the VM20 approximation, using a different region of the [$n \times r/R_{\rm e}$] domain, and an even finer grid.

We used the same approach as we did in VM20, but this time we performed the numerical deprojection of the S\'ersic profile with {\sc Mathematica 12}, instead of {\sc Python}. Given figure~4 of VM20 and the lower limit of $r/R_{\rm e}$ we needed to attain, we calculated the best VM20 parameters for a new logarithmic spaced region [$n \times r/R_{\rm e}$] limited to $-4 \leq \log(r/R_{\rm e}) \leq 3$ (150 steps) and $0.5 \leq n \leq 3.5$ (100 steps). 
The resulting coefficients $a_{i,j}$, presented in the same way as VM20 (see their Eq.~[28]), can be found online\footnote{\url{https://gitlab.com/eduardo-vitral/vitral_mamon_2020b}.} and we hereafter refer to them as VM20bis.

\subsection{Choice of deprojection approximation}
\label{app-ssec: choice-of-dep}

\begin{figure*}
\centering
\includegraphics[width=0.95\hsize]{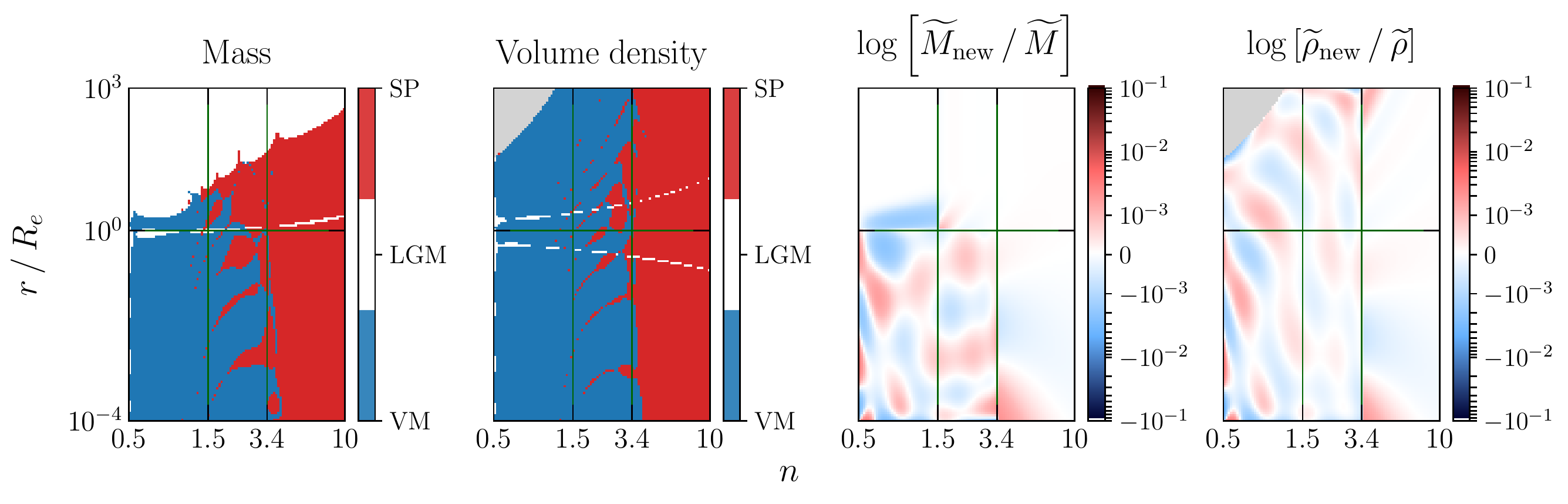}
\caption{Characteristics of approximations to the mass and density profiles of the deprojected S\'ersic model.
\textbf{Left two panels}: Most precise approximation.
SP stands for \protect\cite{Simonneau&Prada04}, LGM stands for \protect\cite{LimaNeto+99} and VM stands for the new VM20bis coefficients applied to the \protect\cite{Vitral&Mamon20} method.
The white curves indicate a thin region preferred by the LGM approximation.
\textbf{Right two panels}: Accuracy of deprojected mass (\emph{left}) and density (\emph{right}) of the hybrid model using VM20bis coefficients, LGM99 and SP04, with respect to the numerical integration done with {\sc Mathematica}. This is the analog of figure~3 of VM20: The color scale is linear for log ratios between --0.001 and 0.001 and logarithmic beyond. 
Both sets of figures employ a [$100 \times 150$] grid of [$\log n \times \log(r/R_{\rm e})$], which is shown in all four panels. The \emph{gray region} in the upper left of each of the density panels is for regions where the numerical integration of {\sc Mathematica} reached the underflow limit because of the very rapid decline of density at large radii for low $n$.}
\label{fig: app-best-approx}
\end{figure*}

The two left panels of Figure~\ref{fig: app-best-approx} display the best fitting approximations in  $100\times 150$ grid in [$\log\,n \times \log\,(r/R_{\rm e})$].
%
We included in \mpo\ a simplified choice of best approximations to the deprojected S\'ersic mass and density profiles. For the mass profile we used
%
\begin{itemize}
    \item $n \times r/R_{\rm e} \in [0.5;1.5] \times [1;10^3]$: LGM99
    \item $n \times r/R_{\rm e} \in [0.5;3.4] \times [10^{-4};1)$: VM20bis
    \item $n \times r/R_{\rm e} \in (3.4;10] \times [10^{-4};1) \cup (1.5;10] \times [1;10^3]$: SP04
\end{itemize}
For the density profile, the division was even simpler:
\begin{itemize}
    \item $n \times r/R_{\rm e} \in (3.4;10] \times [10^{-4};10^3]$: SP04
    \item $n \times r/R_{\rm e} \in [0.5;3.4] \times [10^{-4};10^3]$: VM20bis
\end{itemize}
The outcome of this approximation is highly accurate, as seen in the right panels of Figure~\ref{fig: app-best-approx}  (which are the equivalent of figure~3 in VM20), with the grid lines indicating the divisions adopted  for the 3 approximations. The reader can verify that the relative precision of this hybrid model is on the order of $\sim 0.1\%$.

\section{Handling of field stars}
\label{app: FS}

\begin{figure*}
\centering
\includegraphics[width=0.95\hsize]{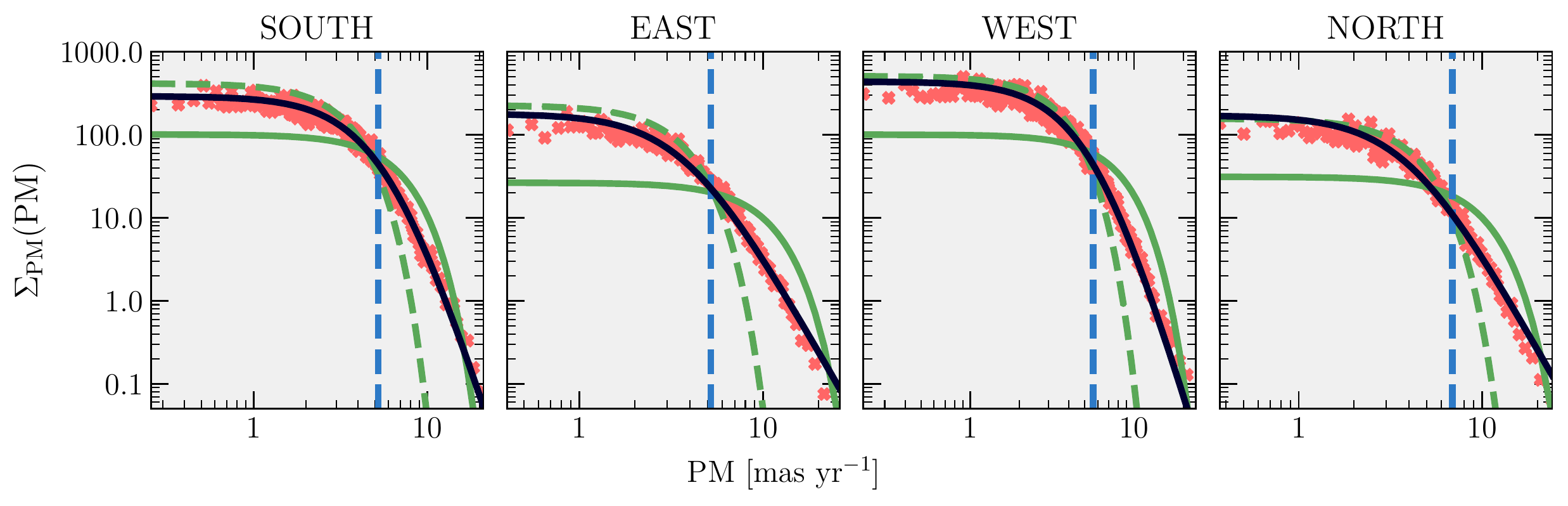}
\caption{
Surface density of  proper motion moduli (defined in eq.~[\ref{eq: PMdef}]) for the four 5 degrees distant regions around  NGC~6397 (for simplicity, we call them SOUTH, EAST, WEST and NORTH), represented by \emph{red crosses}. \emph{Solid green curves} display the best MLE fit for a Gaussian distribution, while \emph{dashed green curves} display the best Gaussian MLE fit when only considering the regions with proper motions inside the limit of the \emph{dashed blue vertical line}. The best MLE fit using Eq.~(\ref{eq: PMpdf}) is displayed as the \emph{black curves}.
}
\label{fig: app-PMpdf}
\end{figure*}

\subsection{Proper motion distribution function}
\label{app: PMpdf}

In this appendix, we justify the need for using wider tails than a Gaussian for 
the distribution of PMs in the field stars, as  mentioned in section~\ref{sssec: pm-filter}.


We downloaded {\sc Gaia DR2} PM data in four regions around NGC~6397 (and also for two other GCs, M4 and NGC 6752, in order to check for generality). These four regions were chosen by doing a cone search, with a $30'$ aperture, for positions 5 degrees distant from the GC centers ($\alpha_{\rm GC}$, $\delta_{\rm GC}$), north, south, east, and west.
%
We applied the same quality flags that we had applied to the \ngc\ data (Sect.~\ref{sssec: Gaia-quality}): the inequalities of Eqs.~(\ref{eq: err_lind18}), (\ref{eq: Gaia_flag1}) and (\ref{eq: Gaia_flag2}).

We then 
fitted the distribution of PM moduli using  both a Gaussian and the form of Eq.~(\ref{eq: PMpdf}). 
We estimated the mean $\mu_{\alpha,*}$ and mean $\mu_{\delta}$ for both distribution functions, a dispersion $\sigma$ for the Gaussian assumption and for Eq.~(\ref{eq: PMpdf}), we estimated the scale radius and outer slope. Finally, we took into account the convolution of both distributions with Gaussian errors.

Figure~\ref{fig: app-PMpdf} shows the distribution of PM moduli, for the four regions  NGC~6397.
One can easily verify that the new expression fits extremely better than a Gaussian in Figure~\ref{fig: app-PMpdf}. Moreover, the calculated {kurtosis excess} of both $\mu_{\alpha,*}$ and $\mu_{\delta}$ always gave huge values (from 16 to over 400, compared to 0 expected for a Gaussian), which clearly implies a non-Gaussian behavior.

To check for robustness of our fits, we also verified that whenever limiting the fitting range to exclude the wider wings of the FS distribution (which cannot be done when considering a GC relatively separated from the FS in PM space), a Gaussian distribution is well fitted. Thus, we decided to employ Eq.~(\ref{eq: PMpdf}) throughout our study in order to account for the entire PM range.

\subsection{Convolution of field stars distribution}
\label{app: conv-MPOPM}

For the GC component, the local Gaussian-like {velocity distribution function} (VDF)
must be convolved by the Gaussian LOS and POS velocity errors, as in Eq.~(\ref{eq: hvi}).
Similarly, since the field star LOS VDF is a Gaussian, the 
LOS velocity errors are added in quadrature to the LOS velocity dispersion.

On the other hand, since the interloper PM distribution function (eq.~[\ref{eq: PMpdf}]) is not Gaussian-like shaped, we need to perform the integral of its convolution.
Calling $R$ and $R_{\rm o}$ the respective true and observed PMs and  
$\epsilon$ the PM error, it is straightforward to follow the recipe of \citeauthor{Binney&Mamon82} (\citeyear{Binney&Mamon82}, appendix
C) for the convolution of two-dimensional data with circular symmetry:
\begin{equation}
   p_{\rm conv}(R_{\rm o}) = {R_{\rm o}\over \epsilon^2}\,\int_0^\infty \d R
 \,  p(R)\,
   \exp \left (-{R^2+R_{\rm o}^2\over 2\epsilon^2}\right)\,I_0\left
   ({R\,R_{\rm o}\over \epsilon^2}\right) \ .
   \label{eq: pconvBM82}
\end{equation}
Mamon \& Vitral (in prep.) provide an approximation for this integral.

\end{appendix}

\end{document}